\documentclass[fleqn,usenatbib]{mnras}

\usepackage{newtxtext,newtxmath}

\usepackage[T1]{fontenc}

\DeclareRobustCommand{\VAN}[3]{#2}
\let\VANthebibliography\thebibliography
\def\thebibliography{\DeclareRobustCommand{\VAN}[3]{##3}\VANthebibliography}

\usepackage{graphicx}	
\usepackage{adjustbox}  
\usepackage{amsmath}	
\usepackage{wasysym}    
\usepackage{bm}         
\usepackage{etoolbox}
\usepackage{float}
\makeatletter 
  \patchcmd{\NAT@citex}
    {\@citea\NAT@hyper@{%
      \NAT@nmfmt{\NAT@nm}%
      \hyper@natlinkbreak{\NAT@aysep\NAT@spacechar}{\@citeb\@extra@b@citeb}%
      \NAT@date}}
    {\@citea\NAT@nmfmt{\NAT@nm}%
    \NAT@aysep\NAT@spacechar\NAT@hyper@{\NAT@date}}{}{}

  \patchcmd{\NAT@citex}
    {\@citea\NAT@hyper@{%
      \NAT@nmfmt{\NAT@nm}%
      \hyper@natlinkbreak{\NAT@spacechar\NAT@@open\if*#1*\else#1\NAT@spacechar\fi}%
        {\@citeb\@extra@b@citeb}%
      \NAT@date}}
    {\@citea\NAT@nmfmt{\NAT@nm}%
    \NAT@spacechar\NAT@@open\if*#1*\else#1\NAT@spacechar\fi\NAT@hyper@{\NAT@date}}
    {}{}
\makeatother

\newcommand\Msun{\text{M}_{\astrosun}} 
\newcommand\Zsun{\text{Z}_{\astrosun}} 
\newcommand\HI{\ion{H}{I}~} 
\newcommand\HII{\ion{H}{II}~} 
\newcommand\HeI{\ion{He}{I}} 
\newcommand\HeII{\ion{He}{II}} 
\newcommand\HeIII{\ion{He}{III}} 
\newcommand\orcid[1]{\href{http://orcid.org/#1}{\adjustbox{trim={-.15\width} {0\height} {-.15\width} {0\height},clip}{\includegraphics[height=10pt]{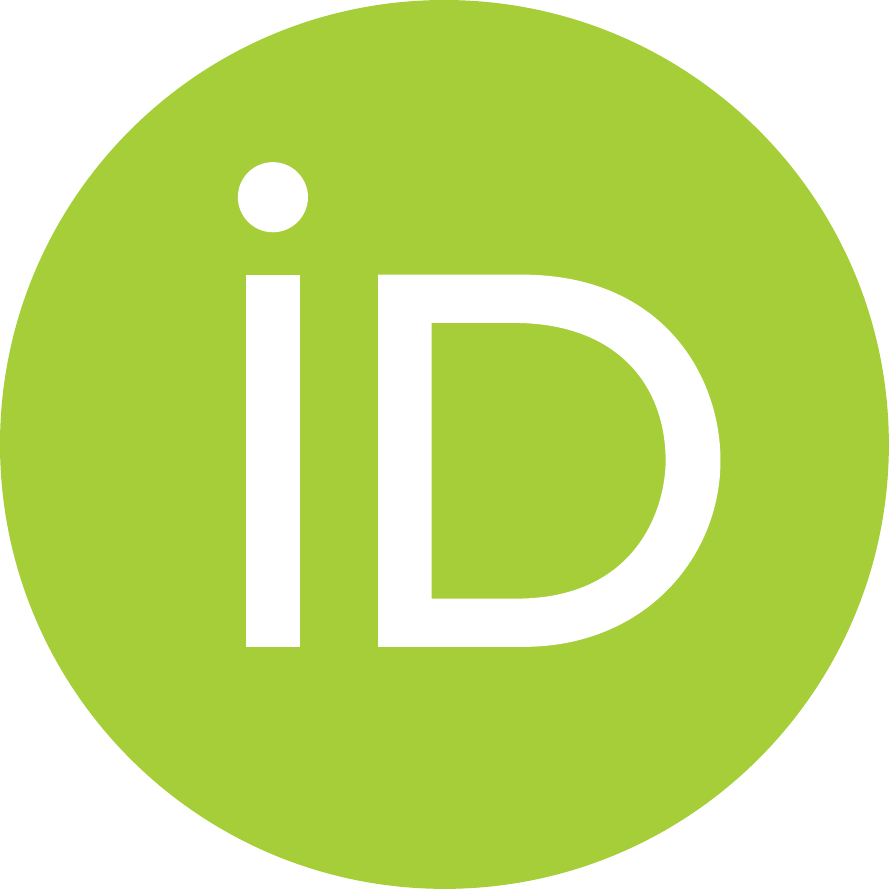}}}}

\title[H$\alpha$ emission in local galaxies]{H$\balpha$ emission in local galaxies: star formation, time variability, and \\the diffuse ionized gas}

\author[Tacchella et al.]{%
Sandro Tacchella\orcid{0000-0002-8224-4505},$^{1}$\thanks{E-mail: \href{mailto:tacchella@unist.ac.kr}{tacchella@unist.ac.kr}}
Aaron Smith\orcid{0000-0002-2838-9033},$^{2}$\thanks{E-mail: \href{mailto:arsmith@mit.edu}{arsmith@mit.edu}; NHFP Einstein Fellow.}
Rahul Kannan\orcid{0000-0001-6092-2187},$^3$\thanks{E-mail: \href{mailto:rahul.kannan@cfa.harvard.edu}{rahul.kannan@cfa.harvard.edu}}
Federico~Marinacci\orcid{0000-0003-3816-7028},$^{4}$
\newauthor
Lars~Hernquist,$^{3}$
Mark~Vogelsberger\orcid{0000-0001-8593-7692},$^{2}$
Paul~Torrey\orcid{0000-0002-5653-0786},$^{5}$
Laura~Sales\orcid{0000-0002-3790-720X}$^{6}$
and
Hui~Li\orcid{0000-0002-1253-2763}$^{7}$\thanks{NHFP Hubble Fellow.}
\\
$^{1}$Department of Physics, Ulsan National Institute of Science and Technology (UNIST), Ulsan 44919, Republic of Korea \\
$^{2}$Center for Astrophysics $\vert$ Harvard \& Smithsonian, 60 Garden St, Cambridge, MA 02138, USA \\
$^{3}$Department of Physics, Massachusetts Institute of Technology, Cambridge, MA 02139, USA \\
$^{4}$Department of Physics and Astronomy ``Augusto Righi'', University of Bologna, via Gobetti 93/2, 40129 Bologna, Italy \\
$^{5}$Department of Astronomy, University of Florida, 211 Bryant Space Sciences Center, Gainesville, FL 32611, USA \\
$^{6}$Department of Physics and Astronomy, University of California, Riverside, 900 University Avenue, Riverside, CA 92521, USA \\
$^{7}$Department of Astronomy, Columbia University, New York, NY 10027, USA
}

\date{Draft version of \today}

\pubyear{2022}

\begin{document}
\label{firstpage}
\pagerange{\pageref{firstpage}--\pageref{lastpage}}
\maketitle

\begin{abstract}
The nebular recombination line H$\alpha$ is widely used as a star-formation rate (SFR) indicator in the local and high-redshift Universe. We present a detailed H$\alpha$ radiative transfer study of high-resolution isolated Milky-Way and Large Magellanic Cloud simulations that include radiative transfer, non-equilibrium thermochemistry, and dust evolution. We focus on the spatial morphology and temporal variability of the H$\alpha$ emission, and its connection to the underlying gas and star formation properties. The H$\alpha$ and H$\beta$ radial and vertical surface brightness profiles are in excellent agreement with observations of nearby galaxies. We find that the fraction of H$\alpha$ emission from collisional excitation amounts to $f_{\rm col}\sim5$--$10\%$, only weakly dependent on radius and vertical height, and that scattering boosts the H$\alpha$ luminosity by $\sim40\%$. The dust correction via the Balmer decrement works well (intrinsic H$\alpha$ emission recoverable within 25\%), though the dust attenuation law depends on the amount of attenuation itself both on spatially resolved and integrated scales. Important for the understanding of the H$\alpha$--SFR connection is the dust and helium absorption of ionizing radiation (Lyman continuum [LyC] photons), which are about $f_{\rm  abs}\approx28\%$ and $f_{\rm He}\approx9\%$, respectively. Together with an escape fraction of $f_{\rm esc}\approx6\%$, this reduces the available budget for hydrogen line emission by nearly half ($f_{\rm H}\approx57\%$). We discuss the impact of the diffuse ionized gas, showing -- among other things -- that the extraplanar H$\alpha$ emission is powered by LyC photons escaping the disc. Future applications of this framework to cosmological (zoom-in) simulations will assist in the interpretation of spectroscopy of high-redshift galaxies with the upcoming \textit{James Webb Space Telescope}.
\end{abstract}




\section{Introduction}
\label{sec:intro}

The Balmer lines are some of the most utilised and observed emission lines in astrophysics since they are in the optical and relatively ubiquitous as they arise from recombination to the $n=2$ level of hydrogen, the most common element. In particular, the H$\alpha$ emission line is one of the prime indicators for the star-formation rate (SFR) of and within galaxies as H$\alpha$ traces the ionized gas of star-forming \HII regions. H$\alpha$ is used to measure the SFRs locally and out to redshift of $z\sim2.5$ both on global \citep[][]{kennicutt83, lee09_UVHa, koyama15, shivaei15_SFR} and spatially resolved scales \citep[e.g.,][]{tacchella15, tacchella18_dust, nelson16_insideout, belfiore18, ellison18}. The upcoming \textit{James Webb Space Telescope} (\textit{JWST}) will extend this redshift limit to $z\sim7$. Furthermore, H$\alpha$ is frequently used to shed light on the variability of the star-formation activity in galaxies since it traces younger stars than other SFR tracers such as the UV continuum \citep[e.g.,][]{weisz12, guo16, caplar19, faisst19, haydon20}.

Although H$\alpha$ is a well-calibrated SFR tracer \citep[e.g.,][]{shivaei15_SFR}, uncertainties remain due to dust attenuation and emission from non-\HII regions. In principle, one can derive the amount of dust attenuation of H$\alpha$ ($A_{\rm H\alpha}$) from the difference of the observed H$\alpha$/H$\beta$ ratio (i.e. the Balmer decrement) relative to the intrinsic H$\alpha$/H$\beta$ ratio, which itself only weakly depends on local conditions (i.e. electron density and temperature of the \HII region) and therefore is well-determined from first principles. Therefore, the Balmer decrement is the method of choice for correcting the observed H$\alpha$ emission for dust attenuation \citep{berman36, calzetti94, groves12, nelson16_balmer}, with the main uncertainty for estimating $A_{\rm H\alpha}$ arising from the adopted attenuation law. 

In addition to aforementioned \HII regions, diffuse ionized gas (DIG; sometimes also called the warm ionized gas) can also emit H$\alpha$. Narrow-band H$\alpha$ imaging surveys suggest that the DIG emission contributes $20$--$60\%$ of the total H$\alpha$ flux in local spiral galaxies \citep{zurita00, oey07, kreckel16, chevance20} and therefore can bias H$\alpha$-based SFRs as well as gas-phase metallicity estimates \citep[e.g.,][]{sanders17, poetrodjojo19, vale-asari19}. Furthermore, the DIG -- as traced by H$\alpha$ -- typically extends vertically above the plane of the disc to scales of about $\sim1$\,kpc \citep{hoyle63, reynolds89, rand90, jo18, levy19}. Two possibilities have been suggested for the origin of this diffuse, extraplanar H$\alpha$ emission. Firstly, the extraplanar H$\alpha$ emission traces extraplanar DIG, which itself is produced by ionizing photons transported through transparent pathways carved out by superbubbles or chimneys \citep{mac-low99, veilleux05}. Secondly, the extraplanar H$\alpha$ emission is caused by dust scattering of the photons originating from \HII regions in the galactic disc \citep{reynolds90, ferrara96}.

Theoretically, it is still challenging to self-consistently model -- on scales of entire galaxies -- the detailed structure of the multi-phase interstellar medium (ISM), including massive stars and supernovae in a radiation hydrodynamical context \citep{rosdahl15, kannan20_mw, vogelsberger20_review}. A handful of theoretical investigations have focused on the production and transport of H$\alpha$ photons \citep[and similar emission lines, e.g.,][]{katz19, wilkins20} with various degrees of sophistication and self-consistency, e.g., while also contending with simulation resolution effects. Previous studies include an estimate of the importance of scattered light in the diffuse H$\alpha$ galactic background \citep{wood99, barnes15}, an investigation of the star formation relation (without dust) from simulations of isolated dwarfs and high-redshift galaxies \citep{kim13_sfr, kim19}, sub-resolution population synthesis for a Milky-Way like galaxy \citep{pellegrini20emp, pellegrini20}, and periodic tall box simulations to explore sub-parsec scale feedback and emission \citep{peters17, kado-fong20}. Our work presented here further expands on this class of detailed emission line modelling. The novelty of our approach is that H$\alpha$ is self-consistently reprocessed from ionizing photons and that LyC and H$\alpha$ are self-consistently scattered/absorbed by dust throughout entire galaxies. In other words, H$\alpha$ is emitted from ionized/excited gas rather than assuming an SFR--H$\alpha$ conversion from \HII regions from young star particles.

Specifically, in this paper we use the high-resolution Milky-Way (MW) and Large Magellanic Cloud (LMC) simulations from \citet{kannan20_mw}, which combines the state-of-the-art \textsc{arepo-rt} code \citep{kannan19_rt} with a non-equilibrium thermochemistry module that accounts for molecular hydrogen (H$_2$) coupled to explicit dust formation and destruction. This is all integrated into a novel stellar feedback framework, the Stars and MUltiphase Gas in GaLaxiEs (SMUGGLE) feedback model \citep{marinacci19}. We employ the Cosmic Ly$\alpha$ Transfer code \citep[\textsc{colt};][]{smith15, smith19} to perform post-processing Monte Carlo radiative transfer (MCRT) calculations for ionizing radiation and Ly$\alpha$, H$\alpha$ and H$\beta$ emission lines. The details of the MCRT calculations and insights into the physics of Ly$\alpha$ escape from disc-like galaxies are presented in our companion study \citep{smith21_rt}.

In this work, we carefully analyse these detailed numerical radiative transfer simulations to shed more light on the emission, absorption, and scattering of Balmer H$\alpha$ and H$\beta$ photons. In particular, we focus on answering questions related to the source properties (collisional excitation versus recombination emission), transport or radiation (dust scattering and absorption), importance and origins of the diffuse ionized gas (including extraplanar emission), and the connection between H$\alpha$ and star formation (including SFR timescales). Furthermore, we are looking forward to future applications of our methodology, including cosmological simulations based on the same framework in conjunction with the \textsc{thesan} project \citep{kannan21, garaldi21, smith21}. This will help improve our understandings of the physics and observational interpretations for star-forming galaxies in the high-redshift Universe.

We introduce the simulations and the performed MCRT calculations in Section~\ref{sec:methods}. We then show in Section~\ref{sec:observations} that our simulation produces realistic spatial distributions of H$\alpha$ and H$\beta$ emission by comparing to measurements of local galaxies. The key results concerning the emission, scattering, and absorption of the Balmer emission lines are described in Section~\ref{sec:balmer_emission_diag}. We focus on H$\alpha$ as a SFR tracer in Section~\ref{sec:sfr} by discussing the Balmer decrement for performing dust attenuation corrections and the timescale of the H$\alpha$ SFR indicator. Finally, we discuss limitations and future prospects in Section~\ref{sec:discussion}, before concluding in Section~\ref{sec:conclusions}.
Throughout this paper, we employ a \citet{chabrier03} initial mass function (IMF). Luminosities, masses, and SFRs are calculated assuming a Planck2015 cosmology \citep{planck-collaboration16}.

\section{Methods}
\label{sec:methods}

\begin{figure*}
\begin{center}
\includegraphics[width=\textwidth]{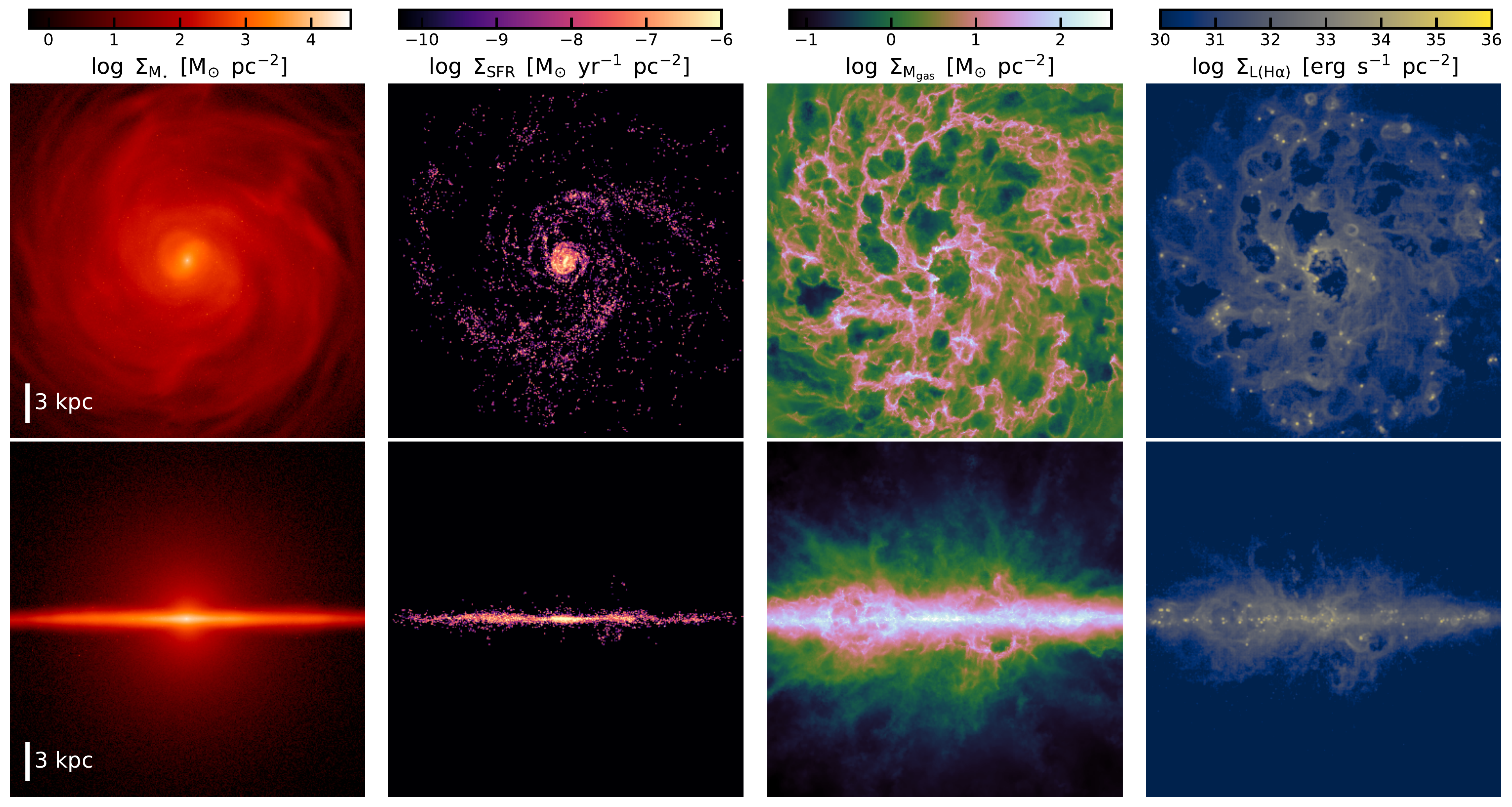}
\caption{MW simulation at 713 Myr. Maps of stellar mass, SFR (averaged over the past 100\,Myr), gas mass and H$\alpha$ luminosity are shown from left to right. The stellar mass, SFR, and H$\alpha$ are smoothed by 50 pc in order to increase visibility. The top and bottom panels show face-on and edge-on projections, respectively. The dimensions of each map are $30\times30$ kpc (the ruler in the bottom left indicates 3 kpc). The star formation is concentrated at the centre and spiral arms, while the H$\alpha$ emission traces the young, compact \HII regions and supernova bubbles. The corresponding LMC-BC03 maps are shown in Appendix~\ref{appsec:LMC_maps}.}
\label{fig:MW_maps}
\end{center}
\end{figure*}

We describe in Section~\ref{subsec:simulation} the simulations, which are based on a novel framework to self-consistently model the effects of radiation fields, dust physics and molecular chemistry (H$_2$) in the ISM of galaxies. In particular, we focus on idealised simulations of MW-like and LMC-like galaxies. Section~\ref{subsec:rt} gives a brief overview of the MCRT calculations employed in order to predict the H$\alpha$ and H$\beta$ emission from the aforementioned simulations. Further details on the MCRT methodology and the Ly$\alpha$ emission of those simulations are presented in our companion paper \citep{smith21_rt}. To illustrate the typical properties of the isolated MW simulation, Fig.~\ref{fig:MW_maps} shows projected images of the MW galaxy at 713 Myr for both face-on and edge-on views of the stellar mass surface density, SFR surface density, gas mass surface density, and H$\alpha$ surface brightness. 

\subsection{Isolated MW and LMC simulations}
\label{subsec:simulation}

\begin{table}
    \setlength{\tabcolsep}{5pt}
    \renewcommand{\arraystretch}{1.5}
    \caption{Median properties of the MW, LMC-BC03 and LMC-BPASS simulations. The $M_{\star}$, $M_{\rm gas}$ and SFR are measured within an aperture of 15 kpc. The SFR$_{50}$ is measured over 50 Myr (stars formed in the past 50 Myr). The effective radius $R_{\rm eff}$ is the half-mass size. }\label{tab:simulation}
    \centering 
        \begin{tabular}{lccccc}
          \hline
          Simulation & $M_{\rm halo}$ & $M_{\star}$ & $M_{\rm gas}$ & SFR$_{50}$ & $R_{\rm eff}$ \\ 
           & $[\text{M}_{\odot}]$ & $[\text{M}_{\odot}]$ & $[\text{M}_{\odot}]$ & $[\text{M}_{\odot}/\mathrm{yr}]$ & [kpc] \\ 
          \hline
          MW  & $1.5\times10^{12}$ & $6.2\times10^{10}$ & $4.2\times10^{9}$ & 2.7 & 4.4 \\ 
          LMC-BC03 & $1.1\times10^{11}$ & $3.5\times10^{9}$ & $5.2\times10^{8}$ & 0.043 & 2.2 \\
          LMC-BPASS & $1.1\times10^{11}$ & $3.5\times10^{9}$ & $5.2\times10^{8}$ & 0.041 & 2.2 \\  \hline
        \end{tabular}
\end{table}

We use the simulations presented in \citet{kannan20_mw} and \citet{kannan21_dust}. Specifically, we focus on high-resolution isolated simulations of a Milky-Way-like galaxy (MW; $M_{\rm halo}=1.5\times10^{12}~\text{M}_{\odot}$) and an LMC-like galaxy (LMC; $M_{\rm halo}=1.1\times10^{11}~\text{M}_{\odot}$). The simulations were performed with \mbox{\textsc{arepo-rt}} \citep{kannan19_rt}, a novel radiation hydrodynamic extension of the moving mesh hydrodynamic code \textsc{arepo} \citep{springel10, weinberger20}\footnote{Public code access and documentation available at \href{https://arepo-code.org}{\texttt{arepo-code.org}}.}. The adopted sub-grid models for star formation and feedback are described in \citet{marinacci19} and \citet{kannan20_mw}. Briefly, gas is allowed to cool down to $10$ K with the cooling function divided into primordial cooling from hydrogen (both molecular and atomic) and helium, metal cooling scaled linearly with the metallicity of the gas and cooling through gas-dust and radiation field interactions, in addition to photoelectric and photoheating from far ultraviolet (FUV) and Lyman continuum photons, respectively. Star particles are probabilistically formed from cold gas above a density threshold of $n=10^{3}~\mathrm{cm}^{-3}$. Additionally, the star forming gas cloud needs to be self-gravitating in order to form stars. There are three feedback mechanisms implemented related to stars: radiative feedback, stellar winds from young O, B and asymptotic giant branch (AGB) stars and supernova feedback. Photoheating, radiation pressure and photoelectric heating are modelled self-consistently through the radiative transfer scheme. Furthermore, the simulations employ a novel self-consistent dust formation and destruction model \citep{mckinnon16, mckinnon17}, which accounts for three distinct dust production channels: SNII, SNIa and AGB stars \citep{dwek98}. The dust is assumed to be dynamically coupled to the gas. The dust mass in the ISM increases due to the gas-phase elements colliding with existing grains \citep{dwek98} and decrease due to shocks from SN remnants \citep{mckee89} and sputtering in high temperature gas \citep{tsai95}. The resulting dust properties are presented in detail in \citet{kannan20_mw}. Briefly, the dust-to-gas ratio of the simulations is $0.01-0.014$ (the canonical MW value is 0.01), with a weak increase towards the central region in the galaxy, towards higher gas density, and towards lower temperature (their Figs. 9 and 10).

Important for this work, we assume the \citet[][hereafter BC03]{bruzual03} model as our fiducial stellar population synthesis (SPS) model. In order to see how this assumption affects our results, we also run the LMC model with the Binary Population and Spectral Synthesis (BPASS) model \citep[v2.2.1;][]{eldridge09, eldridge17}\footnote{See the project website at \href{https://bpass.auckland.ac.nz}{\texttt{bpass.auckland.ac.nz}}.}. We denote these runs as LMC-BC03 and LMC-BPASS. For all SPS models, we assume a \citet{chabrier03} IMF with a high-mass cutoff of $100~\mathrm{M}_{\odot}$. Importantly, we incorporate the different SPS models fully self-consistently, i.e. we take into account both the change of the production rate of ionizing photons with stellar age (Appendix~\ref{appsec:stellar_pop}) and the strength and timing of radiative feedback, stellar winds, and supernova feedback.

\begin{figure*}
\begin{center}
\includegraphics[width=\textwidth]{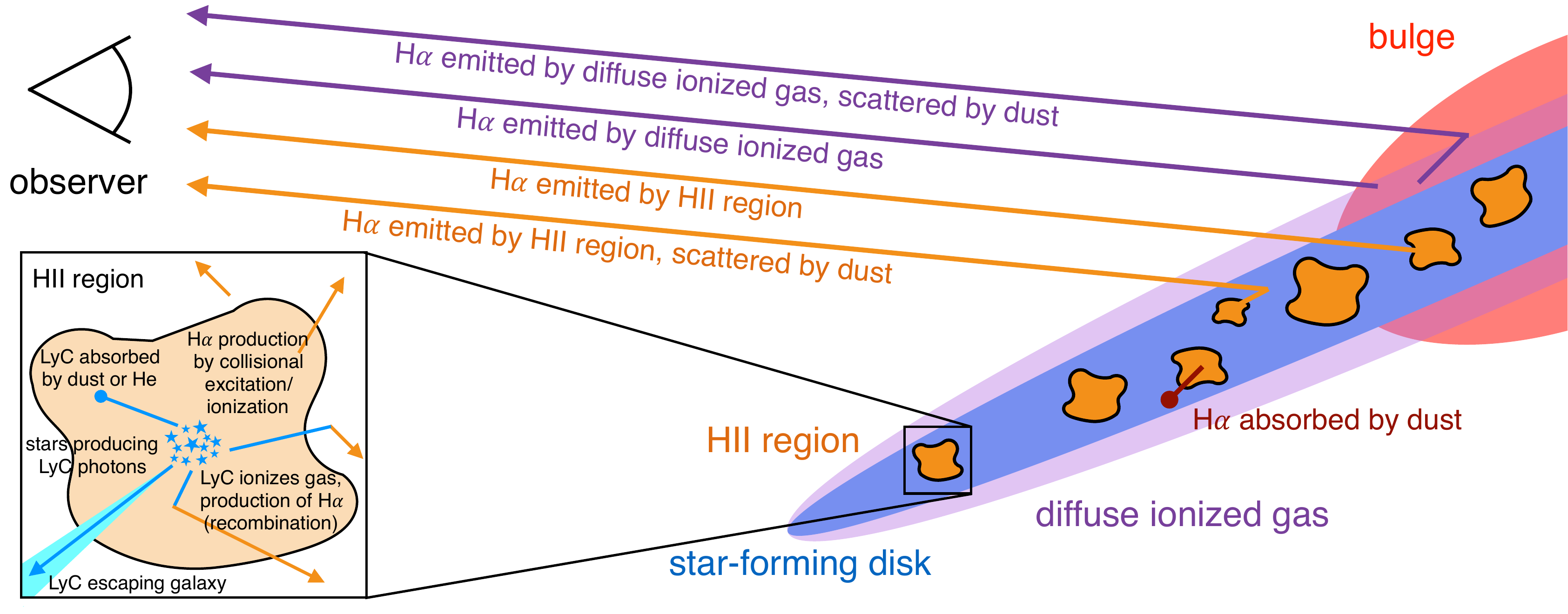}
\caption{Schema of the H$\alpha$ radiative transfer. H$\alpha$ that reaches the observer can either be emitted by nebular \ion{H}{II} regions or diffuse ionized gas (DIG). In this study we differentiate between \ion{H}{II} regions and the DIG with a simple gas density cut at $n_{\rm thresh}=100~\mathrm{cm}^{-3}$, though we also investigate other definitions and their implications (Section~\ref{subsec:DIG_def}). H$\alpha$ photons are produced by collisional excitation, by collisional ionization and by recombination of ionized gas, which is ionized by Lyman continuum (LyC) photons. Both LyC and H$\alpha$ photons can also be scattered or absorbed by dust. Furthermore, some LyC photons are consumed by He, leading to the photoionization of \HeI~ and \HeII. In the case of the MW simulation, on average $\sim28\%$ of the LyC photons are absorbed by dust, $\sim8\%$ ionize \HeI, $\sim1\%$ ionize \HeII, $\sim6\%$ escape the galaxy, implying that only about $f_{\rm H}\approx57\%$ of the produced LyC photons are available for ionizing hydrogen and the production of H$\alpha$ (Section~\ref{subsec:conversion_factor}).}
\label{fig:cartoon}
\end{center}
\end{figure*}

The MW and LMC simulations consist of a dark matter halo, a bulge, and a stellar and gaseous disc setup following the techniques described in \citet{hernquist93} and \citet{springel05}. The dark matter halo is modelled as a static background gravitational field that is not impacted by the baryonic physics. The full setup parameters are listed in Table~2 of \citet{kannan20_mw} but we outline the most relevant details here. The simulation box sizes for the MW and LMC simulations are 600\,kpc and 200\,kpc, respectively. The dark matter halo and the bulge are modelled as Hernquist profiles \citep{hernquist90}. The initial gas and the stellar discs are exponential profiles with effective radii of 6 (2.8) kpc and 3 (1.4) kpc for the MW (LMC) simulation, respectively. The vertical profile of the stellar disc follows a sech$^2$ functional form with a scale height of 300 (140) pc for the MW (LMC) simulation and initial stellar ages are taken to be 5\,Gyr to minimise spurious ionization. The initial gas fractions are 16 (19) per cent for the MW (LMC) and the distributions are computed self-consistently assuming hydrostatic equilibrium. The initial gas temperature is set to $10^4\,\text{K}$ and the initial metallicity to $1\,\Zsun$ ($0.5\,\Zsun$) for the MW (LMC) simulation. The production of new metals is turned off in order to suppress unrealistic gas metallicities, caused by the lack of cosmological gas inflow into the disc. The MW and LMC simulations are run with a stellar mass resolution of $2.8 \times 10^3\,\Msun$ and a gas mass resolution of $1.4 \times 10^3\,\Msun$. The corresponding gravitational softening lengths are $\varepsilon_{\star} = 7.1\,\text{pc}$ and $\varepsilon_\text{gas} = 3.6\,\text{pc}$, respectively. The simulations are each run for 1\,Gyr.

After starting the simulations, they settle into equilibrium after about 200 Myr \citep{kannan20_mw}. In this paper, we disregard these first 200 Myr when computing average or median properties across all snapshots if not otherwise stated. The median values of the key properties of the MW, LMC-BC03 and LMC-BPASS simulations are highlighted in Table~\ref{tab:simulation}. We find a SFR (averaged over the past 50~Myr) for the MW, LMC-BC03 and LMC-BPASS simulations of $2.7~\text{M}_{\odot}~\mathrm{yr}^{-1}$, $0.043~\text{M}_{\odot}~\mathrm{yr}^{-1}$ and $0.041~\text{M}_{\odot}~\mathrm{yr}^{-1}$, respectively. As we will show below, although these stellar and gas properties for the LMC-BC03 and LMC-BPASS model are basically indistinguishable, the Balmer emission from these two models are quite distinct.

\subsection{Radiative transfer of Balmer emission lines}
\label{subsec:rt}

We give a schematic overview of the H$\alpha$ radiative transfer in Fig.~\ref{fig:cartoon}. We employ the Cosmic Ly$\alpha$ Transfer code \citep[\textsc{colt}][]{smith15,smith19}\footnote{For public code access and documentation see \href{https://colt.readthedocs.io}{\texttt{colt.readthedocs.io}}.} to perform post-processing MCRT calculations, briefly summarising the most relevant details here. In particular, the dominant source of H$\alpha$ photons is via cascade recombination of recently ionized hydrogen atoms, with a small contribution of collisional excitation emission. To ensure accurate photon conserving ionization states for line radiative transfer we performed post-processing photoionization equilibrium calculations with \textsc{colt}, which was implemented as an MCRT module mirroring the physics of galaxy formation simulations \citep{rosdahl13_rt,kannan19_rt}. This is also helpful as the simulations do not fully resolve the temperature and density substructure of a fraction of the young \HII regions where line emission is especially strong (see the discussion in our companion paper; \citealt{smith21_rt}). We retain the gas temperature as this is already faithfully modelled but we iteratively recalculate the ionizing radiation field in three bands (\ion{H}{I}, \ion{He}{I}, and \ion{He}{II}) and update the ionization states assuming ionization equilibrium stopping this process when the global recombination emission is converged to within a $0.1$ per cent relative difference. The solver also includes dust absorption and anisotropic scattering, collisional ionization, and a meta-galactic UV background with self-shielding \citep{faucher-giguere09,rahmati13}. We employ $10^8$ photon packets in the MCRT calculations, which is adequate to represent the ionizing radiation field based on sampling from the age and metallicity-dependent stellar SEDs in terms of position, direction, and frequency. The \textsc{colt} output images capture the photon properties at a pixel resolution of 10 pc oriented in face-on and edge-on directions. This resolution was chosen as a compromise to be higher than observations while also not requiring too much data (already $6000^2$ pixels) or degrading the MCRT signal-to-noise ratio.

For the line radiative transfer we calculate the resolved H$\alpha$ and H$\beta$ luminosity caused by radiative recombination as
\begin{equation}
  L_X^\text{rec} = h \nu_X \int P_{\text{B},X}(T,n_e) \alpha_\text{B}(T)\,n_e n_p\,\text{d}V \, ,
\end{equation}
where $X \in \{\text{H}\alpha, \text{H}\beta\}$ denotes the line, $h \nu_X = \{1.89, 2.55\}$\,eV is the energy at line centre, $P_\text{B}$ is the conversion probability per recombination event (e.g., $P_{\text{B},X}(10^4\,\text{K}) \approx \{0.45, 0.12\}$), $\alpha_\text{B}$ is the case B recombination coefficient, and the number densities $n_e$ and $n_p$ are for free electrons and protons, respectively. We also calculate the resolved radiative de-excitation of collisional excitation of neutral hydrogen by free electrons as
\begin{equation}
  L_X^\text{col} = h \nu_X \int q_{\text{col},X}(T)\,n_e n_\text{\ion{H}{I}}\,\text{d}V \, ,
\end{equation}
where $q_{\text{col},X}$ is the collisional rate coefficient and $n_\text{H\,\textsc{i}}$ is the number density of neutral hydrogen.

\begin{figure*}
\begin{center}
\includegraphics[width=\textwidth]{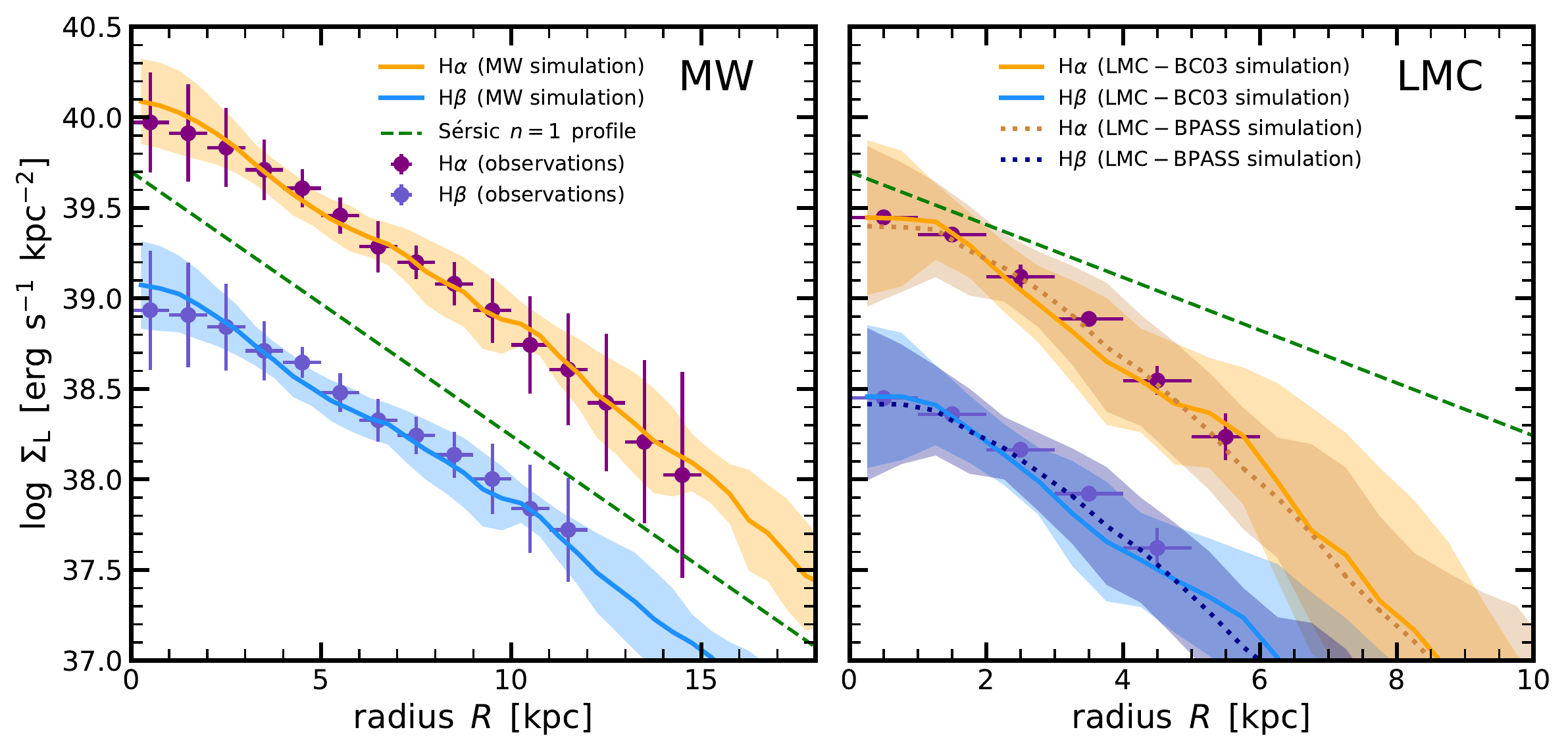}
\caption{Comparison of the predicted H$\alpha$ and H$\beta$ surface brightness profiles from the simulations with the observed ones from the MaNGA survey. The left panel shows the MW simulation run: The solid lines show the median of the simulations, while the points with errorbars mark the observational measurements. The right panel shows the results for the LMC-BC03 (solid lines) and LMC-BPASS (dotted lines) simulations. The shaded regions and the errorbars show the $16^{\rm th}$--$84^{\rm th}$ percentiles. The H$\alpha$ (H$\beta$) profiles have been normalised to a total luminosity of $10^{42}~\mathrm{erg}~\mathrm{s}^{-1}$ ($10^{41}~\mathrm{erg}~\mathrm{s}^{-1}$). In both panels, the green dashed line shows an arbitrarily normalised S\'{e}rsic profile with S\'{e}rsic index $n=1$, indicating that the H$\alpha$ and H$\beta$ profiles are roughly exponential in both observations and simulations. In both the MW and LMC case, the simulations are able to reproduce the observed profiles out to $10$--$15$\,kpc.}
\label{fig:profile_comparison}
\end{center}
\end{figure*}

Stellar continuum spectral luminosities $L_{\lambda,\text{cont}}$ for each line are tabulated by age and metallicity based on the SEDs around the reference wavelengths, which is included to enable self-consistent predictions for line equivalent width measurements. We emphasise that, since the metal enrichment is ignored, all stars have equal metallicity: $1.0~Z_{\odot}$ and $0.5~Z_{\odot}$ in the MW and the LMC case, respectively.
The dust distribution is also self-consistently taken from the simulation such that the local dust absorption coefficient is
\begin{equation}
  k_{\text{d},X} = \kappa_{\text{d},X} \mathcal{D} \rho \, ,
\end{equation}
where the dust opacity is $\kappa_{\text{d},X} = \{6627, 10220\}\,\text{cm}^2/\text{g}$ of dust and the dust-to-gas ratio is $\mathcal{D}$. Dust scattering is modelled based on the albedo $A = \{0.6741, 0.6650\}$ and anisotropic Henyey--Greenstein phase function with asymmetry parameter of $\langle \cos\theta \rangle = \{0.4967, 0.5561\}$, all of which are based on the fiducial Milky Way dust model from \citet{weingartner01}. With the $10^8$ photon packets in the line MCRT calculations, we are able to capture the escaping photon properties and high signal-to-noise images oriented in face-on and edge-on directions. For each of these lines and cameras we also calculate ray-tracing based images of the intrinsic and dust-attenuated line emission based on the adaptive convergence algorithm described in Appendix~A of \citet{yang20}. This method eliminates Monte Carlo noise at the expense of treating dust as purely absorbing; i.e. assuming an albedo of $A = 0$.

\section{Comparison to Observations}
\label{sec:observations}

Before discussing the physics of the predicted H$\alpha$ and H$\beta$ emission from our simulation, we show in this section that the predicted surface brightness profiles are consistent with observations of MW- and LMC-like galaxies in the local Universe. In particular, we compare the radial H$\alpha$ and H$\beta$ emission line profiles to observational measurements of the SDSS/MaNGA survey in Section~\ref{subsec:radial_profile_comparison}, while Section~\ref{subsec:vertical_heights_comparison} compares the vertical scale heights.

\subsection{Radial profiles from SDSS/MaNGA}
\label{subsec:radial_profile_comparison}

We compare our simulated H$\alpha$ maps to SDSS/MaNGA observations \citep{bundy15}. Specifically, we aim at comparing the shapes of the H$\alpha$ and H$\beta$ surface brightness profiles. MaNGA provides high-quality optical ($3600$--$10300$\,\AA\ and a spectral resolution of $R\sim2000$) integral field unit (IFU) spectroscopy for a large ($\sim10\,000$) sample of low-redshift galaxies. Individual galaxies are covered out to a distance of $1.5$--$2.5$ effective radii.

We use DR-15 \citep{aguado19}, which provides the data reduced by the Data Reduction Pipeline \citep[DRP;][]{law16} and stellar and gas properties of the galaxies made available thanks to the Data Analysis Pipeline \citep[DAP;][]{westfall19, belfiore19}. We accessed, inspected, and downloaded the data cubes with \texttt{Marvin} \citep{cherinka19}. The H$\alpha$ and H$\beta$ emission line maps have been corrected for stellar absorption and Milky Way reddening using the \citet{odonnell94} reddening law.

For both the MW and LMC simulations, we construct a sample of galaxies from MaNGA. Our selection is based on both stellar mass and SFR. We use as a rough estimate of the SFR the H$\alpha$-based SFR within the IFU field of view provided by DAP. For the stellar mass, we make use of the NASA Sloan Atlas \citep[NSA;][]{blanton05, blanton11} catalogue, which provides stellar masses derived from elliptical Petrosian photometry. Specifically, for the MW comparison sample, we select galaxies with stellar masses of $\log(M_{\star}/\text{M}_{\odot})=10.6$--$10.9$ and $\mathrm{SFR}/(\text{M}_{\odot}~\mathrm{yr}^{-1})=0.5$--$5.0$, which is consistent with $M_{\star}$ and SFR from our MW simulation (Tab.~\ref{tab:simulation}). For the LMC comparison sample, we select galaxies with stellar masses of $\log(M_{\star}/\text{M}_{\odot})=9.4$--$9.6$ and $\mathrm{SFR}/(\text{M}_{\odot}~\mathrm{yr}^{-1})=0.01$--$0.1$. In both comparison samples, we exclude any galaxies with DAP datacube quality flags cautioning ``do not use'', with two or more ``warning'' flags, as well as objects that are irregular (visual inspection of the H$\alpha$ maps). This yields MW and LMC comparison samples of 52 and 59 galaxies, respectively.

For each MaNGA galaxy, we measure the H$\alpha$ and H$\beta$ surface brightness profiles in the same elliptical apertures in order to account for the effect of inclination on radius measurements. Furthermore, we only consider spaxels with a signal-to-noise ratio of at least 5. Since we aim for a comparison of the shapes, we normalise each individual profile to a total luminosity of $10^{42}~\mathrm{erg}~\mathrm{s}^{-1}$ ($10^{41}~\mathrm{erg}~\mathrm{s}^{-1}$) for the case of H$\alpha$ (H$\beta$). We then compute the median profile and 1$\sigma$ variation as a function of radius. We have also experimented with masking regions of active galactic nuclei (AGN) -- and other excitation mechanisms -- via BPT diagnostics \citep{baldwin81}, finding this had a negligible effect on our median-stacked profile (reduction of 0.2 dex in the centre relative to the outskirts).

In order to compare the observed median H$\alpha$ and H$\beta$ profiles to our simulated ones, we need to account for the point spread function (PSF). MaNGA has a relatively large PSF with a FWHM of $2.5\arcsec$ \citep{bundy15}. This corresponds to an angular distance of $3.9$ and $1.7$\,kpc at the average distance of our MW sample ($\langle z \rangle = 0.078$) and LMC sample ($\langle z \rangle = 0.036$), respectively. We apply a Gaussian PSF to the simulated H$\alpha$ and H$\beta$ face-on maps (see Appendix~\ref{appsec:smoothing}) and then measure the profiles in circular apertures. We compute the median H$\alpha$ and H$\beta$ profiles (again after normalising each profile) of all snapshots after 200\,Myr, which ensures that the simulations have reached a quasi-steady state (see Section~\ref{subsec:simulation}).

In Fig.~\ref{fig:profile_comparison} we compare our simulated H$\alpha$ and H$\beta$ profiles (solid lines; dotted in case for the LMC-BPASS simulation) with the observed MaNGA profiles (points with errorbars). We find excellent agreement in the overall shapes for both the MW simulation out to $10$--$15$\,kpc (left panel) and LMC simulations out to $4$--$6$\,kpc (right panel). To guide the eye, the green dashed lines show a S\'{e}rsic $n=1$ profile, indicating that both the simulated and observed profiles are well described by an exponential function in the case of the MW, while the profiles appear to be steeper in the case of the LMC. An important note is that nuclear regions are not modelled in the simulation and effects due to AGN feedback
are not included \citep[see, e.g.,][]{nelson21}, so we do not expect or judge the agreement there as the consistency could be serendipitous. Furthermore, we find little difference between the surface brightness profile of the LMC-BC03 and the LMC-BPASS model.

\subsection{Vertical scale heights}
\label{subsec:vertical_heights_comparison}

\begin{figure*}
\includegraphics[width=0.5\linewidth]{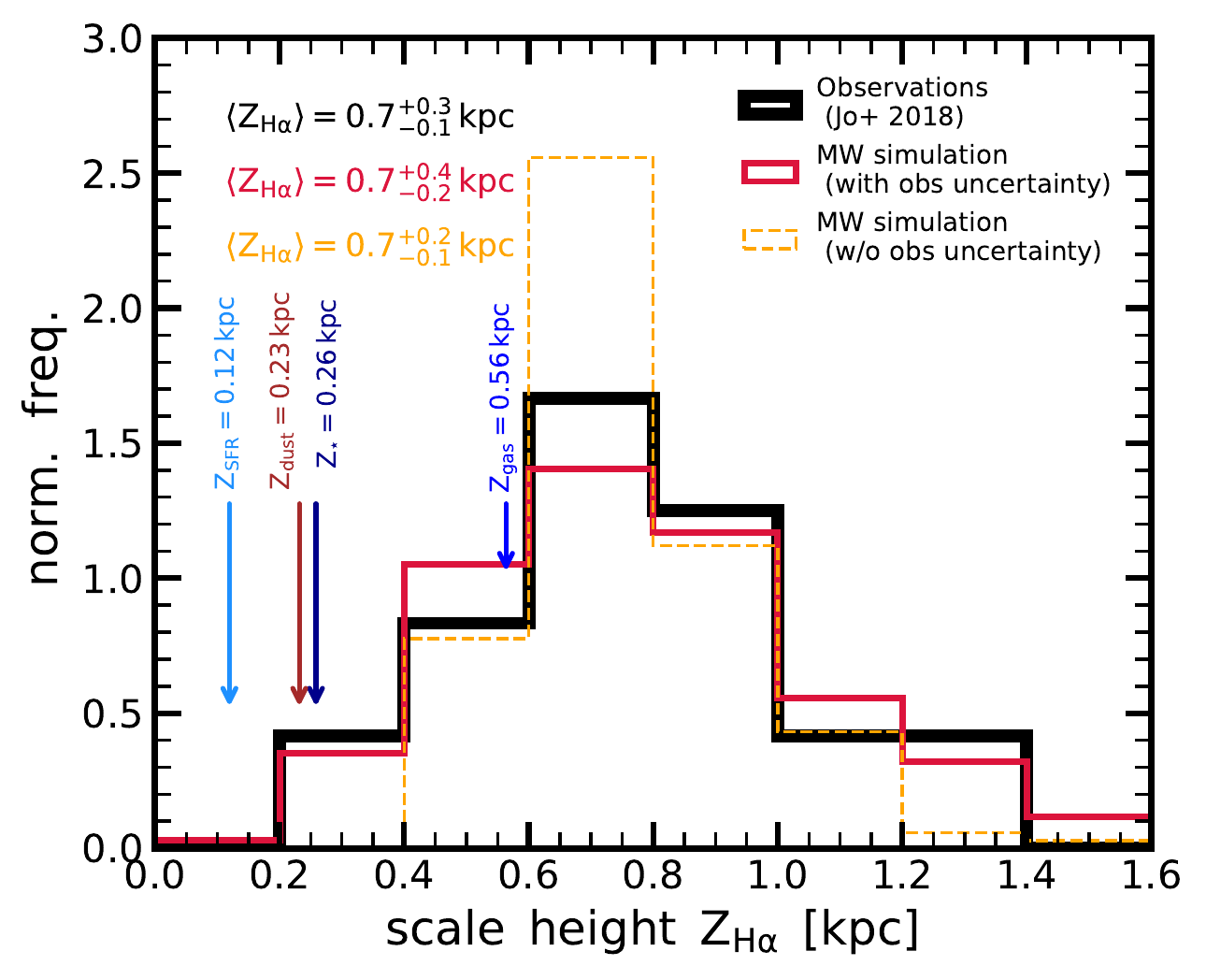}
\includegraphics[width=0.485\linewidth]{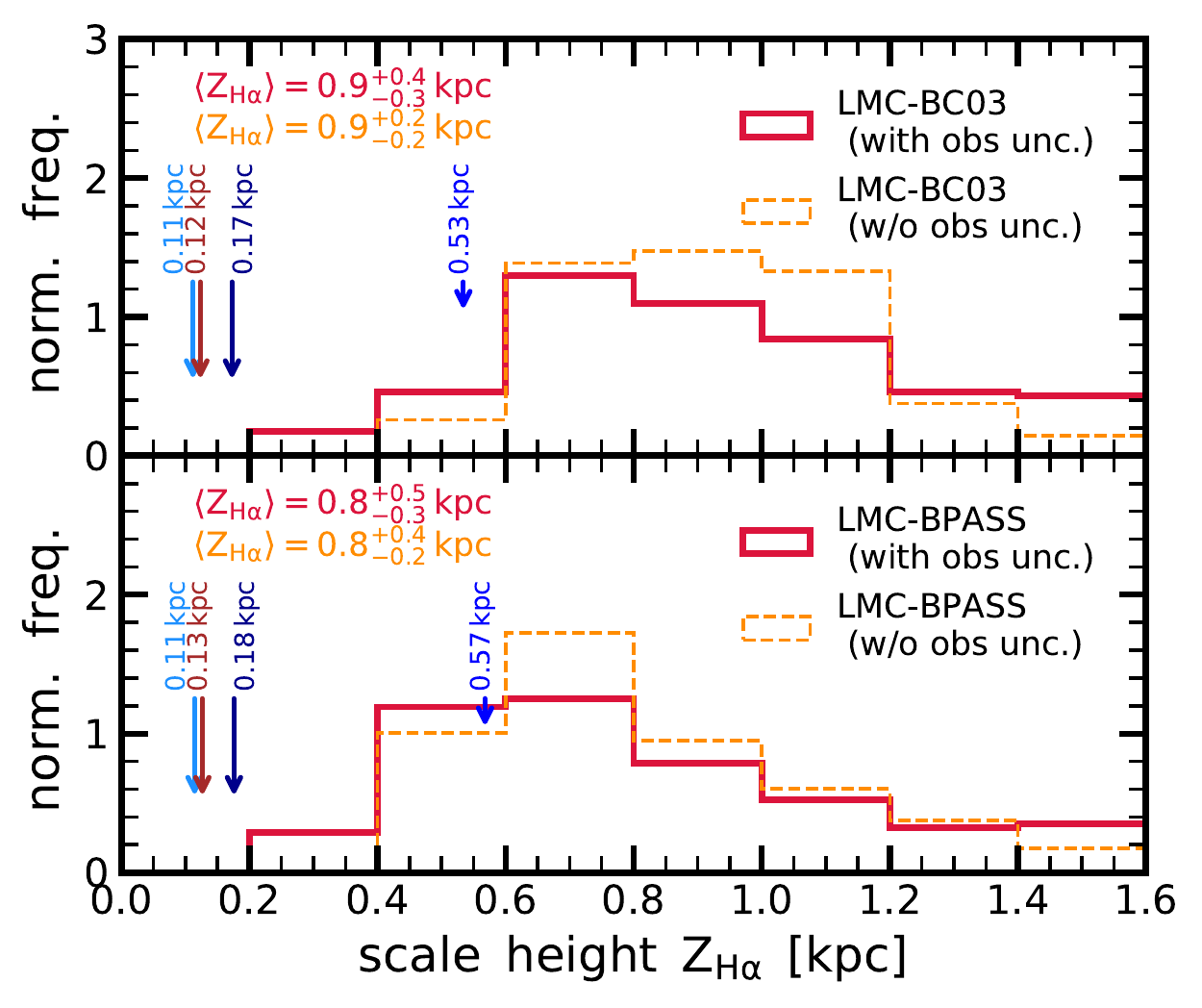}
\caption{Comparison of the vertical H$\alpha$ scale heights Z$_{\rm H\alpha}$ of the simulations with observations. The left panel shows the MW simulation, while the right two panels show the LMC-BC03 (top) and the LMC-BPASS (bottom) simulations. For the LMC, no observational data for a comparison is available. The scale heights in both observations and simulations are measured by fitting an exponential function ($\propto \exp(-z/Z_{\rm H\alpha})$) to the vertical profiles of the H$\alpha$ emission. The observations are taken from scale height measurements by \citet{jo18} of 12 nearby edge-one late-type galaxies. For the MW simulation, we find a typical scale height of Z$_{\rm H\alpha}=0.7~\mathrm{kpc}$ in both observations and simulations. After accounting for the observational uncertainty, also the overall distribution are in good agreement with each other. For reference, the MW scale height for the stellar mass (Z$_{\star}=0.26~\mathrm{kpc}$), dust mass (Z$_{\rm dust}=0.23~\mathrm{kpc}$), and gas mass (Z$_{\rm gas}=0.56~\mathrm{kpc}$) and SFR (Z$_{\rm SFR}=0.12~\mathrm{kpc}$) are indicated. For the LMC, we find the scale heights of H$\alpha$, the SFR and the gas to be comparable to the ones of the MW, while the the scale heights for the stellar mass and dust mass tend to be smaller.}
\label{fig:scaleheight_comparison}
\end{figure*}

The comparison of the vertical profiles from simulations with observations is more challenging than the aforementioned comparison of the radial profiles. The main reason for this is the lack of observational data probing H$\alpha$ of edge-on MW-like galaxies. A recent investigation by \citet{jo18} measured the vertical profiles of the extraplanar H$\alpha$ emission for 38 nearby edge-on late-type galaxies. The data have been taken from six different H$\alpha$ imaging surveys. The galaxies were selected to be within a distance of 30\,Mpc, have no noticeable spiral or asymmetry patterns, and only include data of sufficient quality (based on signal-to-noise ratios). The resulting 38 galaxies span a wide range in SFRs ($\mathrm{SFR}=0.001$--$1.5~\mathrm{M}_{\odot}~\mathrm{yr}^{-1}$), so we sub-select from this observational sample galaxies with $\mathrm{SFR}>0.1~\mathrm{M}_{\odot}~\mathrm{yr}^{-1}$ that also show disturbed discs. These criteria lead to a sample size of 12 galaxies for the MW comparison. We do not perform a comparison with the LMC simulations because of a lack of matching comparison sample of  galaxies. For each galaxy, \citet{jo18} obtained the vertical profiles of the H$\alpha$ emission by horizontally averaging each image and then fitting the profiles with an exponential function ($\propto \exp(-z/Z_{\rm H\alpha})$), where $Z_{\rm H\alpha}$ is the scale height.

We estimate the scale height $Z_{\rm H\alpha}$ in the simulations adopting the same procedure as the observations. Specifically, we compute the H$\alpha$ vertical profiles for each snapshot by horizontally averaging the edge-on projection within the effective radius ($R_{\rm eff}=4.3~\mathrm{kpc}$) and then fitting an exponential function to estimate $Z_{\rm H\alpha}$. We consider only the snapshots 200 Myr after the start of the simulation. In order to compare our $Z_{\rm H\alpha}$ from the simulation with observations, we also add a relative uncertainty of 30\%.

The right panel of Fig.~\ref{fig:scaleheight_comparison} shows the scale height distributions of the observations by \citet{jo18} in black. The median observed scale height is $0.7_{-0.1}^{+0.3}~\mathrm{kpc}$. The scale height distribution for our simulation is shown in red (with observational uncertainty) and in orange (without observational uncertainty). We measure a median scale height in the simulation of $0.7_{-0.2}^{+0.4}~\mathrm{kpc}$, demonstrating that the scale heights from the MW simulation are overall in good agreement with the observations. For reference, the scale height for the stellar mass (Z$_{\star}=0.26~\mathrm{kpc}$), dust mass (Z$_{\rm dust}=0.23~\mathrm{kpc}$), and gas mass (Z$_{\rm gas}=0.56~\mathrm{kpc}$) and SFR (Z$_{\rm SFR}=0.12~\mathrm{kpc}$) are indicated as vertical arrows. This already hints at in-situ ionization of gas, rather than H$\alpha$ photons scattered on dust into the observer's line of sight, being the main driver of this extraplanar H$\alpha$ emission. We will look further into this in Section~\ref{subec:vertical_analysis}. For completeness, the left panels of Fig.~\ref{fig:scaleheight_comparison} show the equivalent plot and numbers for the LMC-BC03 (top) and LMC-BPASS (bottom) simulations. We find that the LMC simulations have comparable scale heights for the gas and H$\alpha$ (independent of the stellar population model) to the MW simulation, while the SFR and stellar mass scale heights of the LMC are smaller than the ones of the MW. We attribute the relatively large gas and H$\alpha$ scale heights of the LMC to the more variable star formation in the LMC than the MW simulation.

\section{Balmer emission, scattering, and absorption}
\label{sec:balmer_emission_diag}

After showing in the previous section that the Balmer (H$\alpha$ and H$\beta$) emission of our MW and LMC simulations are realistic, we now turn towards understanding the Balmer emission, scattering and absorption. Specifically, we present results on the gas temperature and density dependence of the Balmer emission (Section~\ref{subsec:phase}), the collisional excitation and recombination emission (Section~\ref{subsec:fcol}), the importance of scattering (Section~\ref{subsec:scattering}), the radiative transfer of photons from \HII and DIG regions (Section~\ref{subsec:HII_DIG}), different DIG definitions (Section~\ref{subsec:DIG_def}), and the origin of the extraplanar Balmer emission (Section~\ref{subec:vertical_analysis}).

\begin{figure*}
\begin{center}
\includegraphics[width=\textwidth]{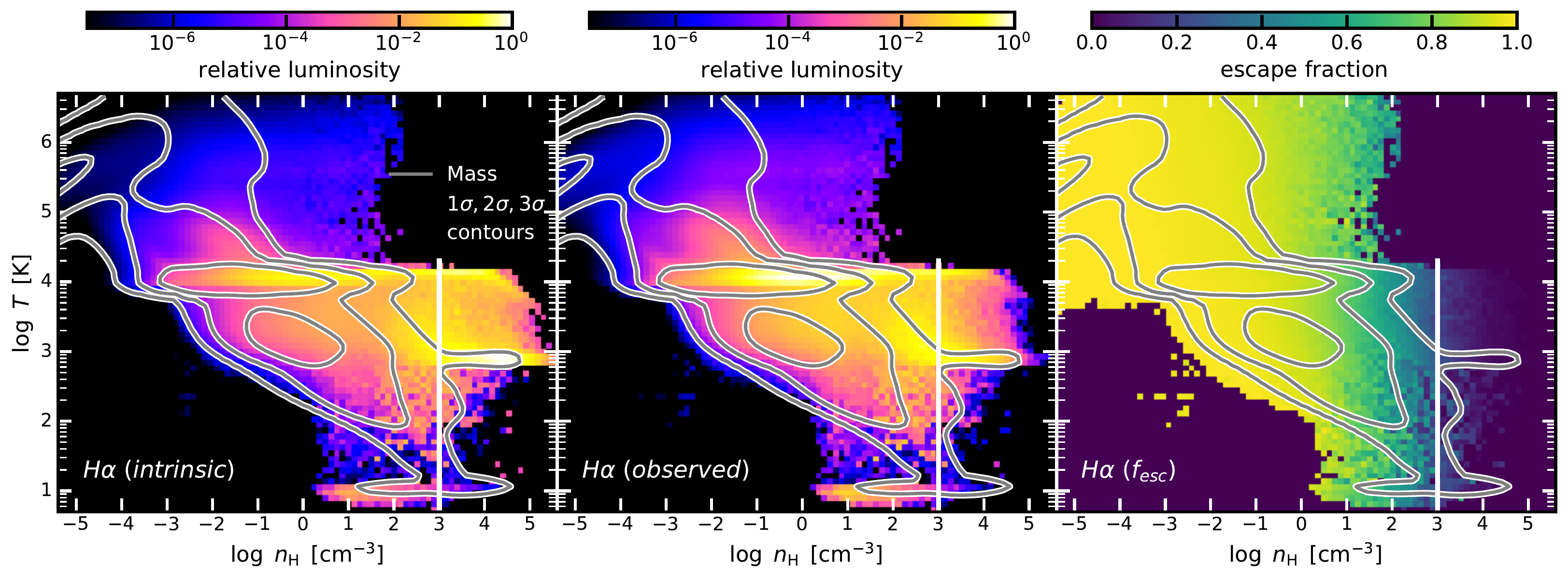}
\caption{Density and temperature dependence of the Balmer H$\alpha$ line emission across all snapshots of the MW simulation. From left to right, we plot the density--temperature phase space diagram colour-coded by the relative intrinsic H$\alpha$ luminosity, the relative observed H$\alpha$ luminosity, and the escape fraction of H$\alpha$ (ratio of observed and intrinsic emission). The contours indicate the mass-weighted distribution of the gas, while the vertical lines show the star-formation threshold. Most of the intrinsic H$\alpha$ emission comes from dense \HII regions with a characteristic temperature of $T\sim 10^3$--$10^4\,\mathrm{K}$, while the observed H$\alpha$ emission stems from $T\sim 10^4\,\mathrm{K}$ gas. The impact of dust absorption can also be seen: the denser the gas, the more pronounced the dust absorption and hence the lower the escape fraction.}
\label{fig:phase}
\end{center}
\end{figure*}

\subsection{Density--temperature dependence of the Balmer emission}
\label{subsec:phase}

In Fig.~\ref{fig:phase}, we show the density and temperature dependence of the Balmer line emission. We have adopted this figure from our companion work \citep{smith21_rt}, which goes into further detail regarding the gas that emits Ly$\alpha$ and Balmer emission. Here, we plot the density--temperature phase space diagram colour-coded by the relative intrinsic H$\alpha$ luminosity, the relative observed H$\alpha$ luminosity, and the escape fraction of H$\alpha$ (ratio of observed and intrinsic emission). In all panels, the contours indicate the mass-weighted distribution of the gas, while the vertical lines show the star-formation threshold. The mass contours highlight the presence of molecular gas ($\sim 10\,\mathrm{K}$) and of collapsing structures prior to heating and disruption via feedback ($\sim 10^3\,\mathrm{K}$). Importantly, although density is the only direct star-formation condition, the temperature enters indirectly through Jeans stability and  converging flows (see \citealt{marinacci19} and \citealt{kannan20_mw} for more details). This ensures that stars only form within cold ($\sim10~\mathrm{K}$) molecular clouds and the warm $\sim10^3~\mathrm{K}$ track is high-density photoheated gas surrounding young stars that has not had time to become a lower density $10^4~\mathrm{K}$ \HII region.

We find that most of the observed H$\alpha$ emission comes from dense \HII regions with a characteristic temperature of $T\sim10^4\,\mathrm{K}$ and density of $n_{\rm H}\sim0.1$--$10\,\mathrm{cm}^{-3}$ (middle panel of Fig.~\ref{fig:phase}). The intrinsic H$\alpha$ emission (left panel of Fig.~\ref{fig:phase}) also shows strong emission features at $T\sim10^3\,\mathrm{K}$, which are caused by both physical and numerical effects. On the numerical side, this feature can be associated with underresolved and consequently underheated \HII regions (see also \citealt{smith21_rt} for an extended discussion of this). On the physical side, as noted above, this feature can be caused by molecular clouds, partial ionization, and transient phenomena. Importantly, although the intrinsic H$\alpha$ emission is bimodal, this second peak at high densities and temperatures of $T\sim10^3\,\mathrm{K}$ diminishes when looking at the observed H$\alpha$ emission (i.e. after scattering and absorption), because the escape fractions from these same dense dusty regions are much lower in comparison to resolved \HII regions at $\sim 10^4\,\mathrm{K}$. This can be clearly seen in the right-hand panel of Fig.~\ref{fig:phase}, which shows the H$\alpha$ escape fraction strongly scales with the density of gas.

\subsection{Emission: collisional excitation and recombination}
\label{subsec:fcol}

\begin{figure*}
\begin{center}
\includegraphics[width=\textwidth]{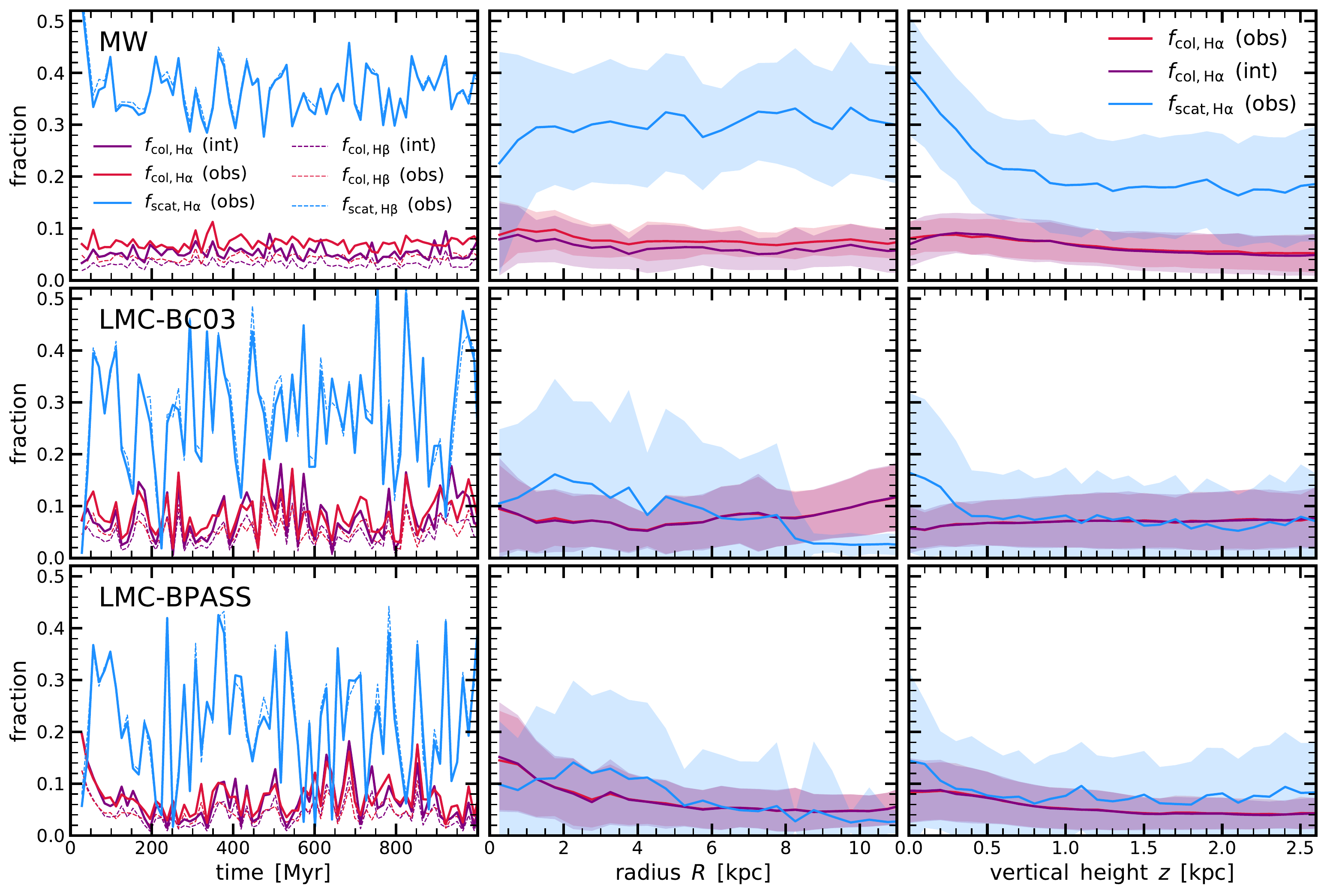}
\caption{Importance of collisional excitation emission ($f_{\rm col}$) and scattered emission ($f_{\rm scat}$). The panel on the left, middle, and right show the time-evolving global value, azimuthally-averaged radial profile, and vertical profile, respectively. The MW, LMC-BC03 and LMC-BPASS simulations are shown from top to bottom, with the LMC simulations showing more variability with time. The solid (dashed) red and purple lines show $f_{\rm col}$ for the intrinsic and observed H$\alpha$ (H$\beta$) emission, respectively. The corresponding blue lines show the fractional increase of observed emission caused by scattering (i.e. $f_{\rm scat}=1-L_{\rm w/o~scattering}/L_{\rm with~scattering}$). We find median values of $f_{\rm col}=5.0_{-0.8}^{+1.7}\%$ ($f_{\rm col}=6.7_{-3.7}^{+6.1}\%$ and $f_{\rm col}=6.0_{-3.2}^{+4.5}\%$) and $f_{\rm col}=7.1_{-1.0}^{+0.9}\%$ ($f_{\rm col}=8.2_{-3.9}^{+3.1}\%$ and $f_{\rm col}=7.1_{-3.3}^{+2.8}\%$) for the intrinsic and observed H$\alpha$ emission for the MW (LMC-BC03 and LMC-BPASS) simulation, respectively, indicating that the Balmer H$\alpha$ and H$\beta$ emission is dominated by radiative recombination as opposed to collisional excitation. Furthermore, scattering is important ($f_{\rm scat}=37_{-5}^{+4}\%$ for the MW, $f_{\rm scat}=28_{-11}^{+11}\%$ for the LMC-BC03 and $f_{\rm scat}=21_{-12}^{+14}\%$ for the LMC-BPASS) in boosting the Balmer emission by allowing Balmer photons to scatter out of dust-obscured regions both in the face-on and edge-on directions. There are only weak radial gradients in both $f_{\rm col}$ and $f_{\rm scat}$ (middle panels), while the $f_{\rm scat}$ and $f_{\rm col}$ decrease with increasing scale height (right panels).}
\label{fig:fcol}
\end{center}
\end{figure*}

As shown in the schema of the H$\alpha$ radiative transfer (Fig.~\ref{fig:cartoon}), recombination, collisional excitation and collisional ionization can lead to Balmer emission. In this section, we focus on the importance of collisional excitation emission since the emission from collisional ionization contributes less than $2\%$ to the total H$\alpha$ emission. In Fig.~\ref{fig:fcol}, we show the fraction of collisional excitation emission $f_{\rm col}$ for the MW (top panels), LMC-BC03 (middle panels) and LMC-BPASS (bottom panels) simulations. We plot the integrated $f_{\rm col}$ as a function of time, the average $f_{\rm col}$ as a function of radial distance, and the average $f_{\rm col}$ as a function of vertical scale height in the left, middle, and right panels, respectively. The solid (dashed) purple and red lines show $f_{\rm col}$ for the intrinsic and observed H$\alpha$ (H$\beta$) emission. 

The observed emission typically has a higher $f_{\rm col}$ than the intrinsic emission, since collisional excitation emission is emitted in regions with lower density and therefore suffers less dust absorption. For the MW, we find a median value of $f_{\rm col}=5.0_{-0.8}^{+1.7}\%$ and $f_{\rm col}=7.1_{-1.0}^{+0.9}\%$ for the intrinsic and observed H$\alpha$ emission, respectively. For the LMC-BC03, we find $f_{\rm col}=6.7_{-3.7}^{+6.1}\%$ and $f_{\rm col}=8.2_{-3.9}^{+3.1}\%$ for the intrinsic and observed H$\alpha$ emission, while the LMC-BPASS returns $f_{\rm col}=6.0_{-3.2}^{+4.5}\%$ and $f_{\rm col}=7.1_{-3.3}^{+2.8}\%$, indicating that changes to the stellar population models lead to small changes in $f_{\rm col}$. These slightly larger values for the intrinsic and observed $f_{\rm col}$ can be explained by the more turbulent and bursty nature of the LMC relative to the MW, which gives rise to more shocks and therefore collisionally excited gas. The variability of $f_{\rm col}$ occurs with the same cadence as star formation fluctuations (see Section~\ref{subsec:time_evolution}), which is due to a combined effect of having a higher recombination rate during a starburst and higher collisional emission due to feedback. This is also confirmed by the rather large variability of $f_{\rm col}$ in the case of the LMC (the confidence intervals quoted above indicate the $16^{\rm th}$ and $84^{\rm th}$ percentiles).

For the H$\beta$ emission, we find the same overall trends as for H$\alpha$. We obtain lower $f_{\rm col}$ values than for the H$\alpha$ emission. This can be explained by the lower collisional rate coefficients of H$\beta$ compared to H$\alpha$ (see Appendix~A of \citealt{smith21_rt}).

The middle and right panels of Fig.~\ref{fig:fcol} show the median radial and vertical profiles of $f_{\rm col}$. Overall, we do not find a strong gradient in $f_{\rm col}$. If anything, there is a weak trend for the MW where $f_{\rm col}$ is slightly higher in the centre than the outskirts and within the disc rather than above the midplane.

In summary, we find that only a small fraction ($5$--$10\%$) of the total Balmer emission stems from collisionally excited gas. Most of the emission stems from radiative recombination. This is consistent with previous theoretical findings by \citet{peters17}, who investigated a small region of a galactic disc with solar neighbourhood-like properties in the stratified disc approximation. They found that collisional excitation emission makes up $1$--$10\%$ of the total Balmer emission.

\subsection{Scattering of H\texorpdfstring{$\balpha$}{α} photons}
\label{subsec:scattering}

Fig.~\ref{fig:fcol} also allows us to investigate the importance of scattering of the Balmer emission. In all of the panels, the solid (dashed) blue lines show the fractional increase of observed H$\alpha$ (H$\beta$) emission caused by scattering, $f_{\rm scat} \equiv 1 - L_{\rm w/o~scattering} / L_{\rm with~scattering}$. There is essentially no difference between H$\alpha$ and H$\beta$. We find scattering to be important with, on average, $f_{\rm scat}=37_{-5}^{+4}\%$ for the MW, $f_{\rm scat}=28_{-11}^{+11}\%$ for the LMC-BC03, and $f_{\rm scat}=21_{-12}^{+14}\%$ for the LMC-BPASS. Furthermore, $f_{\rm scat}$ also does not depend significantly on the viewing angle, i.e. face-on versus edge-on projections lead to differences of less than $1$--$2\%$. This means that if scattering is \textit{not} considered, one would underestimate the luminosity by $20$--$40\%$. The reason for this is that scattering allows Balmer photons to diffuse out of dust-obscured regions. Although the azimuthally-averaged radial gradients are weak (middle panels of Fig.~\ref{fig:fcol}), we find in the next section that scattering is particularly prominent around \HII regions. We find that scattering is high at low scale heights, i.e. within the disc, and then gradually decreases towards larger vertical heights (right panels of Fig.~\ref{fig:fcol}).

\begin{figure*}
\begin{center}
\includegraphics[width=\textwidth]{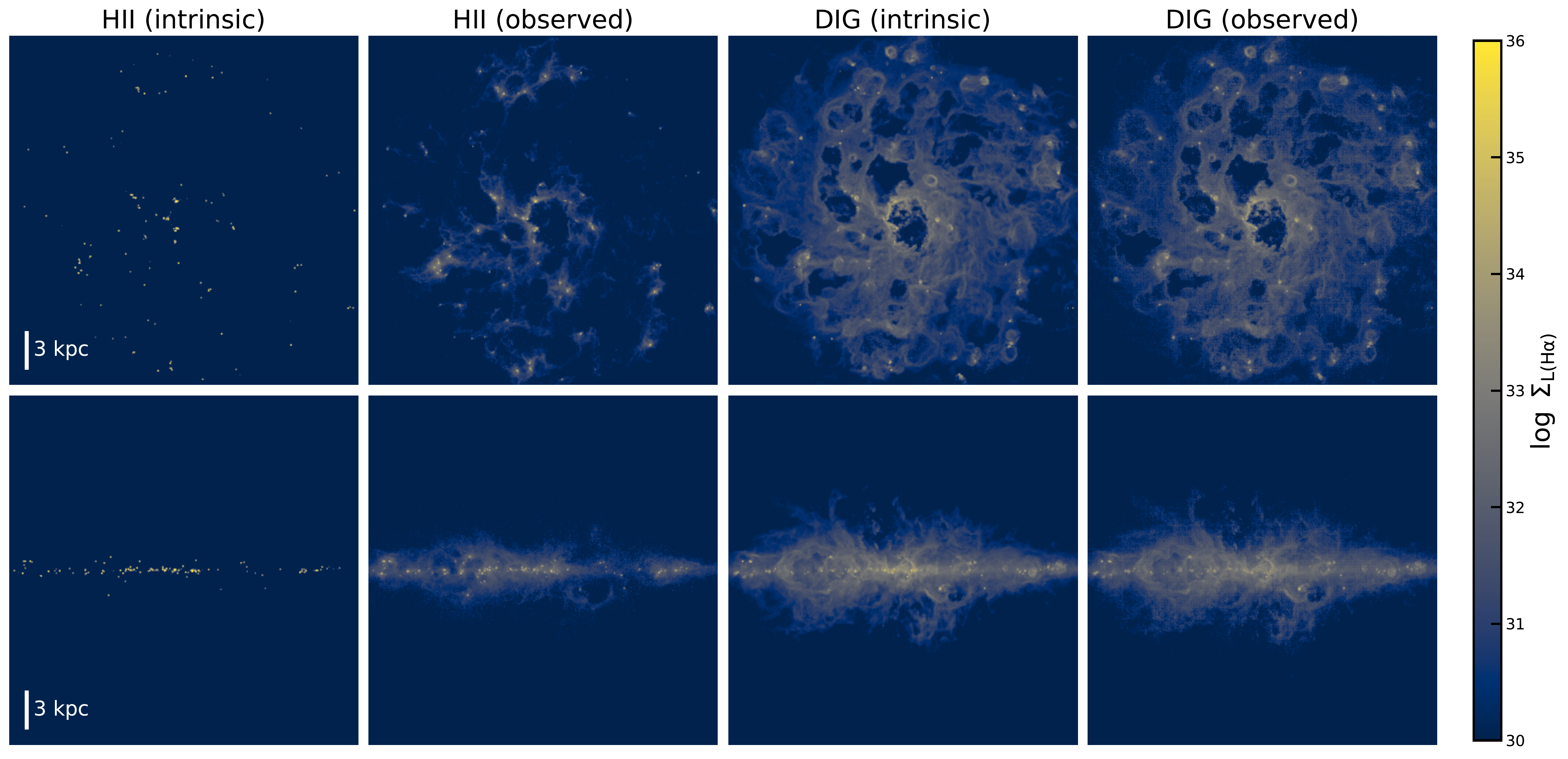}
\caption{H$\alpha$ emission maps for face-on (upper panels) and edge-on (lower panels) sightlines with dimensions of $30\times30$\,kpc. The photons are categorised according to the gas density of the emission region: \HII photons (emitted in regions with $n>100~\mathrm{cm}^{-3}$) are shown in the left panels and diffuse ionized gas (DIG) photons (emitted in regions with $n<100~\mathrm{cm}^{-3}$) are shown in the right panels. For both \HII and DIG photons we show the intrinsically emitted photons and the observed (after attenuation and scattering) photons. The \HII photons are produced on small, confined regions, and only through scattering are able to occupy a significant area. On the other hand, the DIG photons are emitted on diffuse scales, extending significantly ($>3~\mathrm{kpc}$) above the plane of the disc.}
\label{fig:Halpha_map}
\end{center}
\end{figure*}

We investigate in detail the probability that H$\alpha$ photons scatter at least $N$ times in Fig.~22 of our companion paper \citep{smith21_rt}. Averaged over all photon trajectories from all snapshots and sightlines, we find that 76\% (39\%) of all H$\alpha$ photons scatter at least once (at least 10 times) for the MW simulation. When only considering the observed (i.e. escaped) H$\alpha$ photons, these fractions reduce to 36\% and 0.3\%. On average, an intrinsic and observed H$\alpha$ photon scatters 11.7 and 0.8 times, respectively.

\subsection{H\texorpdfstring{$\balpha$}{α} photons originating from \ion{H}{II} regions and the DIG}
\label{subsec:HII_DIG}

\begin{figure*}
\begin{center}
\includegraphics[width=\textwidth]{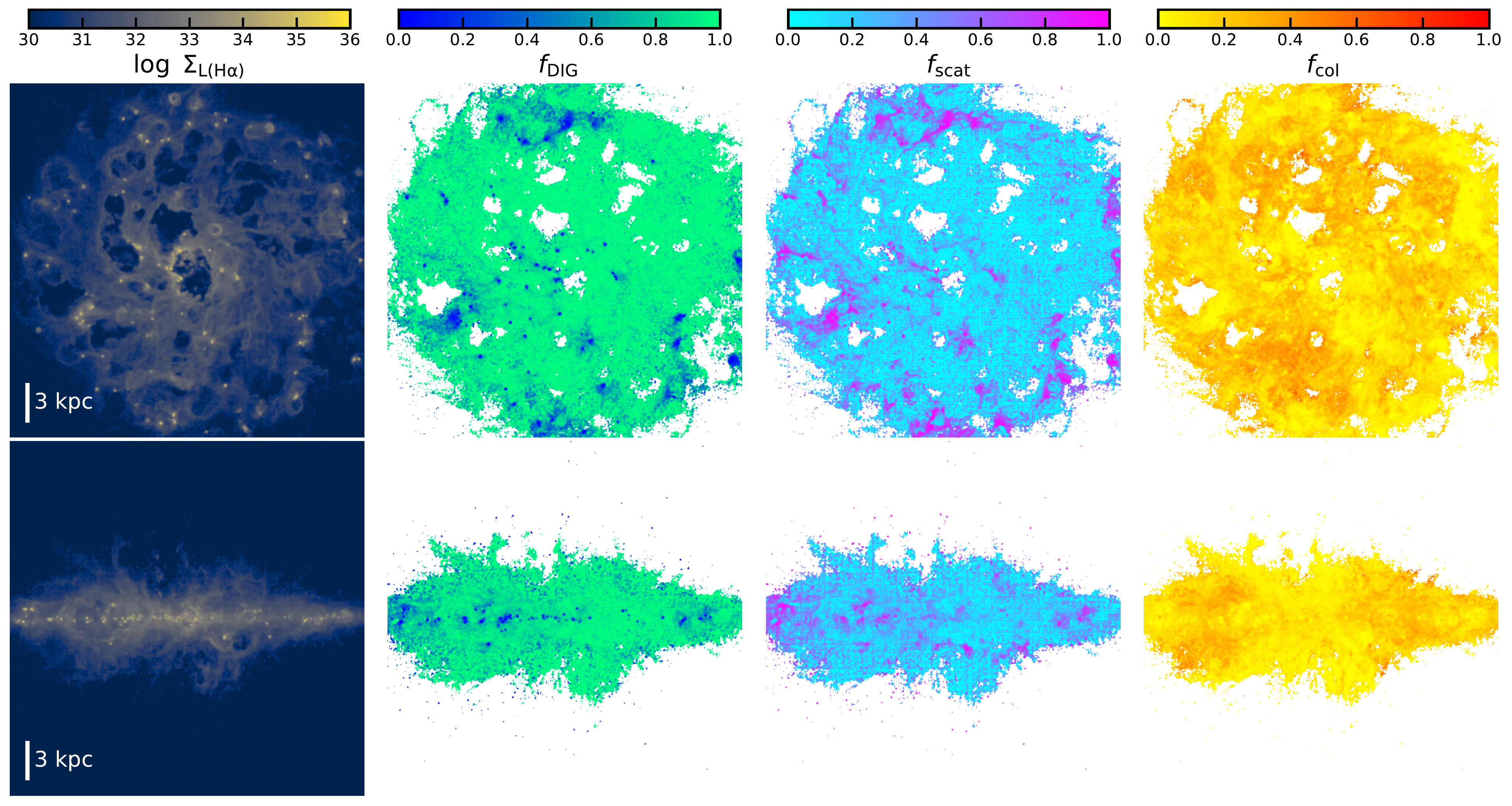}
\caption{Maps of the observed H$\alpha$ emission (left panels), the fraction of DIG photons (second from the left), the scattering factor (second from the right; $f_{\rm scat} \equiv 1 - L_{\rm w/o~scattering} / L_{\rm with~scattering}$), and the fraction of collisionally excited emission (right panels) for face-on (top panels) and edge-on (bottom panels) views. The scattering factor quantifies the fractional increase of observed H$\alpha$ emission caused by scattering. We find that the small-scale, bright regions are dominated by \HII photons, while most of the area is dominated by DIG photons. Scattering can boost the luminosity by a factor of $\gtrsim3$ in the bright \HII regions.}
\label{fig:Halpha_fraction}
\end{center}
\end{figure*}

H$\alpha$ can be emitted from \ion{H}{II} and DIG regions. Observationally as well as theoretically, there are no clear definitions of what is meant by DIG and various works adopt different definitions. In this work, we use a threshold in gas density to differentiate between \ion{H}{II} and the DIG. Specifically, we adopt a gas density threshold of $n_{\rm thresh}=100~\mathrm{cm}^{-3}$. Balmer photons emitted above this threshold are called \ion{H}{II} photons, while photons emitted from gas below $n_{\rm thresh}$ are called DIG photons. Although the choice of $n_{\rm thresh}$ is rather arbitrary, it is motivated by being an order-of-magnitude below the density threshold for star formation, which is $n=10^3~\mathrm{cm}^{-3}$ in our simulations (Section~\ref{subsec:simulation}). Looking at the density--temperature phase diagram (Fig.~\ref{fig:phase}), the DIG is dominated by emission from gas with a temperature of roughly $T\sim10^4~\mathrm{K}$. We compare different DIG definitions in Section~\ref{subsec:DIG_def}.

In Fig.~\ref{fig:Halpha_map} we illustrate the distribution of the H$\alpha$ emission, split into \ion{H}{II} and DIG photons for the same MW snapshot as shown in Fig.~\ref{fig:MW_maps}. The upper and lower panels plot the face-on and edge-on views, respectively. The intrinsic \ion{H}{II} photons, observed \ion{H}{II} photons, intrinsic DIG photons, and observed DIG photons are shown from left to right. The intrinsic \ion{H}{II} photons are emitted from very localised regions, which extend only a few tens of pc. The observed \ion{H}{II} photons are significantly more extended than the intrinsic \ion{H}{II} photons, which can be explained by scattering (see below). The intrinsic and observed DIG photons have similar distributions and are both more extended than the intrinsic and observed \ion{H}{II} photon distributions. This is expected, since the DIG photons are emitted from less dense gas that is spatially more extended and the dust reprocessing is minimal. Furthermore, there are several ring-like features in the DIG map, which are shell overdensities created by supernovae bubbles.

Fig.~\ref{fig:Halpha_fraction} follows the same layout as Fig.~\ref{fig:Halpha_map}, but shows from left to right maps of the total observed H$\alpha$ emission, fraction of DIG photons ($f_{\rm DIG}$), scattering factor ($f_{\rm scat}$), and fraction of photons emitted from collisionally excited gas ($f_{\rm col}$). The map of the observed H$\alpha$ emission clearly shows the point-like sources, which are the \ion{H}{II} regions. Consistent with Fig.~\ref{fig:Halpha_map}, the $f_{\rm DIG}$ map (second from the left) indicates small localised regions with low $f_{\rm DIG}$, which are the \ion{H}{II} regions. The rest is dominated by the DIG, including the regions above the plane of the disc.

The second panel from the right in Fig.~\ref{fig:Halpha_fraction} shows the fractional increase of observed H$\alpha$ emission caused by scattering (see Section~\ref{subsec:scattering}). Comparing it with $f_{\rm DIG}$, we find that the $f_{\rm scat}$ is high where $f_{\rm DIG}$ is low. This highlights that scattering boosts the H$\alpha$ luminosity more significantly in \ion{H}{II} regions than in DIG regions. Specifically, averaged over all MW snapshots, the \ion{H}{II} luminosity is boosted by $f_{\rm scat, HII}=64_{-2}^{+2}\%$, while DIG luminosity only increases by $f_{\rm scat, DIG}=18_{-2}^{+2}\%$. For comparison, the total boost-factor due to scattering is $f_{\rm scat}=38_{-5}^{+4}\%$ (see also Fig.~\ref{fig:fcol}). This trend is even more apparent when looking at the LMC simulations, where we find 
$f_{\rm scat, HII}=51_{-9}^{+10}\%$ ($f_{\rm scat, HII}=52_{-13}^{+6}\%$) and
$f_{\rm scat, DIG}=5_{-1}^{+3}\%$ ($f_{\rm scat, DIG}=5_{-2}^{+4}\%$) for the LMC-BC03 (LMC-BPASS) model.

The panel on the right of Fig.~\ref{fig:Halpha_fraction} shows the map of the fraction of photons from collisionally excited gas. There are only a few localised regions with $f_{\rm col}$, which are typically at the boundary to low surface brightness regions. It seems $f_{\rm col}$ is more determined by the overall large scale motion of the gas; i.e. regions with elevated or reduced $f_{\rm col}$ are typically a few kpc in size.

\subsection{Implications of different DIG definitions}
\label{subsec:DIG_def}

\begin{figure*}
\includegraphics[width=\textwidth]{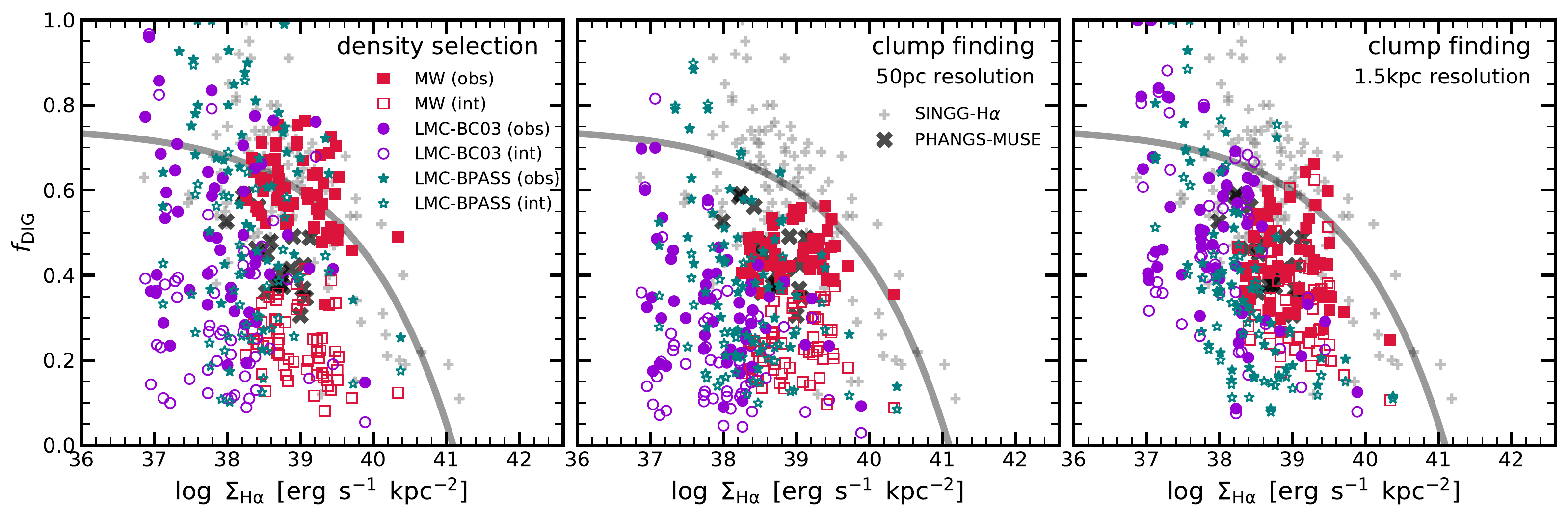}
\caption{Fraction of Balmer emission originating from DIG $f_{\rm DIG}$ as a function of the H$\alpha$ surface brightness $\Sigma_{\rm H\alpha}$ (Eq.~\ref{eq:SHa}) for different DIG selection methods. In the left panel shows our fiducial physical DIG definition that is based on a density cut of the emitting gas, while the middle and right panels use a DIG definition based clump finding on the 2d face-on projection. In all panels, the red squares, purple circles, and teal stars mark our MW, LMC-BC03 and LMC-BPASS simulations, while the filled and empty symbols indicate the observed and intrinsic $f_{\rm DIG}$, respectively. The grey plus and cross symbols are the observational measurements from the SINGG H$\alpha$ survey (\citealt{oey07}; with a median of $f_{\rm DIG}=0.66_{-0.12}^{+0.13}$ in the MW's $\Sigma_{\rm H\alpha}$ range) and the PHANGS-MUSE survey (\citealt{belfiore22}; with a median of $f_{\rm DIG}=0.42_{-0.05}^{+0.11}$). The grey solid line shows the best fitting function (Eq.~\ref{eq:fDIG}; \citealt{sanders17}) of the form $f_{\rm DIG} \propto \Sigma_{\rm H\alpha}^{1/3}$, as suggested by Oey et al. on theoretical arguments. The following are the two key conclusions from this figure: ($i$) $f_{\rm DIG}$ values depend significantly on the selection technique; ($ii$) $f_{\rm DIG}$ from the clump finding method depends heavily on the spatial resolution of the data; and ($iii$) the intrinsic $f_{\rm DIG}$ is typically smaller (factor of 2-3) than the observed $f_{\rm DIG}$ due to the comparatively higher escape fractions.}
\label{fig:f_DIG}
\end{figure*}

\begin{figure}
\includegraphics[width=\linewidth]{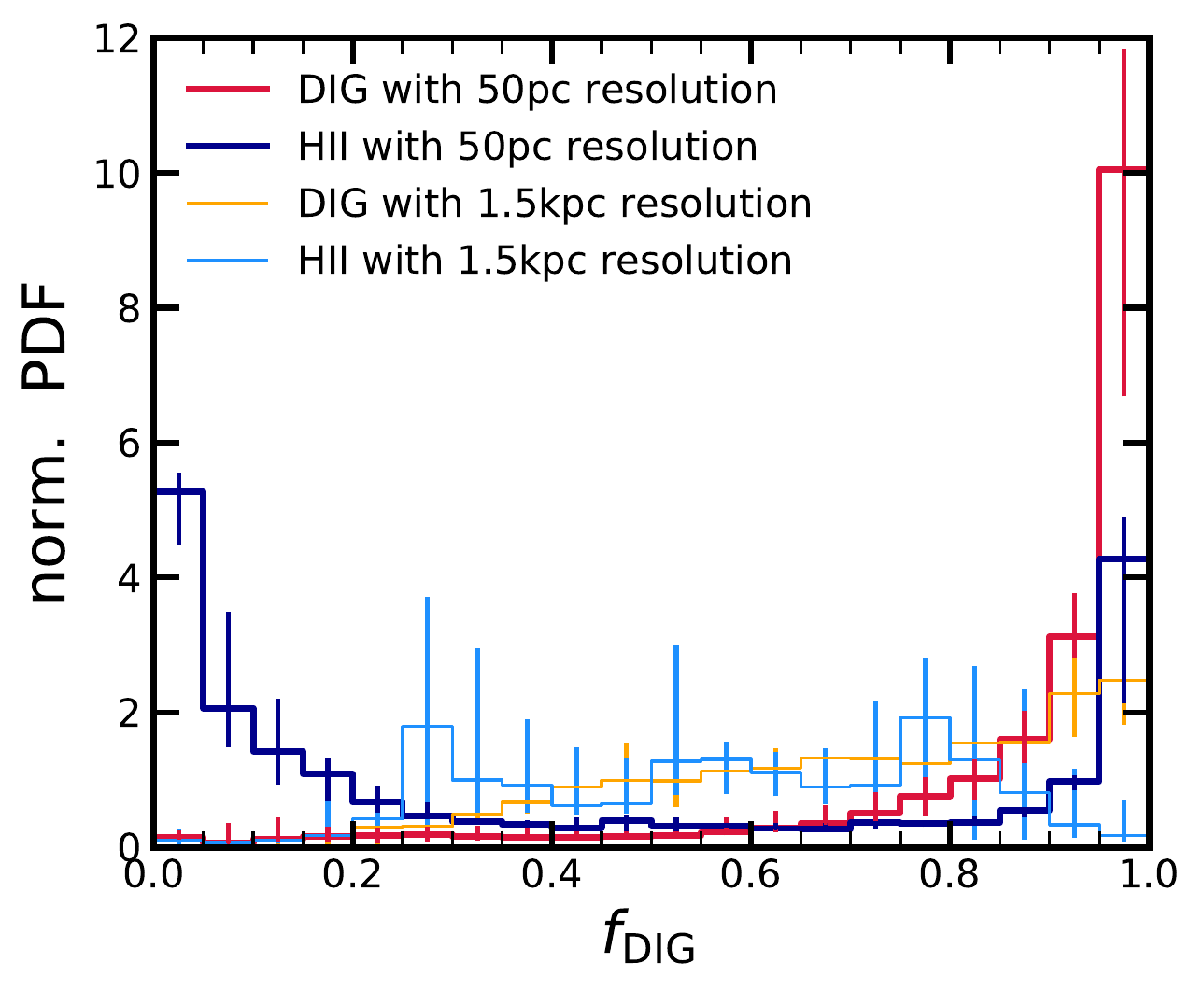}
\caption{Comparison of the different DIG selection methods on sptially resolved scales. We plot the normalised distribution of $f_{\rm DIG}$, as defined by our fiducial density threshold method, for DIG and \HII regions as identified by the clump finding method on images with 50\,pc (thick red and blue lines) and 1.5\,kpc (thin orange and bright blue lines) resolution. The errorbars indicate the $16$--$84^\text{th}$ percentile of the distribution when considering all snapshots. The diffuse regions (non-clumps) indeed correspond to regions of high $f_{\rm DIG}$. However, some clumpy features that are identified as \HII regions actually have a high $f_{\rm DIG}$, which leads to the overall lower $f_{\rm DIG}$ in the clump finding method in comparison with the density selection method as shown in Fig.~\ref{fig:f_DIG}. Lowering the resolution of the clump finding method leads to more mixing.}
\label{fig:compare_DIG_method}
\end{figure}

\begin{figure*}
\includegraphics[width=\textwidth]{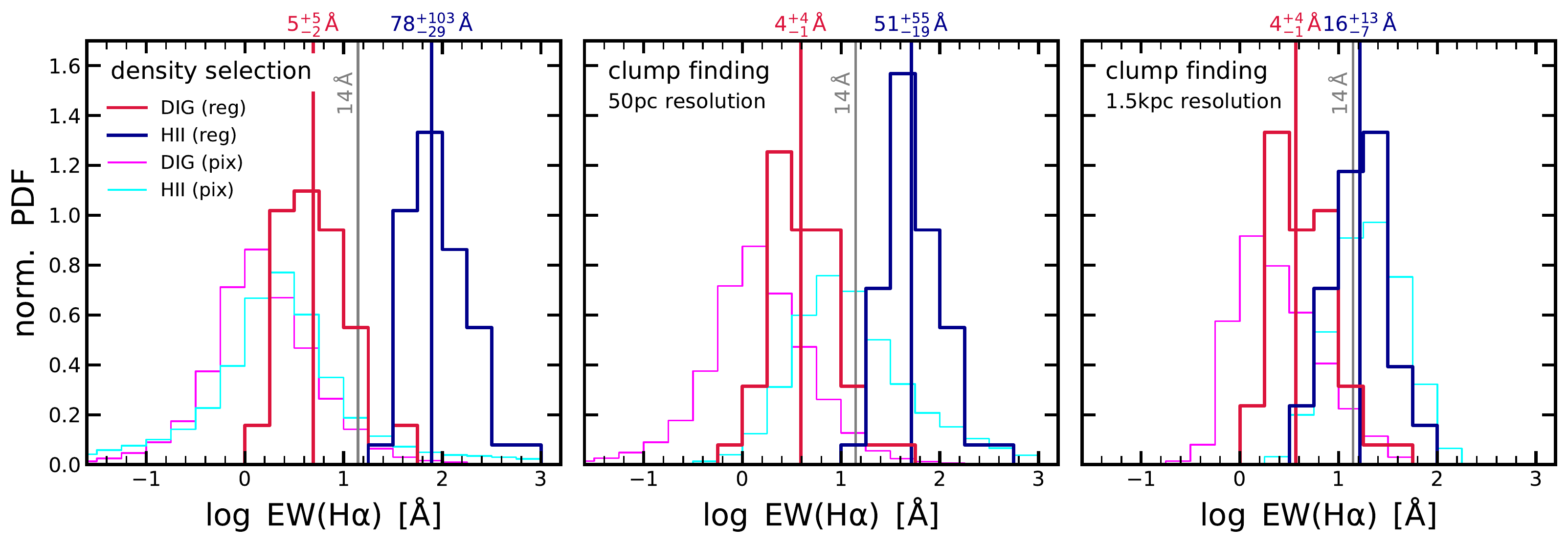}
\caption{H$\alpha$ equivalent width (EW) distributions for the DIG (red) and \HII (blue) regions as identified by the different selection methods. From left to right we show the results for our fiducial physical DIG definition that is based on a density cut of the emitting gas, and clump finding on images with 50\,pc and 1.5\,kpc resolution. These distributions depend not only on the DIG definition, but also on the aperture over which the EW are calculated: the red and blue histograms (together with the vertical lines indicating the median and $16$--$84^\text{th}$ percentile of the distributions) estimate the EW over the entire DIG and \HII regions (``reg''), while the magenta and cyan histograms show the EW of all the individual $10\mathrm{pc}\times10\mathrm{pc}$ pixels classified as either DIG or \HII (``pix''). The vertical grey lines indicate 14\,\AA, suggested by \citet{lacerda18} as a cut to select pure \HII regions.}
\label{fig:EW_Halpha}
\end{figure*}

As mentioned in the previous section, our fiducial identification of the DIG is based on a density cut of the emitting gas: DIG photons are emitted by gas with a density of $n<100~\mathrm{cm}^{-3}$. We now explore an additional DIG identification method, which is based on finding clumps in images. Specifically, high surface brightness regions (clumps) are classified as \HII regions, while the remaining diffuse emission is classified as DIG. This is similar to what is done in some observational data \citep[e.g.,][]{thilker00, oey07, barnes21}. As images, we use the spatially resolved face-on projections with two different resolutions: 50\,pc to mimic MUSE-like resolution and 1.5\,kpc to model MANGA-like resolution of nearby galaxies. The 50\,pc H$\alpha$ surface brightness maps are presented in Fig.~\ref{fig:MW_maps}, while the effect of lowering the resolution is shown in Appendix~\ref{appsec:smoothing}. The clumps are identified with \texttt{astrodendro} \citep{robitaille19}, with a surface brightness threshold of $10^{32}~\mathrm{erg}~\mathrm{s}^{-1}~\mathrm{pc}^{-2}$, a minimum height for a leaf to be considered an independent entity of 5, and a minimum number of pixels for a leaf to be considered an independent entity of 10 (pixel size of 10\,pc). Varying these parameters by an order of magnitude did not affect our results significantly (i.e. $f_{\rm DIG}$ changed by less than 10\%), i.e. the clump finding with this method is robust.  

In Fig.~\ref{fig:f_DIG} we plot the fraction of the H$\alpha$ Balmer emission originating from DIG, $f_{\rm DIG}$, as a function of the H$\alpha$ surface brightness, which is defined as
\begin{equation}
    \Sigma_{\rm H\alpha}=\frac{L_{\rm H\alpha}^{\rm tot}}{2 \pi R^2_{\rm half,H\alpha}},
\label{eq:SHa}
\end{equation}
where $L_{\rm H\alpha}^{\rm tot}$ is the total H$\alpha$ luminosity and $R_{\rm half,H\alpha}$ is the half-light radius of the galaxy's H$\alpha$ emission. We measure $R_{\rm half,H\alpha}$ by determining the radius from the face-on projected maps that encloses half of the total H$\alpha$ flux.

Fig.~\ref{fig:f_DIG} compares the aforementioned different DIG selection methods. Specifically, the left panel shows the result for our fiducial density cut, while the middle and right panels show the results for the clump finding applied to 50\,pc and 1.5\,kpc resolution images. In all panels, the red, purple and teal symbols represent the MW, LMC-BC03 and LMC-BPASS simulations, respectively. The filled symbols show the observed $f_{\rm DIG}$ values, while the open symbols show the intrinsic $f_{\rm DIG}$ values.

The first thing to notice is -- regardless of the exact definition of the DIG -- that the MW values are more clustered together, while the values for the LMC show more scatter, which is largely driven by the more variable star-formation activity. Furthermore, there is a stark difference between the observed and intrinsic $f_{\rm DIG}$: the observed $f_{\rm DIG}$ is about a factor of $2$--$3$ larger than the intrinsic $f_{\rm DIG}$. This makes sense physically since the DIG H$\alpha$ photons are emitted in less dense regions, which allows those photons to escape more easily, while the \HII photons are more likely to be absorbed by dust locally. Investigating the effects of different DIG selection methods, we find that the clump finding approach typically leads to smaller observed $f_{\rm DIG, obs}$ values, but similar intrinsic $f_{\rm DIG, int}$ values, in comparison with our fiducial density-based method. Decreasing the resolution of the images to which the clump finding is applied to leads to an increase of the DIG fraction since high-surface brightness clumps are more smeared out. The exact quantitative values can be found in Tab.~\ref{tab:fdig}.

\begin{table*}
    \setlength{\tabcolsep}{5pt}
    \renewcommand{\arraystretch}{1.5}
    \caption{DIG fraction of the Balmer H$\alpha$ emission for different DIG selection method: our fiducial density selection selection versus clump finding with different resolutions. We also differentiate between the observed (obs) and intrinsic (int) emission.}\label{tab:fdig}
    \centering
        \begin{tabular}{llcclcclcc}
          & & \multicolumn{2}{l}{density selection} & & \multicolumn{2}{l}{clump finding (50 pc)} & & \multicolumn{2}{l}{clump finding (1.5 kpc)}  \\
          \cline{3-4} \cline{6-7} \cline{9-10} 
          Simulation & & $f_{\rm DIG, obs}$ & $f_{\rm DIG, int}$ & & $f_{\rm DIG, obs}$ & $f_{\rm DIG, int}$ & & $f_{\rm DIG, obs}$ & $f_{\rm DIG, int}$ \\ 
          \cline{1-1} \cline{3-4} \cline{6-7} \cline{9-10}
          MW        & & $0.59_{-0.08}^{+0.10}$ & $0.24_{-0.07}^{+0.09}$ & & $0.44_{-0.04}^{+0.08}$ & $0.23_{-0.07}^{+0.09}$ & & $0.42_{-0.10}^{+0.12}$ & $0.37_{-0.13}^{+0.11}$  \\
          LMC-BC03  & & $0.49_{-0.17}^{+0.22}$ & $0.26_{-0.13}^{+0.21}$ & & $0.32_{-0.13}^{+0.12}$ & $0.19_{-0.09}^{+0.11}$ & & $0.50_{-0.21}^{+0.18}$ & $0.45_{-0.19}^{+0.23}$  \\
          LMC-BPASS & & $0.66_{-0.23}^{+0.19}$ & $0.39_{-0.17}^{+0.23}$ & & $0.43_{-0.17}^{+0.15}$ & $0.42_{-0.10}^{+0.12}$ & & $0.37_{-0.19}^{+0.32}$ & $0.33_{-0.18}^{+0.33}$  \\  \hline
        \end{tabular}\\
\end{table*}

In addition, we perform an approximate comparison to observations from the SINGG H$\alpha$ survey \citep{oey07} and the PHANGS-MUSE survey (\citealt{emsellem21, belfiore22}) in Fig.~\ref{fig:f_DIG}. We emphasise that this is only a ``rough'' comparison because it is difficult to mimic the same method as adopted in the observations, such as modelling the exact noise properties of the observational data and the data processing (i.e., binning of pixels in order to achieve a homogeneous signal-to-noise ratio). \citet{oey07} showed that $f_{\rm DIG}$ decreases with increasing $\Sigma_{\rm H\alpha}$. Specifically, they argue that \ion{H}{II} regions occupy a larger fraction of the ionized ISM volume as star formation becomes more concentrated, predicting a dependence of $f_{\rm DIG} \propto \Sigma_{\rm H\alpha}^{1/3}$. This agrees well with data from the SINGG H$\alpha$ survey (grey plus symbols in Fig.~\ref{fig:f_DIG}). \citet{sanders17}, using the data from \citet{oey07}, fitted $f_{\rm DIG}$ as a function of $\Sigma_{\rm H\alpha}$, obtaining:
\begin{equation}
    f_{\rm DIG}=-1.5\times10^{-14}\times\Sigma_{\rm H\alpha}^{1/3} + 0.748 \, ,
\label{eq:fDIG}
\end{equation}
where $\Sigma_{\rm H\alpha}$ is in units of $\mathrm{erg}~\mathrm{s}^{-1}~\mathrm{kpc}^{-2}$. This best-fit is shown as the solid grey curves in Fig.~\ref{fig:f_DIG}. Recently, the higher resolution data from the PHANGS-MUSE survey \citep{belfiore22} point towards a lower DIG fraction than what \citet{oey07} reported at fixed $\Sigma_{\rm H\alpha}$. Those observations are shown as dark grey crosses in Fig.~\ref{fig:f_DIG}. Specifically, in the MW range of $\Sigma_{\rm H\alpha}$ \citet{oey07} reports a median of $f_{\rm DIG}=0.66_{-0.12}^{+0.13}$, while \citet{belfiore22} finds $f_{\rm DIG}=0.42_{-0.05}^{+0.11}$ (see also \citealt{chevance20}, who estimate the DIG fraction of a subsample of PHANGS galaxies using H$\alpha$ narrow-band data, finding overall consistent results with \citealt{belfiore22}). This observational difference indicates that there is not yet a consensus and different observational methods and datasets can easily lead to differences on the 10--20\% level. Nevertheless, we find that our MW simulation agrees very well with the more recent PHANGS-MUSE observations when adopting the clump finding method with 50\,pc resolution data ($f_{\rm DIG, obs}0.44_{-0.04}^{+0.08}$ in the simulation versus $0.42_{-0.05}^{+0.11}$ in the observations). The LMC-BC03 and LMC-BPASS simulations are consistent with the observations as well, but it is more difficult to make a more conclusive statement due to the scarcity of the observational data and the overall large scatter (in both observations and simulations). Obviously, a more rigorous comparison between simulations and observations is needed in the future. Furthermore, investigating other DIG selection methods (for example based on the emission line [\ion{S}{II}]) is also of great interest (Section~\ref{subsec:future}).

Looking further into the different DIG selection methods at hand in this work, it is clear that the clump finding identifies some DIG region photons as stemming from \HII regions, which leads to a lower DIG fraction with respect to the physical density cut. We show this explicitly in Fig.~\ref{fig:compare_DIG_method} for the MW simulation, which plots the normalised distribution of $f_{\rm DIG}$ for DIG and \HII regions as identified by the clump finding method on images with 50\,pc (thick red and blue lines) and 1.5\,kpc (thin orange and bright blue lines) resolution. Here, $f_{\rm DIG}$ refers to our fiducial density-based method and has been estimated for each spatial region separately. We only consider spatial regions above a surface brightness of $10^{30}~\mathrm{erg}/\mathrm{s}/\mathrm{kpc}^{2}$, motivated by observations \citep[e.g.,][]{belfiore22}. The errorbars indicate the $16$--$84^\text{th}$ percentile of the distribution when considering all snapshots. As clearly shown in the case for the high-resolution analysis, some clumpy features that are identified as \HII regions actually have a high $f_{\rm DIG}$, while the diffuse regions (non-clumps) indeed correspond to regions of high $f_{\rm DIG}$. This explains the aforementioned difference (see also Fig.~\ref{fig:f_DIG}). When lowering the resolution, there is much more mixing, leading to \HII regions with high $f_{\rm DIG}$.

Finally, we focus on the H$\alpha$ equivalent width (EW) distribution of the DIG and \HII emission. Some observational studies \citep[e.g.,][]{lacerda18, vale-asari19} define regions with low EW of H$\alpha$ emission (e.g., $\mathrm{EW(H\alpha)}<3$\,\AA) as DIG. The motivation from this stems form the fact that EW(H$\alpha$) can distinguish between ionization due to hot low-mass evolved stars (``HOLMES''; low- to intermediate-mass stars with $0.8$--$8~\mathrm{M_{\odot}}$ in all stages of stellar evolution subsequent to the asymptotic giant branch) from that of star formation (and AGN). The exact boundary between HOLMES and star formation depends on metallicity, IMF and stellar evolution tracks, but it has been put forward that HOLMES typically produce EW(H$\alpha$) of $0.5$--$2.5$\,\AA\ \citep[e.g.,][]{cid-fernandes14, byler19}. Inspired both by theoretical and empirical considerations, \citet{lacerda18} suggested a three-tier scheme: regimes where $\mathrm{EW(H\alpha)}<3$\,\AA\ are dominated by HOLMES (i.e. DIG), regions where $\mathrm{EW(H\alpha)}>14$\,\AA\ trace star-formation complexes (i.e. \HII regions), and the intermediate regions with $\mathrm{EW(H\alpha)}=3$--$14$\,\AA\ reflect a mixed regime.

An important consideration in the discussion of the EW(H$\alpha$) on spatially resolved scales is -- in addition to spatial resolution -- the size of the aperture. In particular, is the EW calculated over individual pixels or binned pixels (according to some signal-to-noise criterion or some classification schema, such as DIG versus \HII regions)? Since the EW is a ratio quantity, calculating the EW for individual or binned pixel can have an important effect, as we highlight now. 

Fig.~\ref{fig:EW_Halpha} plots the normalised EW(H$\alpha$) distributions for the DIG and \HII regions in red and blue, respectively. In the left panel, we use our fiducial physical DIG definition that is based on a density cut of the emitting gas to identify DIG regions in the face-on projection ($f_{\rm DIG}>0.5$). In the middle and right panel, we identified \HII region with the clump finding method described above. These distributions depend not only on the DIG definition, but also on the aperture over which the EW are calculated: the red and blue histograms (together with the vertical lines indicating the median and $16$--$84^\text{th}$ percentile of the distributions) estimate the EW over the entire DIG and \HII regions (``reg'', i.e. binned pixels), while the magenta and cyan histograms show the EW of all the individual $10\mathrm{pc}\times10\mathrm{pc}$ pixels classified as either DIG or \HII (``pix''). We find that this pixel binning affects the distributions significantly, in particular in case of high-resolution data. Binning individual pixels together leads to a bias towards the brightest pixels, hence higher EWs. This effect makes it difficult to perform a robust comparison to observations, since observers typically bin pixels in a non-homogeneous manner in order to achieve an optimal arrangement of signal-to-noise ratios. Quantitatively, we find, when calculating the EW over entire DIG and \HII regions, $\mathrm{EW(H\alpha)}=5_{-2}^{+5}$\,\AA\ for the DIG and $\mathrm{EW(H\alpha)}=78_{-29}^{+103}$\,\AA\ for the \HII regions when performing our fiducial density selection (left panel of Fig.~\ref{fig:EW_Halpha}). Adopting the clump finding method on 50 pc data (middle panel) leads to similar EWs (DIG $\mathrm{EW(H\alpha)}=4_{-1}^{+4}$\,\AA\ for DIG and 
$\mathrm{EW(H\alpha)}=51_{-19}^{+55}$\,\AA\ for \HII regions). In both cases, a value of 14\,\AA\ (vertical grey line) seems to work well for selecting pure \HII regions, which is consistent with \citet{lacerda18}. Furthermore, the 50 pc clump finding results are in excellent agreement with the quoted values from the PHANGS-MUSE survey \citep{belfiore22}: $\mathrm{EW(H\alpha)}=5.3_{-3.0}^{+12.1}$\,\AA\ for the DIG and $\mathrm{EW(H\alpha)}=48.7_{-34.0}^{+98.4}$\,\AA\ for the \HII regions. Lowering the resolution (right panel) leads to more mixing and a smaller difference in the EWs of the DIG and \HII regions. Specifically, the EW threshold to identify \HII region needs to be lowered. 

Focusing on the effect of binning, we find that EW distributions are significantly lower when evaluating individual pixels. This effect is particularly drastic when considering high resolution data: $\mathrm{EW(H\alpha)}=1.5_{-0.4}^{+0.7}$\,\AA\ for the DIG and $\mathrm{EW(H\alpha)}=2.1_{-0.2}^{+0.5}$\,\AA\ for the \HII regions when performing our fiducial density selection, while $\mathrm{EW(H\alpha)}=1.5_{-0.4}^{+0.5}$\,\AA\ for the DIG and $\mathrm{EW(H\alpha)}=12.5_{-1.9}^{+4.0}$\,\AA\ for the \HII regions when adopting the 50pc clump finding. The reason for this is that the faint H$\alpha$ regions ($\mathrm{EW(H\alpha)}<1$\,\AA; regions where stars are present but nearly no H$\alpha$ emission) are binned into larger regions, where their contribution is insignificant. The overall effect is that the long tail to low EWs is binned up and high EW regions dominate. Additionally, adopting the clump finding method leads to a larger difference on the individual pixels than for the density selection, with \HII regions having an EW(H$\alpha$) that is roughly an order-of-magnitude larger than the DIG regions. This is not surprising, because the clump finding method by definition classifies regions of high H$\alpha$ surface brightness as \HII regions, which typically also have a higher EW(H$\alpha$).

In summary, although different selection techniques lead to quantitative different DIG fractions, we find that the observed DIG fraction is of the order of $\sim50\%$. Importantly, the intrinsic DIG fraction (i.e. before accounting for dust absorption and scattering) is a factor of 2--3 times lower. The H$\alpha$ EW is not a good tracer of whether an H$\alpha$ photon has been emitted in dense gas (\HII region) or diffuse gas (DIG region). A key question that we have not yet addressed in the section is: \textit{What powers the DIG?} We find that leaking radiation from \HII regions can play a major role. As shown in our companion paper (Fig. 25 in \citealt{smith21_rt}), in the MW simulation about 30\% of all the LyC photons travel 0.01--1.0\,kpc before they ionize hydrogen (or undergo dust absorption), with an additional 15\% travelling beyond 1\,kpc\footnote{About 6\% of the LyC radiation escapes the galaxy.}. This radiation is in principle enough to power an intrinsic DIG fraction of 20--30\% as required. Nevertheless, we find that additional processes such as collisional excitation and ionization (related to, e.g., shocks) also contribute to the H$\alpha$ emission on the 5--10\% level and therefore can contribute substantially to the powering of the DIG (Section~\ref{subsec:conversion_factor}). As shown in Appendix~\ref{appsec:stellar_pop}, older stars ($>10^8$\,yr) contribute to $\approx2\%$ the ionizing budget of the MW simulation (this increases to $\approx8\%$ for the LMC simulation assuming the BC03 SPS model) and therefore do not dominate the DIG emission in these simulations. This has implications for the high-ionization lines \citep[e.g.,][]{zhang17_DIG}, which is of great interest for upcoming studies (Section~\ref{subsec:future}).

\subsection{Extraplanar Balmer emission}
\label{subec:vertical_analysis}

\begin{figure*}
\begin{center}
\includegraphics[width=\textwidth]{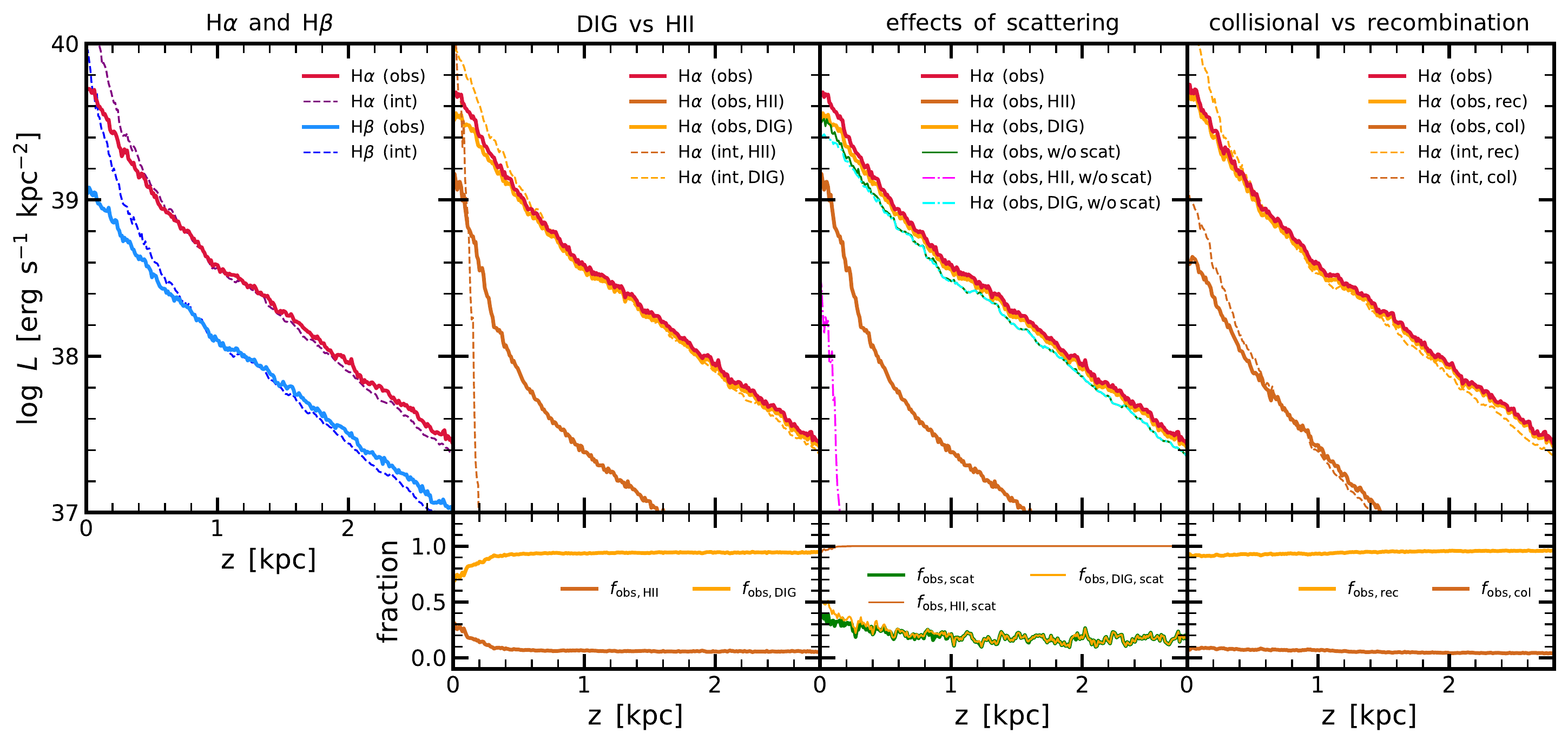}
\caption{Extraplanar Balmer emission of the MW simulation. From left to right, the vertical profiles of H$\alpha$ and H$\beta$, of H$\alpha$ split by DIG and \ion{H}{II} photons, of H$\alpha$ split by scattered and non-scattered photons, and H$\alpha$ split by radiative recombination and collisional excitation. The smaller panels on the bottom show the fractional contribution of the different components. All the vertical profiles are obtained within the effective radius of 4.3 kpc. The key message from this figure is that the Balmer emission at large scale heights ($z>200~\mathrm{pc}$) is mainly emitted in-situ via radiative recombination, emitted in low-density gas (i.e. the DIG), and scattering is significant but not dominant ($f_{\rm scat}\approx 20\%$).}
\label{fig:vertical_diag}
\end{center}
\end{figure*}

We now focus on understanding the origin of the extraplanar Balmer emission, which is directly related to the previously discussion about the DIG. As we have highlighted in the Introduction, there are two possible explanations for this extraplanar emission. On one hand, this emission could be produced by ionizing photons transported through transparent pathways carved out by superbubbles or chimneys. On the other hand, it could be caused by dust scattering of the photons originating from \ion{H}{II} regions in the galactic disc. Motivated by the good agreement between the scale heights of our simulated MW galaxy and the observations (Section~\ref{subsec:vertical_heights_comparison}), we now investigate the physical origin of this extraplanar Balmer emission.

In Fig.~\ref{fig:vertical_diag} we show the vertical profiles within the effective radius of 4.3 kpc for the MW simulation split by different diagnostics. The left panel shows the vertical profiles for H$\alpha$ and H$\beta$ emission. The solid and dashed lines show the observed and intrinsic emission, respectively. As expected, the intrinsic profile is nearly 1\,dex higher than the observed profile within the disc ($z<200~\mathrm{pc}$), which can be explained by the absorption of dust. There is very little difference between the intrinsic and observed emission at high altitudes: the observed emission is slightly above ($<0.2~\mathrm{dex}$) the intrinsic emission because of scattering (see below).

The second panel from the left in Fig.~\ref{fig:vertical_diag} splits the H$\alpha$ emission by photons emitted in \ion{H}{II} regions and the DIG, as defined by the simple gas density cut ($n_{\rm thresh}=100~\mathrm{cm}^{-3}$). We find that the DIG dominates at all scale heights. The fraction of H$\alpha$ photons emitted by the DIG is $f_{\rm DIG}~70\%$ within the disc (see lower panel) and increases towards larger heights to $f_{\rm DIG}>90\%$. There is very little difference between the intrinsic and observed emission for the DIG component. Contrarily, the intrinsic \ion{H}{II} photons are basically all emitted within the disc ($z<200~\mathrm{pc}$), while they are observed out to much higher scale heights ($z>1~\mathrm{kpc}$). This can be explained by dust scattering, as we discuss next.

In the third panel from the left in Fig.~\ref{fig:vertical_diag} we investigate the importance of scattering of H$\alpha$ photons. This shows that indeed all ($\sim100\%$) of the \ion{H}{II} photons at large scale heights have been scattered. Looking at DIG photons, we find that scattering is significant with a fraction of $10$--$30\%$. Since the DIG photons dominate, we find that the majority of extraplanar H$\alpha$ emission is not scattered ($f_{\rm scat}\approx10$--$30\%$).

Finally, in the rightmost panel of Fig.~\ref{fig:vertical_diag} we split the H$\alpha$ emission by the physical emission process, i.e., radiative recombination and collisional excitation. We find that at all scale heights, radiative recombination clearly dominates.

In summary, the extraplanar emission extends to several kpc in our MW simulation. The exponential scale height amounts on average to 0.7\,kpc, consistent with observations (see Fig.~\ref{fig:scaleheight_comparison}). As shown in Fig.~\ref{fig:vertical_diag}, this extraplanar Balmer emission is produced in-situ by ionizing photons emitted from the disc via radiative recombination. Those ionizing photons must be transported through transparent pathways carved out by superbubbles or chimneys (see Fig.~\ref{fig:MW_maps}). This is consistent with our measurement that about 30\% of all the LyC photons travel 0.01--1.0\,kpc before they ionize hydrogen (see Fig. 25 of our companion paper, \citealt{smith21_rt}). Since the gas density is low at large scale heights, we classify all of this Balmer emission as being emitted from the DIG. We find that scattering increases the H$\alpha$ luminosity above the plane of the disc by roughly $f_{\rm scat}\approx10$--$30\%$.

\section{Balmer lines as a star-formation tracer}
\label{sec:sfr}

H$\alpha$ is a prime indicator for the SFR of galaxies, both on global and spatially resolved scales (see Introduction). We now turn to understanding the connection between the SFR and H$\alpha$ in more detail. In particular, we look into the time evolution of H$\alpha$ and SFR (Section~\ref{subsec:time_evolution}), the connection between the intrinsic H$\alpha$ emission and the SFR (``conversion factor''; Section~\ref{subsec:conversion_factor}), dust correction (Section~\ref{subsec:dust}), and star-formation timescale probed by H$\alpha$ (Section~\ref{subsec:timescale}).

\subsection{Time evolution of the H\texorpdfstring{$\balpha$}{α} luminosity and SFR}
\label{subsec:time_evolution}

\begin{figure*}
\includegraphics[width=0.38\linewidth]{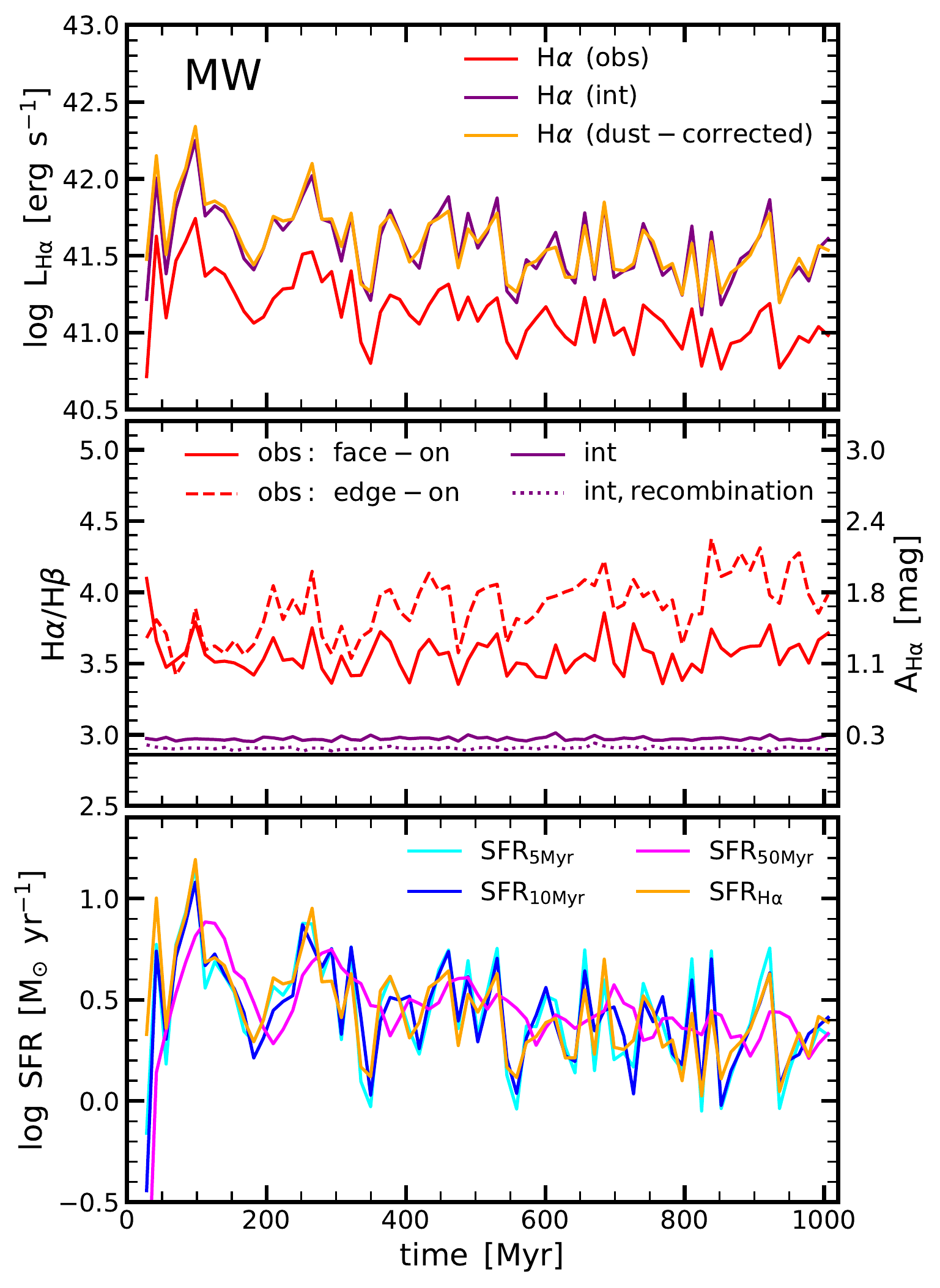}\!\!\includegraphics[width=0.627\linewidth]{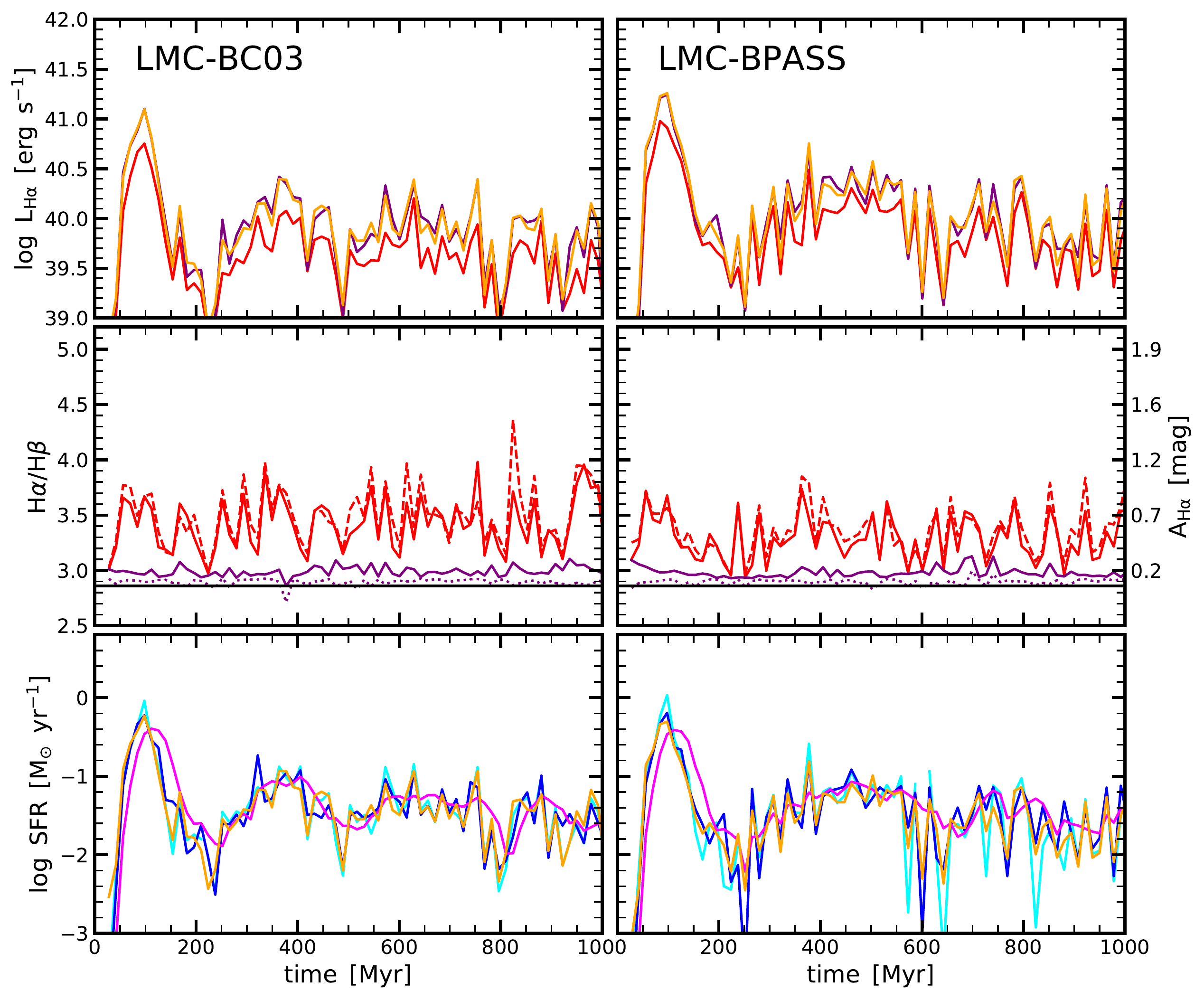}
\caption{Time evolution of the H$\alpha$ luminosity (top panels), Balmer decrement (ratio of the observed H$\alpha$/H$\beta$ in the middle panels), and the SFR (bottom panels) for the MW (left panels), LMC-BC03 (middle panels) and LMC-BPASS (right panels) simulations. All quantities are measured within an aperture of 15\,kpc. The red, purple and orange lines in the top panel show the observed, intrinsic and dust-corrected (using the Balmer decrement) H$\alpha$ luminosities. The Balmer decrement oscillates around a typical value of $\sim3.5$, which corresponds to an attenuation for H$\alpha$ of $\mathrm{A}_{\mathrm{H}\alpha}\approx1.0~\mathrm{mag}$. The SFR is estimated considering stars with ages $t_{\rm age}<5~\mathrm{Myr}$ (cyan), $<10~\mathrm{Myr}$ (blue), and $<50~\mathrm{Myr}$ (pink). The orange line in the bottom panel shows the SFR estimated from the dust-corrected H$\alpha$ luminosity using the Balmer decrement, which closely follows the SFR averaged over short timescales ($<5$--$10~\mathrm{Myr}$).}
\label{fig:timeseries}
\end{figure*}

In Fig.~\ref{fig:timeseries} we show the temporal evolution of the H$\alpha$ luminosity (top panels), Balmer decrement (ratio of the observed H$\alpha$/H$\beta$ in the middle panels), and the SFR (bottom panels) for the MW (left panels), LMC-BC03 (middle panels) and LMC-BPASS (right panels) simulations. All measurements are on global scales; i.e. obtained with an aperture of 15\,kpc. For the H$\alpha$ luminosity, we plot the observed luminosity (red line), intrinsic luminosity (without dust absorption and scatter; purple line), and dust-corrected luminosity (orange line), which we perform via the Balmer decrement as described in Section~\ref{subsec:dust}. As we see, for all simulations, the dust correction works extremely well and is able to recover the intrinsic H$\alpha$ luminosity nearly perfectly.

In the middle panels of Fig.~\ref{fig:timeseries} we show the Balmer decrement for face-on and edge-on views as red solid and dashed lines, respectively. In the MW case, we can clearly see that the edge-on alignment exhibits a larger Balmer decrement, which is consistent with the idea that edge-on integrations lead to higher attenuation (by about 0.5\,mag for H$\alpha$). There is very little difference between the different sightlines in the LMC simulations due to the lower dust content. In these middle panels, we also plot the intrinsic Balmer decrement as solid purple lines. We find on average an intrinsic Balmer decrement of $(\mathrm{H}\alpha/\mathrm{H}\beta)_{\rm int}=2.97_{-0.01}^{+0.01}$, $2.99_{-0.04}^{+0.05}$ and $2.97_{-0.03}^{+0.04}$ for the MW, LMC-BC03 and LMC-BPASS simulations; all values are close to the fiducial value of 2.87 (shown as vertical black line) that assumes an electron temperature of $T_{e}=10^4~\mathrm{K}$, an electron density of $n_e = 10^2~\mathrm{cm}^{-3}$, and Case B recombination conditions \citep{osterbrock06}. Most of the differences can be explained by collisional excitation emission, which intrinsically has a higher Balmer decrement (the dotted purple line shows the Balmer decrement for the intrinsic radiative recombination radiation), while the rest is due to the more realistic range of different gas temperatures and densities (see Fig.~\ref{fig:phase}).

In the bottom panels of Fig.~\ref{fig:timeseries} we plot the SFR estimated from H$\alpha$ (orange line) assuming an effective conversion factor (see next section) and direct measurements of the SFR by considering all stars born in the past 5\,Myr (cyan line), 10\,Myr (blue line), or 50\,Myr (pink line). The H$\alpha$-based SFR employs the dust-corrected H$\alpha$ luminosity and an \textit{effective} H$\alpha$--SFR conversion factor, which we discuss in detail in the next section. We see that the H$\alpha$-based SFR accurately traces the $<5$--$10$\,Myr SFR, something we explore more in Section~\ref{subsec:timescale}.

\subsection{The H\texorpdfstring{$\balpha$}{α}--SFR conversion factor}
\label{subsec:conversion_factor}

\begin{figure*}
\begin{center}
\includegraphics[width=\textwidth]{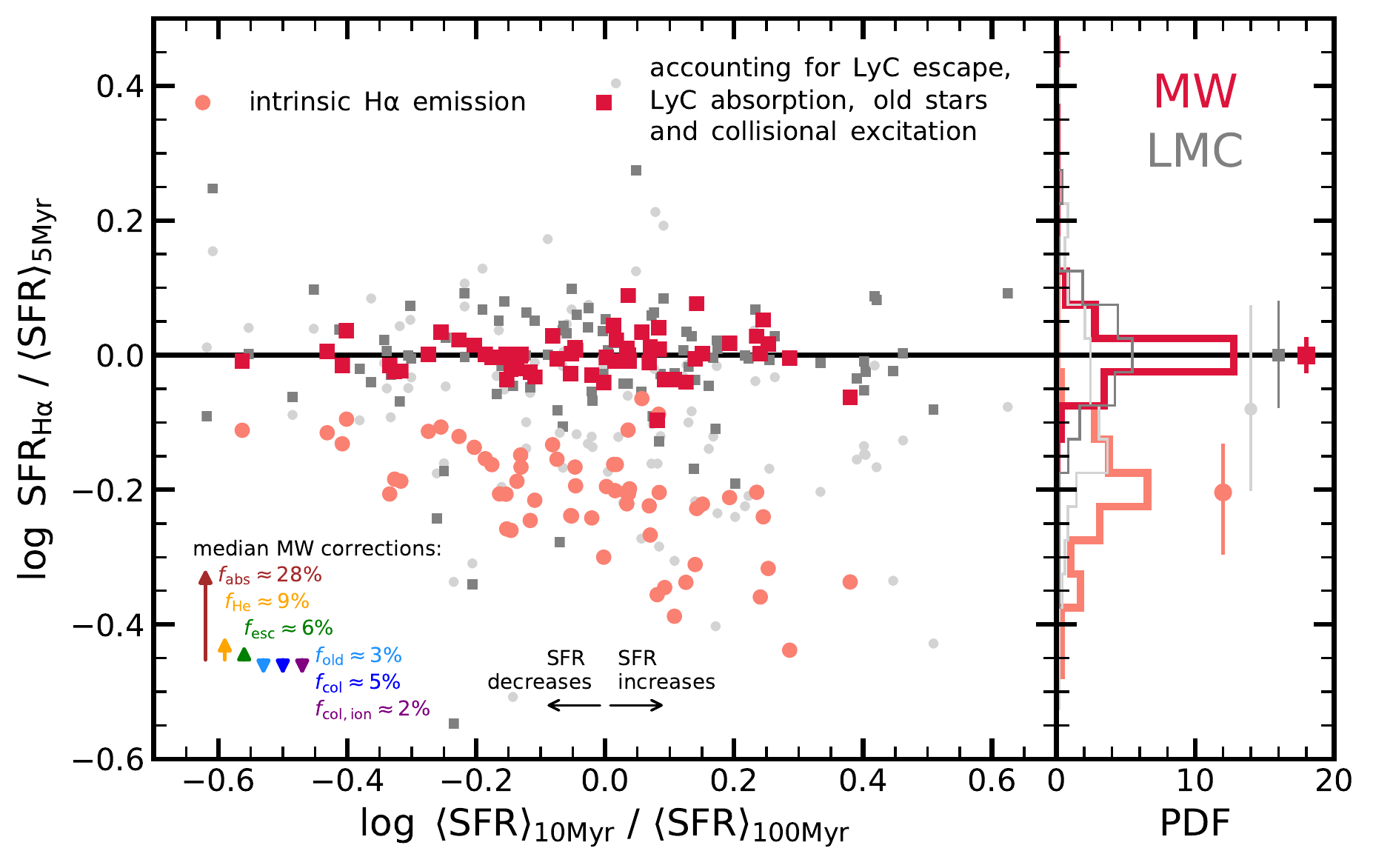}
\caption{H$\alpha$--SFR connection. We compute the SFR from the intrinsic H$\alpha$ emission (i.e. no dust absorption) by multiplying it with the \citet{kennicutt98} conversion factor (Eq.~\ref{eq:SFR_conversion}) and denote it by SFR$_{\rm H\alpha}$. We plot the ratio of SFR$_{\rm H\alpha}$ and the true SFR (averaged over 5\,Myr; $\langle\mathrm{SFR}\rangle_{\rm 5Myr}$) as a function of $\langle\mathrm{SFR}\rangle_{\rm 10Myr}/\langle\mathrm{SFR}\rangle_{\rm 100Myr}$, which indicates whether the SFR is increasing or decreasing over the past 100\,Myr. The salmon and red symbols are from the MW simulations, while the grey symbols are from both the LMC-BC03 and LMC-BPASS simulations. We find that the simple H$\alpha$-based SFR underpredicts the true SFR by $0.20^{+0.09}_{-0.07}$ ($0.05^{+0.12}_{-0.12}$ and $0.12^{+0.11}_{-0.24}$) dex in the MW (LMC-BC03 and LMC-BPASS) case. For the MW simulation, this can be explained by the dust and helium absorption of ionizing LyC photons ($f_{\rm abs}\approx28\%$ and $f_{\rm He}\approx9\%$), while effects related to the escape of LyC photons and H$\alpha$ emission collisionally excited gas roughly cancel each other out ($f_{\rm esc}\approx6\%$ versus $f_{\rm col}\approx5\%$), see arrows in the lower left for the average strengths of these processes. Contributions from older stars ($f_{\rm old}\approx3\%$) and collisionally ionized gas ($f_{\rm col,ion}\approx2\%$) are both small. Since the $f_{\rm abs}$ depends on the recent SFH, these corrections are actually dependent on the SFH itself.}
\label{fig:Ha_SFR}
\end{center}
\end{figure*}

In order to derive a SFR from the observed H$\alpha$ luminosity, two steps are necessary. Firstly, the observed H$\alpha$ luminosity needs to be corrected for dust attenuation to derive the intrinsic H$\alpha$ luminosity. Secondly, the intrinsic H$\alpha$ luminosity needs to be converted to the SFR via an H$\alpha$-to-SFR conversion factor. We now focus on the H$\alpha$-to-SFR conversion (i.e. focusing on the relation between the intrinsic H$\alpha$ emission and its relation with star formation) and discuss the dust correction in the next section. For this section, by ``SFR'' we mean the SFR considering all stars born in the past 5\,Myr (i.e. $\langle\mathrm{SFR}\rangle_{\rm 5Myr}$; Section~\ref{subsec:timescale}).

The intrinsic (or dust-corrected) H$\alpha$ luminosity $L(\mathrm{H}\alpha)_{\rm int}$ can be converted to a SFR via
\begin{equation}
    \mathrm{SFR} = C \times L(\mathrm{H}\alpha)_{\rm int},
\label{eq:SFR_conversion}
\end{equation}
\noindent
where $C$ is the H$\alpha$-to-SFR conversion factor. This factor typically assumes that: ($i$) the star formation has been roughly constant over the timescale probed (in case of H$\alpha$ a few tens of Myr), ($ii$) the stellar IMF is known, and ($iii$) the stellar IMF is fully sampled. Furthermore, in the case of nebular lines such as H$\alpha$, values for the electron temperature and density also need to be assumed. Not surprisingly, there is a significant variation among published calibrations ($\sim30\%$), with most of the dispersion reflecting differences in the stellar evolution and atmosphere models. We adopt for the MW simulation $C=4.5\times10^{-42}\,(\text{M}_{\odot}~\mathrm{yr}^{-1})/(\mathrm{erg}~\mathrm{s}^{-1})$, which is consistent -- after converting from a \citet{salpeter55} to a \citet{chabrier03} IMF using the conversion presented in \citet{driver13} -- with the widely used conversion factor of \citet{kennicutt98}. The conversion factor of the LMC-BC03 and LMC-BPASS simulations (both $0.5~\mathrm{Z}_{\odot}$) are taken to be 4.2 and $2.1\times10^{-42}\,(\text{M}_{\odot}~\mathrm{yr}^{-1})/(\mathrm{erg}~\mathrm{s}^{-1})$, respectively. 

We now convert the intrinsic H$\alpha$ luminosities to SFRs via Eq.~(\ref{eq:SFR_conversion}) with the aforementioned standard \citet{kennicutt98} conversion factor. We compare these SFRs ($\mathrm{SFR}_{\rm H\alpha}$) with the SFR obtained by considering all stars born in the previous 5\,Myr; i.e. the true SFR averaged over the previous 5\,Myr; $\langle\mathrm{SFR}\rangle_{\rm 5 Myr}$. Specifically, in Fig.~\ref{fig:Ha_SFR} we show the ratio of the H$\alpha$-based SFR and the true SFR of the simulation as a function of the ratio of the true SFR averaged over the previous 10\,Myr and the previous 100 Myr; i.e. the $x$-axis shows whether the star-formation history (SFH) is increasing or decreasing over the past 100\,Myr.

The salmon-coloured (bright grey) points in Fig.~\ref{fig:Ha_SFR} are obtained by adopting the H$\alpha$-based SFR of the MW (LMC-BC03 and LMC-BPASS) simulation with the simple conversion described above. We clearly see that these points lie below the black vertical line, indicating that the SFRs obtained from the intrinsic H$\alpha$ luminosity are systematically underestimating the true SFR averaged over the previous 5\,Myr. In fact, there is a trend with the SFH evolution in which snapshots where the SFH over the past 100\,Myr was increasing have H$\alpha$-based SFRs that are $0.2$--$0.4$ dex lower than the true SFR. This difference is only about 0.1 dex when the SFH is decreasing. The median difference for the MW, LMC-BC03 and LMC-BPASS simulations is $-0.20_{-0.09}^{+0.07}$, $-0.05_{-0.12}^{+0.12}$ and $-0.12_{-0.11}^{+0.24}$, respectively. The difference is typically less for the LMC than for the MW simulation. What is the reason for this difference?

As shown in Fig.~\ref{fig:cartoon}, there are several different physical processes that can reduce or boost the H$\alpha$ luminosity relative to the SFR; i.e. the formation rate of young stars. Firstly, ionizing LyC photons from young massive stars can be absorbed by dust and helium before ionizing hydrogen that recombines to emit H$\alpha$ photons. This reduction of H$\alpha$ production relative to the SFR is significant in our simulations: we find (median over all snapshots) that $f_{\rm abs}\approx28\%$ of LyC photons are absorbed by dust, while $f_{\rm He}\approx9\%$ of LyC photons ionizing helium\footnote{Most of the LyC photons absorbed by helium are actually absorbed by \HeI, producing \HeII. We track \HeII\ absorption as well (producing \HeIII), though this fraction is about 8 times lower than \HeI\ absorption.}. Secondly, some ionizing LyC photons escape the galaxy and do not interact with either gas or dust. The fraction of escaping LyC photons is low with an average of $f_{\rm esc}\approx6\%$. Thirdly, as discussed in Section~\ref{subsec:fcol}, collisionally excited gas actually boosts the H$\alpha$ emission relative to the SFR, though this fraction is low with an average of $f_{\rm col}\approx5\%$. Directly related to this is the collisionally ionized gas which contributes negligibly with $f_{\rm col,ion}\approx2\%$. Fourthly, older stars with ages of $>10$\,Myr only weakly boost the H$\alpha$ luminosity on the level of $f_{\rm old}\approx3\%$. Finally, one more channel that is self-consistently modelled is photoionization by the UV background, which produce very little recombination emission ($<1\%$), which we therefore do not report here.

These corrections are different for the LMC simulations: $(f_{\rm abs}, f_{\rm He}, f_{\rm esc}, f_{\rm col}, f_{\rm col,ion}, f_{\rm old})$ are on average $(12\%, 11\%, 5\%, 7\%, 4\%, 8\%)$ and $(15\%, 6\%, 12\%, 6\%, 2\%, 9\%)$ for the LMC-BC03 and LMC-BPASS simulation, respectively. This means that the budget of H-ionizing LyC photons is reduced to $f_{\rm H}\approx57\%$ (i.e. 43\% of the produced LyC photons are absorbed by dust or helium, or escape the galaxy) in the MW simulation, to $f_{\rm H}\approx72\%$ in the LMC-BC03, and to $f_{\rm H}\approx67\%$ in the LMC-BPASS. As expected, the LyC photon absorption by dust is less important for the LMC, while the contribution of older stars is more prominent since the specific SFR of the LMC is lower than that of the MW. Comparing BC03 with BPASS, we find that the LyC escape fraction of the LMC-BPASS model is about twice as high as for the LMC-BC03 model, which can be attributed to the LyC emission of slightly older stellar populations in the former model, which are typically located in less obscured regions of the galaxy. 

After correcting each individual snapshot for these effects -- the typical arrows of corrections for the MW simulation are shown in the bottom left of Fig.~\ref{fig:Ha_SFR} -- we find that the H$\alpha$-based SFR agrees well with the true SFR. The corrected values for the MW (LMC-BC03 and LMC-BPASS) simulation are shown as red-coloured (dark grey) squares in Fig.~\ref{fig:Ha_SFR}. We find the log difference to be $\log(\mathrm{SFR}_{\rm H\alpha, int}/\langle\mathrm{SFR}\rangle)=0.00_{-0.03}^{+0.03}$, $0.00_{-0.04}^{+0.07}$, and $0.00_{-0.09}^{+0.18}$ for the MW, LMC-BC03, and LMC-BPASS simulations, respectively. Importantly, we find that these aforementioned corrections depend on the SFH, i.e. the ratio of the SFR averaged over 10\,Myr and 100\,Myr (x-axis of Fig.~\ref{fig:Ha_SFR}). This makes it difficult to correct for those effects in practice. Nevertheless, we can define an \textit{effective} conversion factor, which takes into account these corrections on average. Specifically, we find $C_{\rm eff}=7.1$, 4.7 and $2.7\times10^{-42}\,(\text{M}_{\odot}~\mathrm{yr}^{-1})/(\mathrm{erg}~\mathrm{s}^{-1})$ for the MW, LMC-BC03 and LMC-BPASS simulation, respectively. These effective conversion factors lead to $\log(\mathrm{SFR}_{\rm H\alpha, int}/\langle\mathrm{SFR}\rangle)=0.00_{-0.09}^{+0.07}$, $0.00_{-0.12}^{+0.12}$ and $0.00_{-0.11}^{+0.24}$, implying that there is little bias in estimated SFRs though the dispersion increases significantly.

In summary, to reproduce the true SFR averaged over 5\,Myr in the simulation from the intrinsic H$\alpha$ emission, we need to include a correction factor to account for the dust and helium absorption, the escape of LyC photons, collisional excitation and ionization emission, and the contribution from older stars. This correction can be achieved by increasing the H$\alpha$-to-SFR conversion factor by $\sim50\%$ (10--30\%) in the MW (LMC) simulation. Absorption of LyC photons is challenging to measure observationally, but it has been highlighted as a caveat in the literature \citep[e.g.,][]{puglisi16, tacchella18_dust}. \citet{salim16}, fitting $\sim700\,000$ local galaxies with CIGALE, found a fraction of LyC photons absorbed by dust to be 0.3 in order to match the observed equivalent widths of the main optical lines (H$\alpha$, H$\beta$, [OII], [OIII]) from SDSS spectra. Consistently, previous studies on nearby spiral galaxies found similar LyC photon absorption fractions of $0.3$--$0.5$ \citep{inoue01, hirashita03, iglesias-paramo04}. This is in good agreement with our inferred absorption and escape of a total of $\sim43\%$ ($f_{\rm abs}\approx28\%$ due to dust absorption, $f_{\rm He}\approx9\%$ due to helium absorption, and $f_{\rm esc}\approx6\%$ due to escape), which highlights that the absorption of LyC photons is significant ($\sim0.1-0.2$ dex) and needs to be accounted for when estimating the SFRs from Balmer emission lines. On the theoretical side, using a periodic tall box simulation to explore sub-parsec scale feedback and emission in the solar neighbourhood, \citet{kado-fong20} find approximately half of ionizing photons are absorbed by gas and half by dust, i.e. $f_{\rm abs}\approx0.5$. This is roughly in the same ballpark as our estimate.

\subsection{Balmer decrement and attenuation law}
\label{subsec:dust}

\begin{figure}
\includegraphics[width=\linewidth]{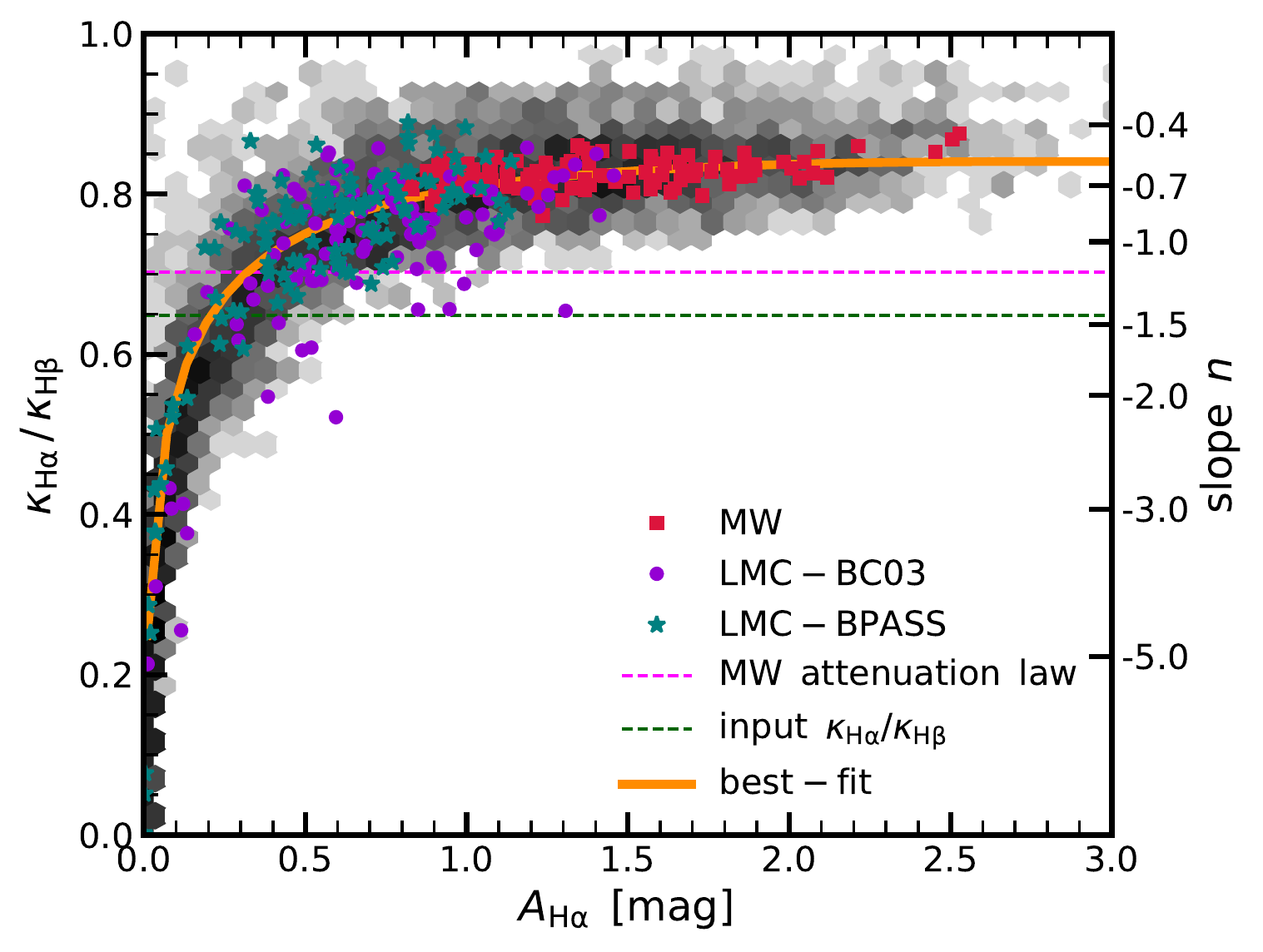}
\caption{Inferred dust attenuation law for Balmer lines (ratio of the attenuation law function at H$\alpha$ and H$\beta$ [$\kappa_{\rm H\alpha}/\kappa_{\rm H\beta}$] on the left and the power-law slope $n$ on the right) as a function of the attenuation at H$\alpha$ ($\mathrm{A}_{\rm H\alpha}$). We solve for the dust attenuation law by inverting Eq.~(\ref{eq:balmer}). The red squares, purple circles and teal stars show the measurements for the MW, LMC-BC03 and LMC-BPASS simulations, respectively. The hex-binned histogram indicate the measurements on spatially resolved scales from random 1\,kpc apertures of the MW simulation. The pink and green dashed lines show the MW attenuation law and the input dust opacities (Section~\ref{subsec:rt}). The orange line indicates the best fit of Eq.~(\ref{eq:slope_tau}). We find that individual snapshots as well as the spatially resolved measurements follow the trend of steeper attenuation laws for less attenuated regions. Furthermore, for the MW simulation, we find a typical power-law attenuation law with an index of $n\approx-0.63$.}
\label{fig:attenuation}
\end{figure}

\begin{figure}
\includegraphics[width=\linewidth]{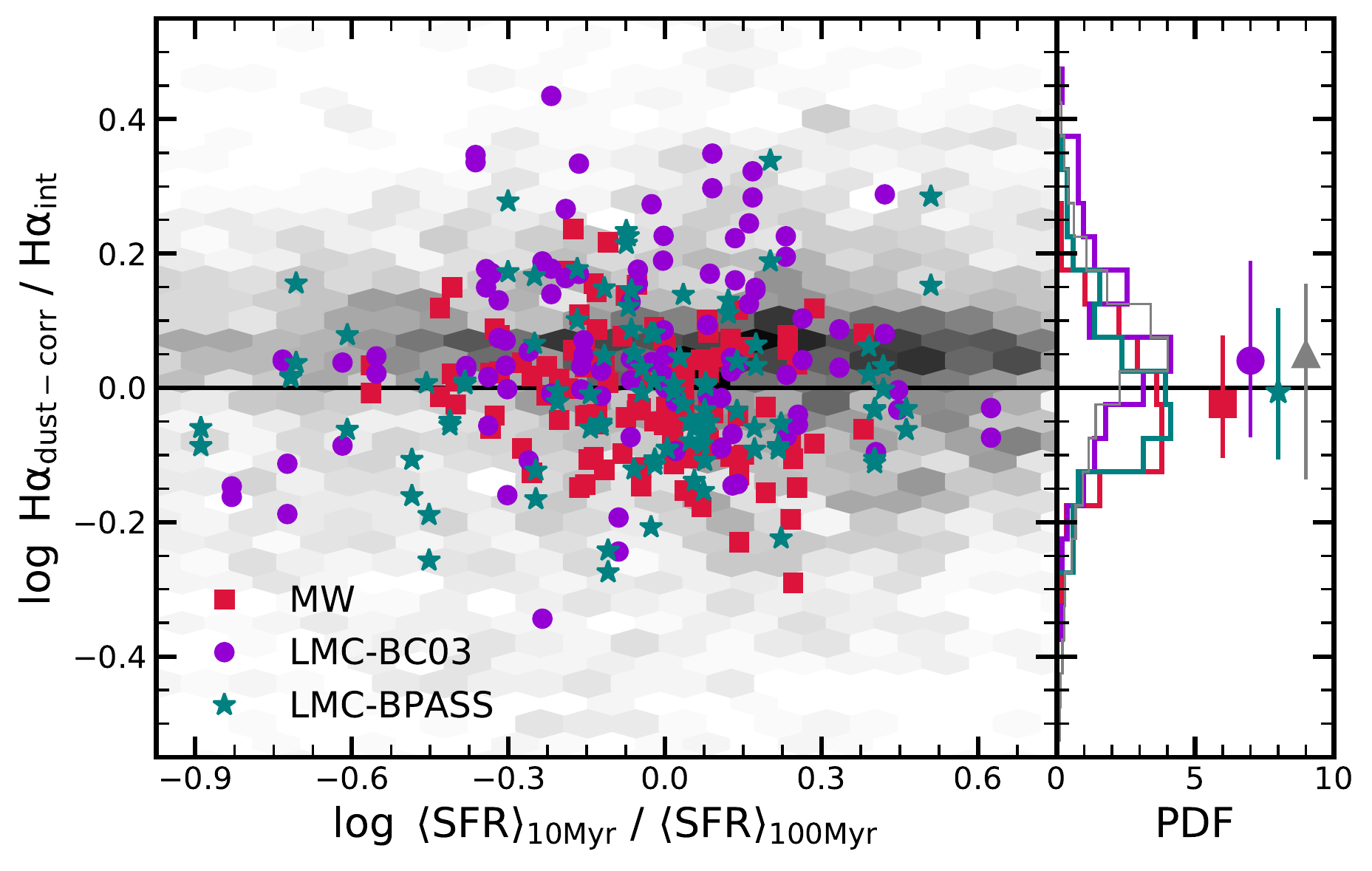}
\caption{Testing the Balmer decrement dust correction. On the left, we plot the log difference of the H$\alpha$ flux corrected for dust via the Balmer decrement ($\mathrm{H}\alpha_{\rm dust-corr}$; Eqs.~\ref{eq:balmer} and \ref{eq:slope_tau}) and the intrinsic H$\alpha$ flux ($\mathrm{H}\alpha_{\rm int}$) as a function of $\langle\mathrm{SFR}\rangle_{\rm 10Myr}/\langle\mathrm{SFR}\rangle_{\rm 100Myr}$, which indicates whether the SFR is increasing or decreasing over the past 100 Myr. The red squares, purple circles and teal stars show the measurements for the MW, LMC-BC03 and LMC-BPASS simulations, respectively. The hex-binned histogram indicates the measurements on spatially resolved scales from random 1\,kpc apertures of the MW simulation. The right panel shows the vertical histogram of $\log(\mathrm{H}\alpha_{\rm dust-corr}/\mathrm{H}\alpha_{\rm int})$, with the medians and the $16$--$84^\text{th}$ percentile indicated with the corresponding symbols and errorbars. The grey triangle shows the median of the random 1\,kpc apertures. The Balmer dust correction works well, independent of the SFH, though there is a significant amount of scatter of $\sim0.1$--$0.2$\,dex.}
\label{fig:dust_correction}
\end{figure}

The dust attenuation towards star-forming regions is most directly probed using Balmer recombination line flux ratios, because as dust attenuation is wavelength dependent, its effects can be measured by comparing the observed and intrinsic Balmer decrements \citep[e.g.,][]{calzetti97}. Specifically, dust will preferentially absorb the shorter wavelength H$\beta$ $\lambda4861$\,\AA\ line rather than the longer wavelength H$\alpha$ $\lambda6563$\,\AA\ line, increasing the observed value of the Balmer decrement $(\mathrm{H}\alpha/\mathrm{H}\beta)_{\rm obs}$. The Balmer decrement can therefore be used to derive the attenuation towards H$\alpha$ in the following way:
\begin{equation}
    A_{\rm H\alpha} = \frac{\kappa(\lambda_{\rm H\alpha})}{\kappa(\lambda_{\rm H\beta})-\kappa(\lambda_{\rm H\alpha})} \times 2.5 \log\left(\frac{(\mathrm{H}\alpha/\mathrm{H}\beta)_{\rm obs}}{(\mathrm{H}\alpha/\mathrm{H}\beta)_{\rm int}}\right),
\label{eq:balmer}
\end{equation}
where $\kappa(\lambda)$ is the attenuation curve. The intrinsic Balmer decrement $(\mathrm{H}\alpha/\mathrm{H}\beta)_{\rm int}$ is not constant for and within galaxies since it depends on both the electron temperature and density and collisional excitation. It is usually assumed to be $(\mathrm{H}\alpha/\mathrm{H}\beta)_{\rm int}=2.87$, which is valid for an electron temperature of $T_{e}=10^4~\mathrm{K}$ and an electron density of $n_e = 10^2~\mathrm{cm}^{-3}$, under Case B recombination conditions \citep{osterbrock06}. However, with higher temperature or higher density, we expect that $(\mathrm{H}\alpha/\mathrm{H}\beta)_{\rm int}$ increases since higher temperature or higher density introduces more collisional mixing of electron states, therefore increasing the ability to populate the $n=4$ state (and thus $\mathrm{H}\beta$). For typical star-forming regions, \citet[][see also \citealt{dopita03}]{osterbrock06} find that increasing $T_{e}$ from $5\times10^3~\mathrm{K}$ to $2\times10^4~\mathrm{K}$ reduces $(\mathrm{H}\alpha/\mathrm{H}\beta)_{\rm int}$ from 3.05 to 2.76 at constant $n_e = 10^2~\mathrm{cm}^{-3}$. At constant $T_{e}=10^4~\mathrm{K}$, $(\mathrm{H}\alpha/\mathrm{H}\beta)_{\rm int}$ only reduces from 2.85 to 2.81 for a change in $n_e$ from $10^4~\mathrm{cm}^{-3}$ to $10^6~\mathrm{cm}^{-3}$. Therefore, $(\mathrm{H}\alpha/\mathrm{H}\beta)_{\rm int}$ is expected to be in the rather narrow range of 2.76 to 3.05. 

As described in Section~\ref{subsec:time_evolution}, the middle panels of Fig.~\ref{fig:timeseries} show the observed and intrinsic Balmer decrement as a function of time. We find on average an intrinsic Balmer decrement of $(\mathrm{H}\alpha/\mathrm{H}\beta)_{\rm int}=2.97_{-0.01}^{+0.01}$ for the MW and $2.98_{-0.02}^{+0.05}$ for the LMC. This is slightly higher than the aforementioned fiducial value of 2.87. Most of this difference can be explained by collisional excitation emission, which intrinsically has a higher Balmer decrement, while the rest is due to a different gas temperature and gas density than the fiducial one ($T_{e}=10^4~\mathrm{K}$ and $n_e = 10^2~\mathrm{cm}^{-3}$; Fig.~\ref{fig:phase}).

In addition to $(\mathrm{H}\alpha/\mathrm{H}\beta)_{\rm int}$, the second important factor in deriving $A_{\rm H\alpha}$ from Eq.~(\ref{eq:balmer}) is the attenuation curve $\kappa(\lambda)$. The attenuation curve for the nebular emission is not well-constrained observationally. Our simulations allow us to infer the attenuation law by inverting Eq.~(\ref{eq:balmer}) and measuring in the simulations $A_{\rm H\alpha}$ and $(\mathrm{H}\alpha/\mathrm{H}\beta)_{\rm obs}$, while assuming $(\mathrm{H}\alpha/\mathrm{H}\beta)_{\rm int}$. In the following, we assume the fiducial value of $(\mathrm{H}\alpha/\mathrm{H}\beta)_{\rm int}=2.87$.

Fig.~\ref{fig:attenuation} shows the inferred shape of the attenuation law (ratio of $\kappa(\lambda_{\rm H\alpha})/\kappa(\lambda_{\rm H\beta})$ on the left and power-law index $n$ on the right) as a function of the measured amount of optical depth (attenuation $A_{\rm H\alpha}$). The integrated measurements for the MW and LMC simulations are shown as red squares and purple circles, respectively. The hex-binned histogram indicate the measurement on spatially resolved scales from random, 1-kpc apertures. The pink dashed line indicates the MW attenuation law, which assumes the \citet{cardelli89} Galactic extinction curve, with an update in the near-UV from \cite{odonnell94}. The green dashed line marks the ratio of the input dust opacities for H$\alpha$ and H$\beta$ (Section~\ref{subsec:rt}).

It is apparent from Fig.~\ref{fig:attenuation} that the integrated measurements follow the spatially resolved measurements for all simulation setups: the lower the attenuation, the steeper the attenuation law. We quantify this trend with the following equation:
\begin{align}
        \log(\kappa_{\rm H\alpha}/\kappa_{\rm H\beta}) &= n\,\log(\lambda_{\rm H\alpha}/\lambda_{\rm H\beta}) \notag \\
         &= -0.09\,\log(A_{\rm H\alpha})^2 + 0.08 \,\log(A_{\rm H\alpha}) - 0.09 \, ,
\label{eq:slope_tau}
\end{align}
where $n$ is the power-law index from $\kappa(\lambda)\propto\lambda^n$. The best-fit line is shown as orange curve in Fig.~\ref{fig:attenuation}.

This trend is consistent with the findings of \citet{chevallard13}, who analysed a diverse series of theoretical attenuation laws and showed that all the studies predict a relationship between the optical depth and attenuation law slope \citep[see also][]{narayanan18, trayford19, salim20}. The physical origin for this optical depth-slope relationship was put forward by \citet{chevallard13}: red light scatters more isotropically than blue. Red photons emitted in the equatorial plane of the galaxy will be more likely to escape, while blue photons have a comparatively increased likelihood of remaining, and subsequently being absorbed. In the low optical depth limit, this corresponds to a steepening of the attenuation curve for face-on directions. 

For the MW simulation, we find a typical slope of the attenuation law of $n=-0.63\pm0.07$ ($\kappa(\lambda_{\rm H\alpha})/\kappa(\lambda_{\rm H\beta})=0.83\pm0.02$) and attenuation of $A_{\rm H\alpha}=1.1_{-0.2}^{+0.3}$ mag. This is consistent with the power-law exponents for the diffuse ISM and natal birth clouds found by \citet{charlot00}. For the LMC simulation, we find a larger variation in both the amount of extinction as well as the slope of the attenuation curve. On average, the LMC has typically a steeper attenuation law and a smaller optical depth than the MW with $A_{\rm H\alpha}=0.6\pm0.3$ mag and $\kappa(\lambda_{\rm H\alpha})/\kappa(\lambda_{\rm H\beta})=0.77_{-0.9}^{+0.05}$, which corresponds to a slope of $n=-0.87^{+0.22}_{-0.42}$. These numbers do not significantly change when considering the LMC-BPASS run.

We emphasise that although the measurements shown in Fig.~\ref{fig:attenuation} are calculated from the face-on projection, there will be little difference when changing the inclination. Specifically, the extreme case of an edge-on projection for the MW leads -- as expected -- to more attenuation with $A_{\rm H\alpha}=1.7\pm0.4$ mag. Accordingly, the attenuation law is slightly flatter with $n=-0.65^{+0.07}_{-0.06}$ ($\kappa(\lambda_{\rm H\alpha})/\kappa(\lambda_{\rm H\beta})=0.82\pm0.02$), but as shown in Fig.~\ref{fig:attenuation}, the slope of the attenuation law roughly remains constant above $A_{\rm H\alpha}>1$ mag. This universal optical depth--slope relationship is independent of inclination, which is consistent with the theoretical analysis by \citet{chevallard13} and observations by \citet[][see also \citealt{salim20} for a review]{salim18_curves}. Similarly, the optical depth--slope relation also holds for spatially resolved scales, as shown by the grey hex-bin histogram in Fig.~\ref{fig:attenuation}. This is consistent with observations of local galaxies that show intriguing variations within galaxies, with regions of high attenuation exhibiting a shallower attenuation curve \citep{decleir19}.

We can use the optical depth--slope relationship (Eq.~\ref{eq:slope_tau}) and iteratively solve for $A_{\rm H\alpha}$ and $n$ (i.e. $\kappa(\lambda_{\rm H\alpha})/\kappa(\lambda_{\rm H\beta})$)  to obtained the dust-corrected H$\alpha$ luminosity. Fig.~\ref{fig:dust_correction} plots the ratio of the dust-corrected H$\alpha$ luminosity and the intrinsic H$\alpha$ luminosity as a function of $\langle\mathrm{SFR}\rangle_{\rm 10Myr}/\langle\mathrm{SFR}\rangle_{\rm 100Myr}$, which indicates whether the SFR is increasing or decreasing over the past 100 Myr. The red squares, purple circles and teal stars show the measurements for the MW, LMC-BC03 and LMC-BPASS simulations, respectively. The hex-binned histogram indicates the measurements on spatially resolved scales from random 1\,kpc apertures of the MW simulation. We find that the Balmer dust correction works well, independent of the SFH, though there is a significant amount of scatter with $\sim0.1$--$0.2$\,dex. Specifically, the median and the $16$--$84^\text{th}$ percentile of the log difference between $\mathrm{H}\alpha_{\rm dust-corr}$ and $\mathrm{H}\alpha_{\rm int}$ are $-0.02_{-0.08}^{+0.10}$, $0.04_{-0.13}^{+0.20}$ and $-0.01_{-0.10}^{+0.13}$ for the MW, LMC-BC03 and LMC-BPASS simulations, respectively. For the spatially resolved data of the MW, we find $\log(\mathrm{H}\alpha_{\rm dust-corr}/\mathrm{H}\alpha_{\rm int})=0.05_{-0.19}^{+0.11}$. These numbers are indicated in the right panel of Fig.~\ref{fig:dust_correction}.

\subsection{Timescale of the H\texorpdfstring{$\balpha$}{α} SFR tracer}
\label{subsec:timescale}

\begin{figure*}
\includegraphics[width=\linewidth]{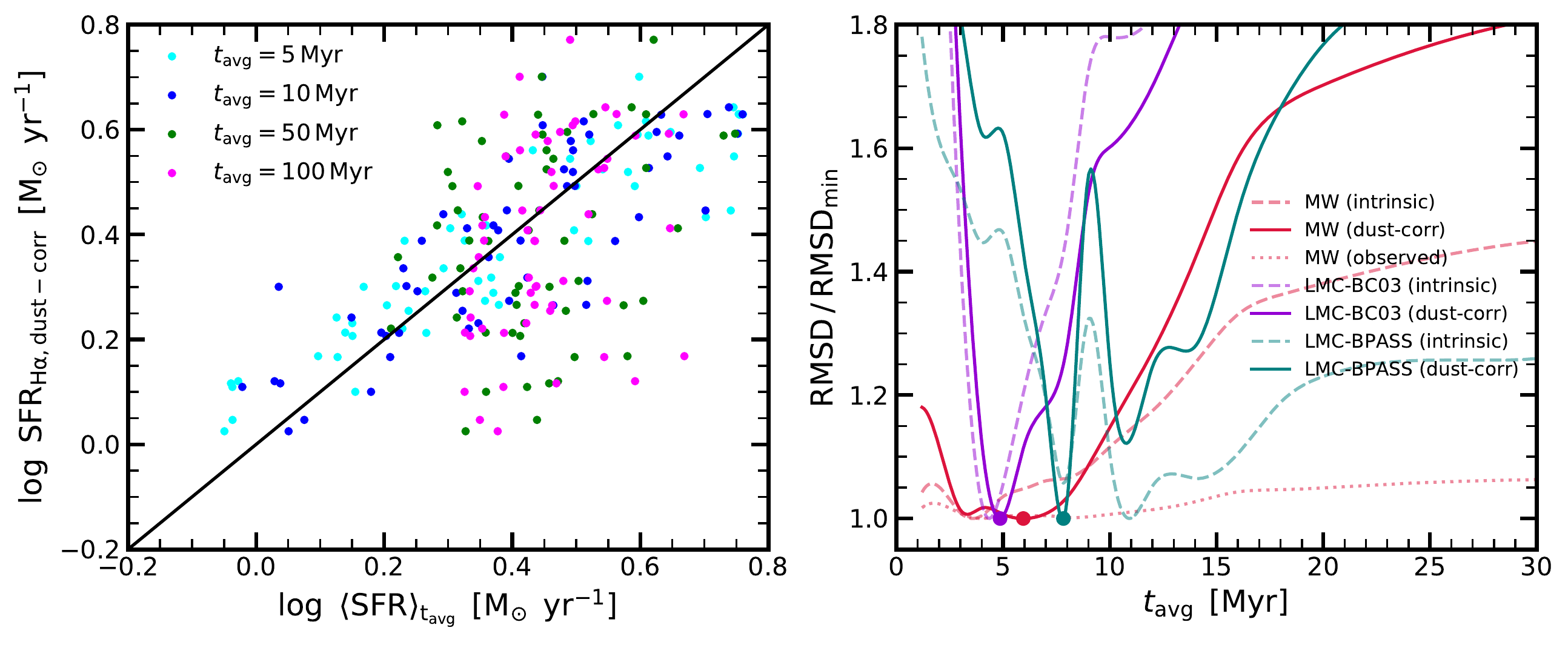}
\caption{Inferring the timescale of the H$\alpha$ SFR indicator. \textit{Left:} Scatter plot of the dust-corrected H$\alpha$ SFR ($\mathrm{SFR}_{\rm H\alpha,dust-corr}$) versus the true SFR ($\langle \mathrm{SFR} \rangle_{\rm t_{\rm avg}}$), averaged over 5 Myr (cyan), 10 Myr (blue), 50 Myr (green) and 100 Myr (pink). As expected from Fig.~\ref{fig:timeseries}, averaging over shorter timescales ($5-10~\mathrm{Myr}$) leads a smaller scatter than averaging over longer timescales. \textit{Right:} Normalised root mean square deviation (RMSD; Eq.~\ref{eq:DMSE}) between the true SFR of the simulation averaged over different timescales $t_{\rm avg}$ and the SFR estimated from the intrinsic H$\alpha$ emission (dashed lines), the observed H$\alpha$ emission (dotted line; only shown for the MW case), and the dust-corrected (via Balmer decrement) H$\alpha$ emission (solid lines). The MW, LMC-BC03 and LMC-BPASS simulations are shown in red, purple and teal, respectively. The solid dots indicate the minima for the dust-corrected H$\alpha$ SFR indicator, which minimises the scatter in the 1-to-1 relation (as shown on the left). This averaging timescale depends on the stellar population model, the star-formation variability and the line-of-sight projection (Tab.~\ref{tab:tavg}).}
\label{fig:Halpha_timescale}
\end{figure*}

\begin{table*}
    \setlength{\tabcolsep}{5pt}
    \renewcommand{\arraystretch}{1.5}
    \caption{Star-formation timescale ($t_{\rm avg, min}$) and $1\sigma$ scatter of the measured-to-time-averaged-true-SFR ratio ($\sigma_{\rm R}$) for the MW and LMC simulations. The following SFR indicators are shown from left to right: dust-corrected H$\alpha$, intrinsic H$\alpha$, observed (w/o dust correction) H$\alpha$, and edge-on dust-corrected H$\alpha$ (instead of our default face-on project).}\label{tab:tavg}
    \centering
        \begin{tabular}{llcclcclcclcc}
          & & \multicolumn{2}{l}{H$\alpha$ (dust-corr)} & & \multicolumn{2}{l}{H$\alpha$ (int)} & & \multicolumn{2}{l}{H$\alpha$ (obs)} & & \multicolumn{2}{l}{H$\alpha$ (dust-corr; edge-on)}  \\
          \cline{3-4} \cline{6-7} \cline{9-10} \cline{12-13} 
          Simulation & & $t_{\rm avg, min}$ & $\sigma_{\rm R}$ & & $t_{\rm avg, min}$ & $\sigma_{\rm R}$ & & $t_{\rm avg, min}$ & $\sigma_{\rm R}$ & & $t_{\rm avg, min}$ & $\sigma_{\rm R}$ \\ 
          \cline{1-1} \cline{3-4} \cline{6-7} \cline{9-10} \cline{12-13} 
          MW & & 5.9$_{-3.4}^{+2.3}$ & 0.10 & & 3.6$_{-1.3}^{+2.7}$ & 0.07 & & 4.0$_{-3.0}^{+16.0}$ & 0.13 & & 7.7$_{-1.5}^{+2.4}$ & 0.11 \\
          LMC-BC03 & & 4.9$_{-0.5}^{+0.6}$ & 0.12 & & 4.5$_{-0.5}^{+0.5}$ & 0.12 & & 7.5$_{-2.4}^{+0.7}$ & 0.12 & & 4.7$_{-0.5}^{+0.5}$ & 0.12 \\
          LMC-BPASS & & 7.8$_{-0.3}^{+0.2}$ & 0.17 & & 11.0$_{-0.7}^{+1.2}$ & 0.18 & & 11.1$_{-0.5}^{+0.7}$ & 0.14 & & 7.8$_{-0.3}^{+0.2}$ & 0.18 \\  \hline
        \end{tabular}\\
\end{table*}

As shown in the bottom panels of Fig.~\ref{fig:timeseries}, the SFR averaged over short timescales ($5$--$10$\,Myr) is in better agreement with the H$\alpha$-based SFR than the SFR averaged over longer timescales. Understanding the star-formation timescale that H$\alpha$ traces is of great importance since by comparing H$\alpha$-based SFRs to SFRs obtained with other indicators (for example from the UV continuum), we are able to learn about the variability of star formation \citep[e.g.,][]{sparre17_bursty, caplar19, emami19}. The variability of star formation (sometimes also called ``burstiness'') is directly related to how star formation is regulated and it informs us as to what timescales are important for the evolution of galaxies \citep{iyer20, tacchella20, wang20_p1}.

The goal of this section is to determine the timescale of the H$\alpha$ SFR indicator. We follow the approach outlined in \citet{caplar19} and \citet{flores-velazquez21}. Specifically, we find the best-fitting averaging timescale $t_{\rm avg, min}$ that minimises the difference between the true SFR $\langle\mathrm{SFR}\rangle_{\rm t_{\rm avg}}$ of the simulation and the observed SFR $\mathrm{SFR}_{\rm ind}$. The timescale $t_{\rm avg, min}$ is the width of the boxcar integration over which one would have to average the true SFR of the simulation to match the indicated SFR (in our case H$\alpha$) at the time of an observation. To find $t_{\rm avg, min}$, we minimise the follow root mean square deviation estimate:
\begin{equation}
    \mathrm{RMSD}(t_{\rm avg}) = \sqrt{\frac{\sum_i \left(\log\mathrm{SFR}_{\rm ind}^{i} - \log\langle \mathrm{SFR}\rangle_{t_{\rm avg}}^{i}  \right)^{2}}{N}}
    \label{eq:DMSE}
\end{equation}
where $i$ is the index running over the $N$ snapshots. 

In the left panel of Fig.~\ref{fig:Halpha_timescale} we show the dust-corrected H$\alpha$ SFR ($\mathrm{SFR}_{\rm H\alpha,dust-corr}$) versus the true SFR ($\langle \mathrm{SFR} \rangle_{\rm t_{\rm avg}}$), averaged over 5 Myr (cyan), 10 Myr (blue), 50 Myr (green) and 100 Myr (pink). As expected from Fig.~\ref{fig:timeseries}, averaging over shorter timescales ($5$--$10~\mathrm{Myr}$) leads a smaller scatter and a better alignment with the one-to-one relation than averaging over longer timescales. The RMSD introduced above (see Eq.~\ref{eq:DMSE}) corresponds to the scatter with respect to this one-to-one line.

In the right panel of Fig.~\ref{fig:Halpha_timescale} we plot the normalised RMSD as a function of the averaging timescale. The MW, LMC-BC03 and LMC-BPASS simulations are shown in red, purple and teal, respectively. The dashed, solid, and dotted (only shown for the MW) lines show the results for the SFR estimated from the intrinsic H$\alpha$ emission, the dust-corrected (via Balmer decrement with the optical depth--slope relationship given by Eq.~\ref{eq:slope_tau} and the effective conversion factor discussed in Section~\ref{subsec:conversion_factor}) H$\alpha$ emission, and the observed H$\alpha$ emission. The timescales that minimise the RMSD for the case of the dust-corrected H$\alpha$ SFR are indicated with the circles and we tabulate these timescales in Tab.~\ref{tab:tavg}. The uncertainties given in Tab.~\ref{tab:tavg} are estimated by considering the RMSD 5\% above the its minimum.

As summarised in Tab.~\ref{tab:tavg}, the timescale of the H$\alpha$ SFR indicator depends on several factors, including the star-formation variability itself (MW versus LMC), the amount of dust attenuation (face-on versus edge-on) and the stellar population model (BC03 versus BPASS). This timescale also depends on the IMF, though we have fixed the IMF in our current investigation (see also Section~\ref{subsec:caveats}). We find that, for the MW simulation, the SFR estimated from the intrinsic H$\alpha$ luminosity traces the SFR on a timescale of $3.6_{-1.3}^{+2.7}$\,Myr, which is shorter than the timescale of $4.0_{-3.0}^{+16.0}$\,Myr for the observed H$\alpha$-based SFR (not corrected for dust), though the latter has a significant tail to much longer timescales. This can be explained by the fact that the youngest stars are still embedded within molecular clouds, which leads to higher dust absorption of H$\alpha$ photons that trace those youngest stars. When correcting for dust, the tail towards long timescales of observed H$\alpha$-based SFR disappears. The numbers for the face-on and edge-on case are 5.9\,Myr and 7.7\,Myr, respectively, indicating that the amount of dust attenuation plays an important role as well.

Comparing these results to the LMC simulations (both LMC-BC03 and LMC-BPASS), we find qualitatively similar trends, though there are quantitative differences. Firstly, the adjustment of the SFR based on the intrinsic H$\alpha$ luminosity to the one based on the dust-corrected and observed H$\alpha$ luminosity is smaller. This is because there is less dust attenuation in the LMC simulations in comparison with the MW simulation. Secondly, the averaging timescale for the intrinsic H$\alpha$-based SFR is longer for the LMC simulation ($4.5_{-0.5}^{+0.5}$\,Myr for the LMC-BC03 and $11.0_{-0.7}^{+1.2}$\,Myr for the LMC-BPASS) with respect to the MW simulation ($3.6_{-1.3}^{+2.7}$\,Myr). The difference in the LMC-BC03 can be explained by the larger fractional contribution of older stars (i.e. lower specific SFR), while the difference in LMC-BPASS model additionally stems from the fact that intermediate age stars ($10$--$100$\,Myr) in the BPASS model produce significantly more ionizing photons than the ones in the BC03 model (Appendix~\ref{appsec:stellar_pop}). In summary, we find that H$\alpha$ traces the SFR on timescales of 5--8\,Myr.

This is overall consistent with quoted values in the literature. \citet{flores-velazquez21} found 5\,Myr by convolving the SFHs of the FIRE simulation with the H$\alpha$ response function of the BPASS model (i.e. without radiative transfer). Similarly, using a family of variable SFHs drawn from a power spectrum density, \citet{caplar19} quote values of 2--6\,Myr, depending on the burstiness of the SFH itself. \citet{haydon20} used an idealised hydrodynamical disc galaxy simulation without radiative transfer (no dust attenuation and scatter) and inferred an H$\alpha$ timescale of $4.32_{-0.23}^{+0.09}$\,Myr, which is close to our intrinsic timescale (Tab.~\ref{tab:tavg}). \citet{haydon20b} revisited results from \citet{haydon20} by quantify the impact of dust extinction, find that extinction mostly decreases the SFR tracer emission timescale (factor of a few for H$\alpha$). Observationally, correlating maps of different light tracers such as H$\alpha$, CO and 24$\mu$m on spatially resolved scales to empirically constrain the giant molecular cloud lifecycle \citep[see also][]{kruijssen19, chevance20}, \citet{kim21} found a duration of H$\alpha$ emitting phase of 5.5--8.9\,Myr for nearby spiral galaxies, which is consistent with our dust-corrected timescale estimates. 

Finally, $\mathrm{SFR}_{\rm ind}$ is not exactly equal to $\langle \mathrm{SFR}\rangle_{t_{\rm avg}}$ in the general case. Therefore, we now address the question: how well does the measured SFR match the true SFR averaged over $t_{\rm avg, min}$? Following \citet{flores-velazquez21}, we investigate the scatter of the measured-to-time-averaged-true-SFR ratio ($R_{t_{\rm avg}}^{\rm ind} = \mathrm{SFR}_{\rm ind} / \langle\mathrm{SFR}\rangle_{t_{\rm avg}}$) and provide in Tab.~\ref{tab:tavg} the standard deviation of $\log R_{t_{\rm avg}}^{\rm ind}$ ($\equiv\sigma_{\rm R}$). We find $\sigma_{\rm R}$ for the dust-corrected H$\alpha$-based SFR to be 0.10, 0.12 and 0.17 for the MW, LMC-BC03 and LMC-BPASS, respectively. The values for $\sigma_{\rm R}$ only marginally decrease when switching to the intrinsic H$\alpha$-based SFR, highlighting that this scatter is caused by both the uncertainty of the H$\alpha$-to-SFR conversion (Section~\ref{subsec:conversion_factor}) and the dust correction (Section~\ref{subsec:dust}).

\section{Limitations and future prospects}
\label{sec:discussion}

\subsection{Current limitations and caveats}
\label{subsec:caveats}

An important result regarding the H$\alpha$-to-SFR conversion is the significant effect of the dust absorption of ionizing LyC photons produced by young stars. In the case of the MW simulation, $\sim28\%$ of the LyC photons are absorbed by dust, $\sim8\%$ ionize \HeI, $\sim1\%$ ionize \HeII, $\sim6\%$ escape the galaxy, implying that only about $\sim57\%$ of the produced LyC photons are available for ionizing hydrogen. Although we discuss in Section~\ref{subsec:conversion_factor} that the rather high dust absorption is consistent with observational estimates, we acknowledge here that -- even with parsec-scale resolution -- it remains challenging to resolve dust in ionized \HII regions. While the present treatment is state-of-the-art, dust modelling in general is highly uncertain, especially considering complex multi-scale effects such as spatially dependent grain size distributions and compositions, decoupling of dust and gas kinematics, and various physical processes impacting growth and destruction mechanisms that in principle affect dust attenuation law shapes \citep[e.g.,][]{draine11, aoyama18, mckinnon18, mckinnon19, li19_dust}.

As described in Section~\ref{sec:methods}, we do not model the effects of AGNs, both in regard to the hydrodynamics or the radiative transfer. AGN are an important source of ionizing radiation, in particular near the central regions of some galaxies. Although we find a good agreement with the overall Balmer surface brightness profiles (Section~\ref{sec:observations}), we do not judge the agreement in the centre as the consistency could be serendipitous. Obviously, it is of great interest to understand the impact of AGN feedback on the surrounding gas and how this could be detected in IFU data through broadened emission lines \citep[e.g.,][]{forster-schreiber14, genzel14b}. Along similar lines, we have not included X-ray binaries in the numerical model, though they are of interest for both changing the cooling function \citep[e.g.,][]{kannan16} and as a production site for ionizing raditation \citep[e.g.,][]{schaerer19, senchyna20}.

SPS models are the foundation for interpreting galaxy observations. They account for the light emitted by stars and the reprocessing of that light by dust, and some codes also include nebular emission \citep[see reviews by][]{walcher11, conroy13_rev}. These SPS models have a number of assumed parameters that can alter the amount of ionizing flux significantly \citep[e.g.,][]{eldridge12, levesque12, stanway16}, through stellar atmospheres (winds, opacities) and stellar evolution (mass loss, rotation, binarity). Here, we have investigated how stellar binarity affects some of our conclusions by running the LMC simulation with the BC03 as well as the BPASS stellar models (Section~\ref{sec:methods}). Changing the SPS has two effects. Firstly, it changes the production rate of ionizing photons with stellar age (Appendix~\ref{appsec:stellar_pop}). Secondly, it affects the strength and timing of radiative feedback, stellar winds and supernova feedback, which are all modelled self-consistently. We show, among other things, that ($i$) the SFH is more variable in the LMC-BC03 simulation than in the LMC-BPASS model (Fig.~\ref{fig:timeseries}); ($ii$) there is a higher helium absorption but a lower escape fraction of LyC photons  in the LMC-BC03 model than in the LMC-BPASS model; and ($iii$) in the LMC-BC03 model H$\alpha$ traces the SFR on $\sim5$\,Myr while it is $\sim8$\,Myr in the LMC-BPASS case (Fig.~\ref{fig:Halpha_timescale} and Tab.~\ref{tab:tavg}). In the future (see below), it will be of great interest to better understand how we can constrain the SPS model (by breaking degeneracies with, e.g., IMF, metallicity, and SFH) from galaxy integrated measurements that are based on combining UV-to-IR photometry with measurements of the Balmer (and possibly Paschen) emission lines. 

Finally, we note that the stellar mass resolution of $2.8\times10^3~\text{M}_{\odot}$ per particle is sufficient to resolve the SFH of the LMC and MW simulation. In a 5\,Myr time bin, we will be able to resolve a SFR of $0.6\times10^{-3}~\mathrm{M}_{\odot}/\mathrm{yr}$ with one particle. This implies that the median SFR of the LMC simulation is resolved with $\sim66$ ($\sim13$) particles considering a time in of 5\,Myr (1\,Myr) time interval. Nevertheless, for the LMC (and lower mass dwarfs), stochastic sampling of the IMF will have interesting and important effects \citep[e.g.,][]{fumagalli11, da-silva14}. In particular, regarding the H$\alpha$-to-SFR conversion, even if a galaxy has a constant SFR, at low SFR the H$\alpha$ emission will fluctuate because of the fact that massive OB stars, which dominate the production of LyC photons, form rarely. In the limit of very low SFRs, these massive stars can be so rare that their number fluctuates significantly with time, leading to an additional source of stochasticity. This leads to an increase of scatter and also an overall bias to lower H$\alpha$ emission for a given SFR. 

\subsection{Future applications}
\label{subsec:future}

The presented Balmer line (H$\alpha$ and H$\beta$) radiative transfer study of high-resolution MW-like and LMC-like galaxies is only the first step and will allow us to make further progress in several different directions in the future. As mentioned above, having broad-band UV-IR photometry in addition to the Balmer emission lines will allow us to tackle interesting questions related to the accuracy of SED modelling and degeneracies related to the SPS models. Therefore, we plan to expand our spectral coverage of synthetic observations to include stellar and dust continuum radiative transfer based on state-of-the-art pipelines  \citep[e.g.,][]{camps15, narayanan21}. 

Along these lines, we would like to extend our spatially resolved MCRT framework to include metal ionization for self-consistent nebular line studies. This will allow us to address several interesting issues, including: ($i$) the selection efficiency of the DIG via the strength of the [\ion{S}{ii}] emission line; ($ii$) understanding how the DIG potentially biases strong-line gas-phase metallicity measurements \citep[e.g.,][]{sanders17, poetrodjojo19}; ($iii$) improved selection of star formation, AGN and shock regions in galaxies; and ($iv$) calibrations of ISM pressure and electron density diagnostics \citep[e.g.,][]{kewley19_review}. 

Employing our spatially resolved MCRT methodology with these aforementioned improvements will be key to understanding the physics of emission lines of and within galaxies both at low and high redshifts. In particular, we are currently working on zoom-in simulations based on the SMUGGLE-RT framework as part of the \textsc{thesan} project \citep{kannan21, smith21, garaldi21, kannan21_line}, which are ideally suited for studying the high-redshift Universe. These and similar simulations will allow us to ($i$) better understand current and upcoming observations by being able to accurately estimate uncertainties and be aware of possible biases; and ($ii$) gain inspiration for new observational projects.

\section{Conclusions}
\label{sec:conclusions}

In this work, we shed new light onto the emission, absorption, and scattering of Balmer emission lines (H$\alpha$ and H$\beta$) in MW-like and LMC-like galaxies to better understand the importance of collisionally excited gas, emission of the diffuse ionized gas (including extraplanar emission), and connections between H$\alpha$ emission and star formation rates. We perform a detailed radiative transfer analysis of high-resolution isolated MW and LMC simulations that include radiative transfer, non-equilibrium thermochemistry, and dust evolution. We confirm that the simulations produce realistic H$\alpha$ and H$\beta$ radial and vertical surface brightness profiles that are consistent with observations of local MW-like and LMC-like galaxies (Figs.~\ref{fig:profile_comparison} and \ref{fig:scaleheight_comparison}). The main results and conclusions of this work are as follows:
\begin{enumerate}
    \item H$\alpha$ (and H$\beta$) photons are predominantly produced by recombination of ionized gas, which is ionized by LyC photons (Fig.~\ref{fig:cartoon}). We also find that a small fraction ($5$--$10\%$) of the total Balmer emission stems from collisionally excited gas (Fig.~\ref{fig:fcol}) and an even smaller fraction ($<2\%$) from collisionally ionized gas.
    \item Although different selection techniques lead to quantitatively different DIG fractions, we find that the observed DIG fraction is of the order of $\sim50\%$, roughly consistent with observations (Fig.~\ref{fig:f_DIG}). Importantly, the intrinsic DIG fraction (i.e. before accounting for dust absorption and scattering) is a factor of 2--3 times lower because of dust attenuation effects. Furthermore, we show in Fig.~\ref{fig:EW_Halpha} that the H$\alpha$ EW distribution is not easy to interpret because it is severely affected by spatial binning. We confirm that a value of $\mathrm{EW(H\alpha)}=14$\,\AA\ seems to work well for selecting pure \HII regions.
    \item The extraplanar Balmer emission is produced in-situ by ionizing photons emitted from the disc via radiative recombination (Fig.~\ref{fig:vertical_diag}). Those ionizing photons must be transported through transparent pathways carved out by superbubbles or chimneys. Since the gas density is low at large scale heights, we classify all of this Balmer emission as being emitted from the DIG.
    \item When converting the intrinsic (or dust-corrected) H$\alpha$ luminosity to an SFR, we find that a correction of $\sim10-60\%$ needs to be applied in order to account for the pre-absorption of LyC photons by dust ($f_{\rm abs}\approx28\%$) and helium ($f_{\rm He}\approx9\%$) and for the escape of LyC photons ($f_{\rm esc}\approx6\%$; see Fig.~\ref{fig:Ha_SFR}). This correction depends on the recent star-formation history and therefore cannot straightforwardly be applied in observations without further information. Effects related to H$\alpha$ emission by collisionally excited and ionized gas and related to older stars contribute at the $5\%$-level or less.
    \item The intrinsic H$\alpha$ can be well-recovered (within 25\%) from the observed H$\alpha$ emission by applying the Balmer decrement method (see Fig.~\ref{fig:dust_correction}). We find that the nebular attenuation law depends on the amount of attenuation both on integrated and spatially resolved scales, and that more attenuated regions follow a shallower attenuation law (see Eq.~\ref{eq:slope_tau} and Fig.~\ref{fig:attenuation}).
    \item The star-formation timescale for the H$\alpha$-based SFR indicator is $5$--$10$\,Myr (see Fig.~\ref{fig:Halpha_timescale} and Tab.~\ref{tab:tavg}). We quantify how the exact timescale depends on the variability of the star-formation history (LMC versus MW), the stellar population model, and the amount of attenuation. It is important to consider this variation when measuring the ``burstiness'' of star formation from, for example, the H$\alpha$-to-UV ratio.
\end{enumerate}

In the future, we will apply this framework to cosmological (zoom-in) simulations. Such studies will allow us to more accurately interpret spectroscopic data of high-redshift galaxies with the upcoming \textit{James Webb Space Telescope}. Specifically, drawing from an improved understanding of the H$\alpha$--SFR connection, we will be able to assess the spatially resolved stellar-mass growth and the regulation of star formation via time variability investigations.

\section*{Acknowledgements}

We thank the referee for a thorough report that greatly improved this work. We thank Samir Salim for detailed comments; Ben Johnson, Diederik Kruijssen, and Ryan Sanders for insightful discussions; and Francesco Belfiore for sharing and discussing the results of the PHANGS-MUSE survey. This research made use of NASA's Astrophysics Data System (ADS), the arXiv.org preprint server, the Python plotting library \texttt{matplotlib} \citep{hunter07}, and \texttt{astropy}, a community-developed core Python package for Astronomy \citep{astropycollaboration13, astropycollaboration18}. ST is supported by the 2021 Research Fund 1.210134.01 of UNIST (Ulsan National Institute of Science \& Technology) and by the Smithsonian Astrophysical Observatory through the CfA Fellowship.
AS and HL acknowledge support for Program numbers \textit{HST}-HF2-51421.001-A and HST-HF2-51438.001-A provided by NASA through a grant from the Space Telescope Science Institute, which is operated by the Association of Universities for Research in Astronomy, Incorporated, under NASA contract NAS5-26555.
MV acknowledges support through NASA ATP 19-ATP19-0019, 19-ATP19-0020, 19-ATP19-0167, and NSF grants AST-1814053, AST-1814259, AST-1909831, AST-2007355 and AST-2107724.
FM acknowledges support through the program ``Rita Levi Montalcini'' of the Italian MUR.
PT acknowledges support from NSF AST-1909933, NSF AST-2008490, and NASA ATP Grant 80NSSC20K0502.
LVS acknowledges support from the NSF AST 1817233 and NSF CAREER 1945310 grants.
Computing resources supporting this work were provided by the Extreme Science and Engineering Discovery Environment (XSEDE), at Comet through allocation TG-AST200007 and by the NASA High-End Computing (HEC) Program through the NASA Advanced Supercomputing (NAS) Division at Ames Research Center.

Funding for the Sloan Digital Sky Survey IV has been provided by the Alfred P. Sloan Foundation, the U.S. Department of Energy Office of Science, and the Participating Institutions. SDSS-IV acknowledges support and resources from the Center for High Performance Computing  at the University of Utah. The SDSS website is www.sdss.org. SDSS-IV is managed by the Astrophysical Research Consortium for the Participating Institutions of the SDSS Collaboration including the Brazilian Participation Group, the Carnegie Institution for Science, Carnegie Mellon University, Center for Astrophysics | Harvard \& Smithsonian, the Chilean Participation Group, the French Participation Group, Instituto de Astrof\'isica de Canarias, The Johns Hopkins University, Kavli Institute for the Physics and Mathematics of the Universe (IPMU) / University of Tokyo, the Korean Participation Group, Lawrence Berkeley National Laboratory, Leibniz Institut f\"ur Astrophysik Potsdam (AIP),  Max-Planck-Institut f\"ur Astronomie (MPIA Heidelberg), Max-Planck-Institut f\"ur Astrophysik (MPA Garching), Max-Planck-Institut f\"ur Extraterrestrische Physik (MPE), National Astronomical Observatories of China, New Mexico State University, New York University, University of Notre Dame, Observat\'ario Nacional / MCTI, The Ohio State University, Pennsylvania State University, Shanghai Astronomical Observatory, United Kingdom Participation Group, Universidad Nacional Aut\'onoma de M\'exico, University of Arizona, University of Colorado Boulder, University of Oxford, University of Portsmouth, University of Utah, University of Virginia, University of Washington, University of Wisconsin, Vanderbilt University, and Yale University.

\section*{Data Availability}

Raw data were generated by performing simulations at the NASA Pleiades computer. Derived data supporting the findings of this study are available from the corresponding author ST on request.



\bibliographystyle{mnras}

\begin{thebibliography}{}
\makeatletter
\relax
\def\mn@urlcharsother{\let\do\@makeother \do\$\do\&\do\#\do\^\do\_\do\%\do\~}
\def\mn@doi{\begingroup\mn@urlcharsother \@ifnextchar [ {\mn@doi@}
  {\mn@doi@[]}}
\def\mn@doi@[#1]#2{\def\@tempa{#1}\ifx\@tempa\@empty \href
  {http://dx.doi.org/#2} {doi:#2}\else \href {http://dx.doi.org/#2} {#1}\fi
  \endgroup}
\def\mn@eprint#1#2{\mn@eprint@#1:#2::\@nil}
\def\mn@eprint@arXiv#1{\href {http://arxiv.org/abs/#1} {{\tt arXiv:#1}}}
\def\mn@eprint@dblp#1{\href {http://dblp.uni-trier.de/rec/bibtex/#1.xml}
  {dblp:#1}}
\def\mn@eprint@#1:#2:#3:#4\@nil{\def\@tempa {#1}\def\@tempb {#2}\def\@tempc
  {#3}\ifx \@tempc \@empty \let \@tempc \@tempb \let \@tempb \@tempa \fi \ifx
  \@tempb \@empty \def\@tempb {arXiv}\fi \@ifundefined
  {mn@eprint@\@tempb}{\@tempb:\@tempc}{\expandafter \expandafter \csname
  mn@eprint@\@tempb\endcsname \expandafter{\@tempc}}}

\bibitem[\protect\citeauthoryear{{Aguado} et~al.,}{{Aguado}
  et~al.}{2019}]{aguado19}
{Aguado} D.~S.,  et~al., 2019, \mn@doi [\apjs] {10.3847/1538-4365/aaf651},
  \href {https://ui.adsabs.harvard.edu/abs/2019ApJS..240...23A} {240, 23}

\bibitem[\protect\citeauthoryear{{Aoyama}, {Hou}, {Hirashita}, {Nagamine}  \&
  {Shimizu}}{{Aoyama} et~al.}{2018}]{aoyama18}
{Aoyama} S.,  {Hou} K.-C.,  {Hirashita} H.,  {Nagamine} K.,   {Shimizu} I.,
  2018, \mn@doi [\mnras] {10.1093/mnras/sty1431}, \href
  {https://ui.adsabs.harvard.edu/abs/2018MNRAS.478.4905A} {478, 4905}

\bibitem[\protect\citeauthoryear{{Astropy Collaboration} et~al.,}{{Astropy
  Collaboration} et~al.}{2013}]{astropycollaboration13}
{Astropy Collaboration} et~al., 2013, \mn@doi [\aap]
  {10.1051/0004-6361/201322068}, \href
  {http://adsabs.harvard.edu/abs/2013A26A...558A..33A} {558, A33}

\bibitem[\protect\citeauthoryear{{Astropy Collaboration} et~al.,}{{Astropy
  Collaboration} et~al.}{2018}]{astropycollaboration18}
{Astropy Collaboration} et~al., 2018, \mn@doi [\aj] {10.3847/1538-3881/aabc4f},
  \href {https://ui.adsabs.harvard.edu/abs/2018AJ....156..123A} {156, 123}

\bibitem[\protect\citeauthoryear{{Baldwin}, {Phillips}  \&
  {Terlevich}}{{Baldwin} et~al.}{1981}]{baldwin81}
{Baldwin} J.~A.,  {Phillips} M.~M.,   {Terlevich} R.,  1981, \mn@doi [\pasp]
  {10.1086/130766}, \href {http://adsabs.harvard.edu/abs/1981PASP...93....5B}
  {93, 5}

\bibitem[\protect\citeauthoryear{{Barnes}, {Wood}, {Hill}  \&
  {Haffner}}{{Barnes} et~al.}{2015}]{barnes15}
{Barnes} J.~E.,  {Wood} K.,  {Hill} A.~S.,   {Haffner} L.~M.,  2015, \mn@doi
  [\mnras] {10.1093/mnras/stu2454}, \href
  {https://ui.adsabs.harvard.edu/abs/2015MNRAS.447..559B} {447, 559}

\bibitem[\protect\citeauthoryear{{Barnes} et~al.,}{{Barnes}
  et~al.}{2021}]{barnes21}
{Barnes} A.~T.,  et~al., 2021, \mn@doi [\mnras] {10.1093/mnras/stab2958}, \href
  {https://ui.adsabs.harvard.edu/abs/2021MNRAS.508.5362B} {508, 5362}

\bibitem[\protect\citeauthoryear{{Belfiore} et~al.,}{{Belfiore}
  et~al.}{2018}]{belfiore18}
{Belfiore} F.,  et~al., 2018, \mn@doi [\mnras] {10.1093/mnras/sty768}, \href
  {https://ui.adsabs.harvard.edu/abs/2018MNRAS.477.3014B} {477, 3014}

\bibitem[\protect\citeauthoryear{{Belfiore} et~al.,}{{Belfiore}
  et~al.}{2019}]{belfiore19}
{Belfiore} F.,  et~al., 2019, \mn@doi [\aj] {10.3847/1538-3881/ab3e4e}, \href
  {https://ui.adsabs.harvard.edu/abs/2019AJ....158..160B} {158, 160}

\bibitem[\protect\citeauthoryear{{Belfiore} et~al.,}{{Belfiore}
  et~al.}{2022}]{belfiore22}
{Belfiore} F.,  et~al., 2022, \mn@doi [\aap] {10.1051/0004-6361/202141859},
  \href {https://ui.adsabs.harvard.edu/abs/2022A&A...659A..26B} {659, A26}

\bibitem[\protect\citeauthoryear{{Berman}}{{Berman}}{1936}]{berman36}
{Berman} L.,  1936, \mn@doi [\mnras] {10.1093/mnras/96.9.890}, \href
  {https://ui.adsabs.harvard.edu/abs/1936MNRAS..96..890B} {96, 890}

\bibitem[\protect\citeauthoryear{{Blanton} et~al.,}{{Blanton}
  et~al.}{2005}]{blanton05}
{Blanton} M.~R.,  et~al., 2005, \mn@doi [\aj] {10.1086/429803}, \href
  {https://ui.adsabs.harvard.edu/abs/2005AJ....129.2562B} {129, 2562}

\bibitem[\protect\citeauthoryear{{Blanton}, {Kazin}, {Muna}, {Weaver}  \&
  {Price-Whelan}}{{Blanton} et~al.}{2011}]{blanton11}
{Blanton} M.~R.,  {Kazin} E.,  {Muna} D.,  {Weaver} B.~A.,   {Price-Whelan} A.,
   2011, \mn@doi [\aj] {10.1088/0004-6256/142/1/31}, \href
  {https://ui.adsabs.harvard.edu/abs/2011AJ....142...31B} {142, 31}

\bibitem[\protect\citeauthoryear{{Bruzual} \& {Charlot}}{{Bruzual} \&
  {Charlot}}{2003}]{bruzual03}
{Bruzual} G.,  {Charlot} S.,  2003, \mn@doi [\mnras]
  {10.1046/j.1365-8711.2003.06897.x}, \href
  {http://adsabs.harvard.edu/abs/2003MNRAS.344.1000B} {344, 1000}

\bibitem[\protect\citeauthoryear{{Bundy} et~al.,}{{Bundy}
  et~al.}{2015}]{bundy15}
{Bundy} K.,  et~al., 2015, \mn@doi [\apj] {10.1088/0004-637X/798/1/7}, \href
  {http://adsabs.harvard.edu/abs/2015ApJ...798....7B} {798, 7}

\bibitem[\protect\citeauthoryear{{Byler}, {Dalcanton}, {Conroy}  \&
  {Johnson}}{{Byler} et~al.}{2017}]{byler17}
{Byler} N.,  {Dalcanton} J.~J.,  {Conroy} C.,   {Johnson} B.~D.,  2017, \mn@doi
  [\apj] {10.3847/1538-4357/aa6c66}, \href
  {http://adsabs.harvard.edu/abs/2017ApJ...840...44B} {840, 44}

\bibitem[\protect\citeauthoryear{{Byler}, {Dalcanton}, {Conroy}, {Johnson},
  {Choi}, {Dotter}  \& {Rosenfield}}{{Byler} et~al.}{2019}]{byler19}
{Byler} N.,  {Dalcanton} J.~J.,  {Conroy} C.,  {Johnson} B.~D.,  {Choi} J.,
  {Dotter} A.,   {Rosenfield} P.,  2019, \mn@doi [\aj]
  {10.3847/1538-3881/ab1b70}, \href
  {https://ui.adsabs.harvard.edu/abs/2019AJ....158....2B} {158, 2}

\bibitem[\protect\citeauthoryear{{Calzetti}}{{Calzetti}}{1997}]{calzetti97}
{Calzetti} D.,  1997, \mn@doi [\aj] {10.1086/118242}, \href
  {http://adsabs.harvard.edu/abs/1997AJ....113..162C} {113, 162}

\bibitem[\protect\citeauthoryear{{Calzetti}, {Kinney}  \&
  {Storchi-Bergmann}}{{Calzetti} et~al.}{1994}]{calzetti94}
{Calzetti} D.,  {Kinney} A.~L.,   {Storchi-Bergmann} T.,  1994, \mn@doi [\apj]
  {10.1086/174346}, \href {http://adsabs.harvard.edu/abs/1994ApJ...429..582C}
  {429, 582}

\bibitem[\protect\citeauthoryear{{Camps} \& {Baes}}{{Camps} \&
  {Baes}}{2015}]{camps15}
{Camps} P.,  {Baes} M.,  2015, \mn@doi [Astronomy and Computing]
  {10.1016/j.ascom.2014.10.004}, \href
  {http://adsabs.harvard.edu/abs/2015A%26C.....9...20C} {9, 20}

\bibitem[\protect\citeauthoryear{{Caplar} \& {Tacchella}}{{Caplar} \&
  {Tacchella}}{2019}]{caplar19}
{Caplar} N.,  {Tacchella} S.,  2019, \mn@doi [\mnras] {10.1093/mnras/stz1449},
  \href {https://ui.adsabs.harvard.edu/abs/2019MNRAS.487.3845C} {487, 3845}

\bibitem[\protect\citeauthoryear{{Cardelli}, {Clayton}  \& {Mathis}}{{Cardelli}
  et~al.}{1989}]{cardelli89}
{Cardelli} J.~A.,  {Clayton} G.~C.,   {Mathis} J.~S.,  1989, \mn@doi [\apj]
  {10.1086/167900}, \href {http://adsabs.harvard.edu/abs/1989ApJ...345..245C}
  {345, 245}

\bibitem[\protect\citeauthoryear{{Chabrier}}{{Chabrier}}{2003}]{chabrier03}
{Chabrier} G.,  2003, \mn@doi [\pasp] {10.1086/376392}, \href
  {http://adsabs.harvard.edu/abs/2003PASP..115..763C} {115, 763}

\bibitem[\protect\citeauthoryear{{Charlot} \& {Fall}}{{Charlot} \&
  {Fall}}{2000}]{charlot00}
{Charlot} S.,  {Fall} S.~M.,  2000, \mn@doi [\apj] {10.1086/309250}, \href
  {https://ui.adsabs.harvard.edu/abs/2000ApJ...539..718C} {539, 718}

\bibitem[\protect\citeauthoryear{{Cherinka} et~al.,}{{Cherinka}
  et~al.}{2019}]{cherinka19}
{Cherinka} B.,  et~al., 2019, \mn@doi [\aj] {10.3847/1538-3881/ab2634}, \href
  {https://ui.adsabs.harvard.edu/abs/2019AJ....158...74C} {158, 74}

\bibitem[\protect\citeauthoryear{{Chevallard}, {Charlot}, {Wandelt}  \&
  {Wild}}{{Chevallard} et~al.}{2013}]{chevallard13}
{Chevallard} J.,  {Charlot} S.,  {Wandelt} B.,   {Wild} V.,  2013, \mn@doi
  [\mnras] {10.1093/mnras/stt523}, \href
  {https://ui.adsabs.harvard.edu/abs/2013MNRAS.432.2061C} {432, 2061}

\bibitem[\protect\citeauthoryear{{Chevance} et~al.,}{{Chevance}
  et~al.}{2020}]{chevance20}
{Chevance} M.,  et~al., 2020, \mn@doi [\mnras] {10.1093/mnras/stz3525}, \href
  {https://ui.adsabs.harvard.edu/abs/2020MNRAS.493.2872C} {493, 2872}

\bibitem[\protect\citeauthoryear{{Choi}, {Conroy}  \& {Byler}}{{Choi}
  et~al.}{2017}]{choi17}
{Choi} J.,  {Conroy} C.,   {Byler} N.,  2017, \mn@doi [\apj]
  {10.3847/1538-4357/aa679f}, \href
  {http://adsabs.harvard.edu/abs/2017ApJ...838..159C} {838, 159}

\bibitem[\protect\citeauthoryear{{Cid Fernandes} et~al.,}{{Cid Fernandes}
  et~al.}{2014}]{cid-fernandes14}
{Cid Fernandes} R.,  et~al., 2014, \mn@doi [\aap]
  {10.1051/0004-6361/201321692}, \href
  {http://adsabs.harvard.edu/abs/2014A26A...561A.130C} {561, A130}

\bibitem[\protect\citeauthoryear{{Conroy}}{{Conroy}}{2013}]{conroy13_rev}
{Conroy} C.,  2013, \mn@doi [\araa] {10.1146/annurev-astro-082812-141017},
  \href {http://adsabs.harvard.edu/abs/2013ARA%26A..51..393C} {51, 393}

\bibitem[\protect\citeauthoryear{{Decleir} et~al.,}{{Decleir}
  et~al.}{2019}]{decleir19}
{Decleir} M.,  et~al., 2019, \mn@doi [\mnras] {10.1093/mnras/stz805}, \href
  {https://ui.adsabs.harvard.edu/abs/2019MNRAS.486..743D} {486, 743}

\bibitem[\protect\citeauthoryear{{Dopita} \& {Sutherland}}{{Dopita} \&
  {Sutherland}}{2003}]{dopita03}
{Dopita} M.~A.,  {Sutherland} R.~S.,  2003, {Astrophysics of the diffuse
  universe}.
Springer

\bibitem[\protect\citeauthoryear{{Draine}}{{Draine}}{2011}]{draine11}
{Draine} B.~T.,  2011, \mn@doi [\apj] {10.1088/0004-637X/732/2/100}, \href
  {https://ui.adsabs.harvard.edu/abs/2011ApJ...732..100D} {732, 100}

\bibitem[\protect\citeauthoryear{{Driver}, {Robotham}, {Bland-Hawthorn},
  {Brown}, {Hopkins}, {Liske}, {Phillipps}  \& {Wilkins}}{{Driver}
  et~al.}{2013}]{driver13}
{Driver} S.~P.,  {Robotham} A.~S.~G.,  {Bland-Hawthorn} J.,  {Brown} M.,
  {Hopkins} A.,  {Liske} J.,  {Phillipps} S.,   {Wilkins} S.,  2013, \mn@doi
  [\mnras] {10.1093/mnras/sts717}, \href
  {https://ui.adsabs.harvard.edu/abs/2013MNRAS.430.2622D} {430, 2622}

\bibitem[\protect\citeauthoryear{{Dwek}}{{Dwek}}{1998}]{dwek98}
{Dwek} E.,  1998, \mn@doi [\apj] {10.1086/305829}, \href
  {https://ui.adsabs.harvard.edu/abs/1998ApJ...501..643D} {501, 643}

\bibitem[\protect\citeauthoryear{{Eldridge} \& {Stanway}}{{Eldridge} \&
  {Stanway}}{2009}]{eldridge09}
{Eldridge} J.~J.,  {Stanway} E.~R.,  2009, \mn@doi [\mnras]
  {10.1111/j.1365-2966.2009.15514.x}, \href
  {https://ui.adsabs.harvard.edu/abs/2009MNRAS.400.1019E} {400, 1019}

\bibitem[\protect\citeauthoryear{{Eldridge} \& {Stanway}}{{Eldridge} \&
  {Stanway}}{2012}]{eldridge12}
{Eldridge} J.~J.,  {Stanway} E.~R.,  2012, \mn@doi [\mnras]
  {10.1111/j.1365-2966.2011.19713.x}, \href
  {https://ui.adsabs.harvard.edu/abs/2012MNRAS.419..479E} {419, 479}

\bibitem[\protect\citeauthoryear{{Eldridge}, {Stanway}, {Xiao}, {McClelland },
  {Taylor}, {Ng}, {Greis}  \& {Bray}}{{Eldridge} et~al.}{2017}]{eldridge17}
{Eldridge} J.~J.,  {Stanway} E.~R.,  {Xiao} L.,  {McClelland } L.~A.~S.,
  {Taylor} G.,  {Ng} M.,  {Greis} S.~M.~L.,   {Bray} J.~C.,  2017, \mn@doi
  [\pasa] {10.1017/pasa.2017.51}, \href
  {https://ui.adsabs.harvard.edu/abs/2017PASA...34...58E} {34, e058}

\bibitem[\protect\citeauthoryear{{Ellison}, {S{\'a}nchez}, {Ibarra-Medel},
  {Antonio}, {Mendel}  \& {Barrera-Ballesteros}}{{Ellison}
  et~al.}{2018}]{ellison18}
{Ellison} S.~L.,  {S{\'a}nchez} S.~F.,  {Ibarra-Medel} H.,  {Antonio} B.,
  {Mendel} J.~T.,   {Barrera-Ballesteros} J.,  2018, \mn@doi [\mnras]
  {10.1093/mnras/stx2882}, \href
  {http://adsabs.harvard.edu/abs/2018MNRAS.474.2039E} {474, 2039}

\bibitem[\protect\citeauthoryear{{Emami}, {Siana}, {Weisz}, {Johnson}, {Ma}  \&
  {El-Badry}}{{Emami} et~al.}{2019}]{emami19}
{Emami} N.,  {Siana} B.,  {Weisz} D.~R.,  {Johnson} B.~D.,  {Ma} X.,
  {El-Badry} K.,  2019, \mn@doi [\apj] {10.3847/1538-4357/ab211a}, \href
  {https://ui.adsabs.harvard.edu/abs/2019ApJ...881...71E} {881, 71}

\bibitem[\protect\citeauthoryear{{Emsellem} et~al.,}{{Emsellem}
  et~al.}{2021}]{emsellem21}
{Emsellem} E.,  et~al., 2021, arXiv e-prints, \href
  {https://ui.adsabs.harvard.edu/abs/2021arXiv211003708E} {p. arXiv:2110.03708}

\bibitem[\protect\citeauthoryear{{Faisst}, {Capak}, {Emami}, {Tacchella}  \&
  {Larson}}{{Faisst} et~al.}{2019}]{faisst19}
{Faisst} A.~L.,  {Capak} P.~L.,  {Emami} N.,  {Tacchella} S.,   {Larson} K.~L.,
   2019, \mn@doi [\apj] {10.3847/1538-4357/ab425b}, \href
  {https://ui.adsabs.harvard.edu/abs/2019ApJ...884..133F} {884, 133}

\bibitem[\protect\citeauthoryear{{Faucher-Gigu{\`e}re}, {Lidz}, {Zaldarriaga}
  \& {Hernquist}}{{Faucher-Gigu{\`e}re} et~al.}{2009}]{faucher-giguere09}
{Faucher-Gigu{\`e}re} C.-A.,  {Lidz} A.,  {Zaldarriaga} M.,   {Hernquist} L.,
  2009, \mn@doi [\apj] {10.1088/0004-637X/703/2/1416}, \href
  {https://ui.adsabs.harvard.edu/abs/2009ApJ...703.1416F} {703, 1416}

\bibitem[\protect\citeauthoryear{{Ferrara}, {Bianchi}, {Dettmar}  \&
  {Giovanardi}}{{Ferrara} et~al.}{1996}]{ferrara96}
{Ferrara} A.,  {Bianchi} S.,  {Dettmar} R.-J.,   {Giovanardi} C.,  1996,
  \mn@doi [\apjl] {10.1086/310212}, \href
  {https://ui.adsabs.harvard.edu/abs/1996ApJ...467L..69F} {467, L69}

\bibitem[\protect\citeauthoryear{{Flores Vel{\'a}zquez} et~al.,}{{Flores
  Vel{\'a}zquez} et~al.}{2021}]{flores-velazquez21}
{Flores Vel{\'a}zquez} J.~A.,  et~al., 2021, \mn@doi [\mnras]
  {10.1093/mnras/staa3893}, \href
  {https://ui.adsabs.harvard.edu/abs/2021MNRAS.501.4812F} {501, 4812}

\bibitem[\protect\citeauthoryear{{F{\"o}rster Schreiber} et~al.,}{{F{\"o}rster
  Schreiber} et~al.}{2014}]{forster-schreiber14}
{F{\"o}rster Schreiber} N.~M.,  et~al., 2014, \mn@doi [\apj]
  {10.1088/0004-637X/787/1/38}, \href
  {http://adsabs.harvard.edu/abs/2014ApJ...787...38F} {787, 38}

\bibitem[\protect\citeauthoryear{{Fumagalli}, {da Silva}  \&
  {Krumholz}}{{Fumagalli} et~al.}{2011}]{fumagalli11}
{Fumagalli} M.,  {da Silva} R.~L.,   {Krumholz} M.~R.,  2011, \mn@doi [\apjl]
  {10.1088/2041-8205/741/2/L26}, \href
  {http://adsabs.harvard.edu/abs/2011ApJ...741L..26F} {741, L26}

\bibitem[\protect\citeauthoryear{{Garaldi}, {Kannan}, {Smith}, {Springel},
  {Pakmor}, {Vogelsberger}  \& {Hernquist}}{{Garaldi} et~al.}{2021}]{garaldi21}
{Garaldi} E.,  {Kannan} R.,  {Smith} A.,  {Springel} V.,  {Pakmor} R.,
  {Vogelsberger} M.,   {Hernquist} L.,  2021, arXiv e-prints, \href
  {https://ui.adsabs.harvard.edu/abs/2021arXiv211001628G} {p. arXiv:2110.01628}

\bibitem[\protect\citeauthoryear{{Genzel} et~al.,}{{Genzel}
  et~al.}{2014}]{genzel14b}
{Genzel} R.,  et~al., 2014, \mn@doi [\apj] {10.1088/0004-637X/796/1/7}, \href
  {http://adsabs.harvard.edu/abs/2014ApJ...796....7G} {796, 7}

\bibitem[\protect\citeauthoryear{{Groves}, {Brinchmann}  \& {Walcher}}{{Groves}
  et~al.}{2012}]{groves12}
{Groves} B.,  {Brinchmann} J.,   {Walcher} C.~J.,  2012, \mn@doi [\mnras]
  {10.1111/j.1365-2966.2011.19796.x}, \href
  {http://adsabs.harvard.edu/abs/2012MNRAS.419.1402G} {419, 1402}

\bibitem[\protect\citeauthoryear{{Guo} et~al.,}{{Guo} et~al.}{2016}]{guo16}
{Guo} Y.,  et~al., 2016, \mn@doi [\apj] {10.3847/1538-4357/833/1/37}, \href
  {http://adsabs.harvard.edu/abs/2016ApJ...833...37G} {833, 37}

\bibitem[\protect\citeauthoryear{{Haydon}, {Fujimoto}, {Chevance}, {Kruijssen},
  {Krumholz}  \& {Longmore}}{{Haydon} et~al.}{2020a}]{haydon20b}
{Haydon} D.~T.,  {Fujimoto} Y.,  {Chevance} M.,  {Kruijssen} J.~M.~D.,
  {Krumholz} M.~R.,   {Longmore} S.~N.,  2020a, \mn@doi [\mnras]
  {10.1093/mnras/staa2162}, \href
  {https://ui.adsabs.harvard.edu/abs/2020MNRAS.497.5076H} {497, 5076}

\bibitem[\protect\citeauthoryear{{Haydon}, {Kruijssen}, {Chevance}, {Hygate},
  {Krumholz}, {Schruba}  \& {Longmore}}{{Haydon} et~al.}{2020b}]{haydon20}
{Haydon} D.~T.,  {Kruijssen} J.~M.~D.,  {Chevance} M.,  {Hygate} A. P.~S.,
  {Krumholz} M.~R.,  {Schruba} A.,   {Longmore} S.~N.,  2020b, \mn@doi [\mnras]
  {10.1093/mnras/staa2430}, \href
  {https://ui.adsabs.harvard.edu/abs/2020MNRAS.498..235H} {498, 235}

\bibitem[\protect\citeauthoryear{{Hernquist}}{{Hernquist}}{1990}]{hernquist90}
{Hernquist} L.,  1990, \mn@doi [\apj] {10.1086/168845}, \href
  {http://adsabs.harvard.edu/abs/1990ApJ...356..359H} {356, 359}

\bibitem[\protect\citeauthoryear{{Hernquist}}{{Hernquist}}{1993}]{hernquist93}
{Hernquist} L.,  1993, \mn@doi [\apj] {10.1086/172686}, \href
  {http://adsabs.harvard.edu/abs/1993ApJ...409..548H} {409, 548}

\bibitem[\protect\citeauthoryear{{Hirashita}, {Buat}  \& {Inoue}}{{Hirashita}
  et~al.}{2003}]{hirashita03}
{Hirashita} H.,  {Buat} V.,   {Inoue} A.~K.,  2003, \mn@doi [\aap]
  {10.1051/0004-6361:20031144}, \href
  {https://ui.adsabs.harvard.edu/abs/2003A&A...410...83H} {410, 83}

\bibitem[\protect\citeauthoryear{{Hoyle} \& {Ellis}}{{Hoyle} \&
  {Ellis}}{1963}]{hoyle63}
{Hoyle} F.,  {Ellis} G.~R.~A.,  1963, \mn@doi [Australian Journal of Physics]
  {10.1071/PH630001}, \href
  {https://ui.adsabs.harvard.edu/abs/1963AuJPh..16....1H} {16, 1}

\bibitem[\protect\citeauthoryear{Hunter}{Hunter}{2007}]{hunter07}
Hunter J.~D.,  2007, Computing In Science \& Engineering, 9, 90

\bibitem[\protect\citeauthoryear{{Iglesias-P{\'a}ramo}, {Boselli}, {Gavazzi}
  \& {Zaccardo}}{{Iglesias-P{\'a}ramo} et~al.}{2004}]{iglesias-paramo04}
{Iglesias-P{\'a}ramo} J.,  {Boselli} A.,  {Gavazzi} G.,   {Zaccardo} A.,  2004,
  \mn@doi [\aap] {10.1051/0004-6361:20034572}, \href
  {https://ui.adsabs.harvard.edu/abs/2004A&A...421..887I} {421, 887}

\bibitem[\protect\citeauthoryear{{Inoue}, {Hirashita}  \& {Kamaya}}{{Inoue}
  et~al.}{2001}]{inoue01}
{Inoue} A.~K.,  {Hirashita} H.,   {Kamaya} H.,  2001, \mn@doi [\apj]
  {10.1086/321499}, \href
  {https://ui.adsabs.harvard.edu/abs/2001ApJ...555..613I} {555, 613}

\bibitem[\protect\citeauthoryear{{Iyer} et~al.,}{{Iyer} et~al.}{2020}]{iyer20}
{Iyer} K.~G.,  et~al., 2020, \mn@doi [\mnras] {10.1093/mnras/staa2150}, \href
  {https://ui.adsabs.harvard.edu/abs/2020MNRAS.498..430I} {498, 430}

\bibitem[\protect\citeauthoryear{{Jo}, {Seon}, {Shinn}, {Yang}, {Lee}  \&
  {Min}}{{Jo} et~al.}{2018}]{jo18}
{Jo} Y.-S.,  {Seon} K.-i.,  {Shinn} J.-H.,  {Yang} Y.,  {Lee} D.,   {Min}
  K.-W.,  2018, \mn@doi [\apj] {10.3847/1538-4357/aacbca}, \href
  {https://ui.adsabs.harvard.edu/abs/2018ApJ...862...25J} {862, 25}

\bibitem[\protect\citeauthoryear{{Kado-Fong}, {Kim}, {Ostriker}  \&
  {Kim}}{{Kado-Fong} et~al.}{2020}]{kado-fong20}
{Kado-Fong} E.,  {Kim} J.-G.,  {Ostriker} E.~C.,   {Kim} C.-G.,  2020, \mn@doi
  [\apj] {10.3847/1538-4357/ab9abd}, \href
  {https://ui.adsabs.harvard.edu/abs/2020ApJ...897..143K} {897, 143}

\bibitem[\protect\citeauthoryear{{Kannan}, {Vogelsberger}, {Stinson},
  {Hennawi}, {Marinacci}, {Springel}  \& {Macci{\`o}}}{{Kannan}
  et~al.}{2016}]{kannan16}
{Kannan} R.,  {Vogelsberger} M.,  {Stinson} G.~S.,  {Hennawi} J.~F.,
  {Marinacci} F.,  {Springel} V.,   {Macci{\`o}} A.~V.,  2016, \mn@doi [\mnras]
  {10.1093/mnras/stw463}, \href
  {https://ui.adsabs.harvard.edu/abs/2016MNRAS.458.2516K} {458, 2516}

\bibitem[\protect\citeauthoryear{{Kannan}, {Vogelsberger}, {Marinacci},
  {McKinnon}, {Pakmor}  \& {Springel}}{{Kannan} et~al.}{2019}]{kannan19_rt}
{Kannan} R.,  {Vogelsberger} M.,  {Marinacci} F.,  {McKinnon} R.,  {Pakmor} R.,
    {Springel} V.,  2019, \mn@doi [\mnras] {10.1093/mnras/stz287}, \href
  {http://adsabs.harvard.edu/abs/2019MNRAS.485..117K} {485, 117}

\bibitem[\protect\citeauthoryear{{Kannan}, {Marinacci}, {Vogelsberger},
  {Sales}, {Torrey}, {Springel}  \& {Hernquist}}{{Kannan}
  et~al.}{2020}]{kannan20_mw}
{Kannan} R.,  {Marinacci} F.,  {Vogelsberger} M.,  {Sales} L.~V.,  {Torrey} P.,
   {Springel} V.,   {Hernquist} L.,  2020, \mn@doi [\mnras]
  {10.1093/mnras/staa3249}, \href
  {https://ui.adsabs.harvard.edu/abs/2020MNRAS.499.5732K} {499, 5732}

\bibitem[\protect\citeauthoryear{{Kannan}, {Garaldi}, {Smith}, {Pakmor},
  {Springel}, {Vogelsberger}  \& {Hernquist}}{{Kannan}
  et~al.}{2021a}]{kannan21}
{Kannan} R.,  {Garaldi} E.,  {Smith} A.,  {Pakmor} R.,  {Springel} V.,
  {Vogelsberger} M.,   {Hernquist} L.,  2021a, arXiv e-prints, \href
  {https://ui.adsabs.harvard.edu/abs/2021arXiv211000584K} {p. arXiv:2110.00584}

\bibitem[\protect\citeauthoryear{{Kannan}, {Smith}, {Garaldi}, {Shen},
  {Vogelsberger}, {Pakmor}, {Springel}  \& {Hernquist}}{{Kannan}
  et~al.}{2021b}]{kannan21_line}
{Kannan} R.,  {Smith} A.,  {Garaldi} E.,  {Shen} X.,  {Vogelsberger} M.,
  {Pakmor} R.,  {Springel} V.,   {Hernquist} L.,  2021b, arXiv e-prints, \href
  {https://ui.adsabs.harvard.edu/abs/2021arXiv211102411K} {p. arXiv:2111.02411}

\bibitem[\protect\citeauthoryear{{Kannan}, {Vogelsberger}, {Marinacci},
  {Sales}, {Torrey}  \& {Hernquist}}{{Kannan} et~al.}{2021c}]{kannan21_dust}
{Kannan} R.,  {Vogelsberger} M.,  {Marinacci} F.,  {Sales} L.~V.,  {Torrey} P.,
    {Hernquist} L.,  2021c, \mn@doi [\mnras] {10.1093/mnras/stab416}, \href
  {https://ui.adsabs.harvard.edu/abs/2021MNRAS.503..336K} {503, 336}

\bibitem[\protect\citeauthoryear{{Katz} et~al.,}{{Katz} et~al.}{2019}]{katz19}
{Katz} H.,  et~al., 2019, \mn@doi [\mnras] {10.1093/mnras/stz1672}, \href
  {https://ui.adsabs.harvard.edu/abs/2019MNRAS.487.5902K} {487, 5902}

\bibitem[\protect\citeauthoryear{{Kennicutt}}{{Kennicutt}}{1983}]{kennicutt83}
{Kennicutt} Jr. R.~C.,  1983, \mn@doi [\apj] {10.1086/161261}, \href
  {http://adsabs.harvard.edu/abs/1983ApJ...272...54K} {272, 54}

\bibitem[\protect\citeauthoryear{{Kennicutt}}{{Kennicutt}}{1998}]{kennicutt98}
{Kennicutt} Jr. R.~C.,  1998, \mn@doi [\araa] {10.1146/annurev.astro.36.1.189},
  \href {http://adsabs.harvard.edu/abs/1998ARA26A..36..189K} {36, 189}

\bibitem[\protect\citeauthoryear{{Kewley}, {Nicholls}  \&
  {Sutherland}}{{Kewley} et~al.}{2019}]{kewley19_review}
{Kewley} L.~J.,  {Nicholls} D.~C.,   {Sutherland} R.~S.,  2019, \mn@doi [\araa]
  {10.1146/annurev-astro-081817-051832}, \href
  {https://ui.adsabs.harvard.edu/abs/2019ARA&A..57..511K} {57, 511}

\bibitem[\protect\citeauthoryear{{Kim}, {Krumholz}, {Wise}, {Turk}, {Goldbaum}
  \& {Abel}}{{Kim} et~al.}{2013}]{kim13_sfr}
{Kim} J.-h.,  {Krumholz} M.~R.,  {Wise} J.~H.,  {Turk} M.~J.,  {Goldbaum}
  N.~J.,   {Abel} T.,  2013, \mn@doi [\apj] {10.1088/0004-637X/779/1/8}, \href
  {https://ui.adsabs.harvard.edu/abs/2013ApJ...779....8K} {779, 8}

\bibitem[\protect\citeauthoryear{{Kim}, {Wise}, {Abel}, {Jo}, {Primack}  \&
  {Hopkins}}{{Kim} et~al.}{2019}]{kim19}
{Kim} J.-h.,  {Wise} J.~H.,  {Abel} T.,  {Jo} Y.,  {Primack} J.~R.,   {Hopkins}
  P.~F.,  2019, \mn@doi [\apj] {10.3847/1538-4357/ab510b}, \href
  {https://ui.adsabs.harvard.edu/abs/2019ApJ...887..120K} {887, 120}

\bibitem[\protect\citeauthoryear{{Kim} et~al.,}{{Kim} et~al.}{2021}]{kim21}
{Kim} J.,  et~al., 2021, \mn@doi [\mnras] {10.1093/mnras/stab878}, \href
  {https://ui.adsabs.harvard.edu/abs/2021MNRAS.504..487K} {504, 487}

\bibitem[\protect\citeauthoryear{{Koyama} et~al.,}{{Koyama}
  et~al.}{2015}]{koyama15}
{Koyama} Y.,  et~al., 2015, \mn@doi [\mnras] {10.1093/mnras/stv1599}, \href
  {http://adsabs.harvard.edu/abs/2015MNRAS.453..879K} {453, 879}

\bibitem[\protect\citeauthoryear{{Kreckel}, {Blanc}, {Schinnerer}, {Groves},
  {Adamo}, {Hughes}  \& {Meidt}}{{Kreckel} et~al.}{2016}]{kreckel16}
{Kreckel} K.,  {Blanc} G.~A.,  {Schinnerer} E.,  {Groves} B.,  {Adamo} A.,
  {Hughes} A.,   {Meidt} S.,  2016, \mn@doi [\apj]
  {10.3847/0004-637X/827/2/103}, \href
  {https://ui.adsabs.harvard.edu/abs/2016ApJ...827..103K} {827, 103}

\bibitem[\protect\citeauthoryear{{Kruijssen} et~al.,}{{Kruijssen}
  et~al.}{2019}]{kruijssen19}
{Kruijssen} J.~M.~D.,  et~al., 2019, \mn@doi [\nat]
  {10.1038/s41586-019-1194-3}, \href
  {https://ui.adsabs.harvard.edu/abs/2019Natur.569..519K} {569, 519}

\bibitem[\protect\citeauthoryear{{Lacerda} et~al.,}{{Lacerda}
  et~al.}{2018}]{lacerda18}
{Lacerda} E.~A.~D.,  et~al., 2018, \mn@doi [\mnras] {10.1093/mnras/stx3022},
  \href {https://ui.adsabs.harvard.edu/abs/2018MNRAS.474.3727L} {474, 3727}

\bibitem[\protect\citeauthoryear{{Law} et~al.,}{{Law} et~al.}{2016}]{law16}
{Law} D.~R.,  et~al., 2016, \mn@doi [\aj] {10.3847/0004-6256/152/4/83}, \href
  {https://ui.adsabs.harvard.edu/abs/2016AJ....152...83L} {152, 83}

\bibitem[\protect\citeauthoryear{{Lee} et~al.,}{{Lee}
  et~al.}{2009}]{lee09_UVHa}
{Lee} J.~C.,  et~al., 2009, \mn@doi [\apj] {10.1088/0004-637X/706/1/599}, \href
  {http://adsabs.harvard.edu/abs/2009ApJ...706..599L} {706, 599}

\bibitem[\protect\citeauthoryear{{Levesque}, {Leitherer}, {Ekstrom}, {Meynet}
  \& {Schaerer}}{{Levesque} et~al.}{2012}]{levesque12}
{Levesque} E.~M.,  {Leitherer} C.,  {Ekstrom} S.,  {Meynet} G.,   {Schaerer}
  D.,  2012, \mn@doi [\apj] {10.1088/0004-637X/751/1/67}, \href
  {https://ui.adsabs.harvard.edu/abs/2012ApJ...751...67L} {751, 67}

\bibitem[\protect\citeauthoryear{{Levy} et~al.,}{{Levy} et~al.}{2019}]{levy19}
{Levy} R.~C.,  et~al., 2019, \mn@doi [\apj] {10.3847/1538-4357/ab2ed4}, \href
  {https://ui.adsabs.harvard.edu/abs/2019ApJ...882...84L} {882, 84}

\bibitem[\protect\citeauthoryear{{Li}, {Narayanan}  \& {Dav{\'e}}}{{Li}
  et~al.}{2019}]{li19_dust}
{Li} Q.,  {Narayanan} D.,   {Dav{\'e}} R.,  2019, \mn@doi [\mnras]
  {10.1093/mnras/stz2684}, \href
  {https://ui.adsabs.harvard.edu/abs/2019MNRAS.490.1425L} {490, 1425}

\bibitem[\protect\citeauthoryear{{Ma}, {Quataert}, {Wetzel}, {Hopkins},
  {Faucher-Gigu{\`e}re}  \& {Kere{\v{s}}}}{{Ma} et~al.}{2020}]{ma20}
{Ma} X.,  {Quataert} E.,  {Wetzel} A.,  {Hopkins} P.~F.,  {Faucher-Gigu{\`e}re}
  C.-A.,   {Kere{\v{s}}} D.,  2020, \mn@doi [\mnras] {10.1093/mnras/staa2404},
  \href {https://ui.adsabs.harvard.edu/abs/2020MNRAS.498.2001M} {498, 2001}

\bibitem[\protect\citeauthoryear{{Mac Low} \& {Ferrara}}{{Mac Low} \&
  {Ferrara}}{1999}]{mac-low99}
{Mac Low} M.-M.,  {Ferrara} A.,  1999, \mn@doi [\apj] {10.1086/306832}, \href
  {https://ui.adsabs.harvard.edu/abs/1999ApJ...513..142M} {513, 142}

\bibitem[\protect\citeauthoryear{{Marinacci}, {Sales}, {Vogelsberger}, {Torrey}
   \& {Springel}}{{Marinacci} et~al.}{2019}]{marinacci19}
{Marinacci} F.,  {Sales} L.~V.,  {Vogelsberger} M.,  {Torrey} P.,   {Springel}
  V.,  2019, \mn@doi [\mnras] {10.1093/mnras/stz2391}, \href
  {https://ui.adsabs.harvard.edu/abs/2019MNRAS.489.4233M} {489, 4233}

\bibitem[\protect\citeauthoryear{{McKee}}{{McKee}}{1989}]{mckee89}
{McKee} C.,  1989, in {Allamandola} L.~J.,  {Tielens} A.~G.~G.~M.,  eds,
  Proceedings of the 135th Symposium of the International Astronomical Union
  Vol. 135, Interstellar Dust. p.~431

\bibitem[\protect\citeauthoryear{{McKinnon}, {Torrey}  \&
  {Vogelsberger}}{{McKinnon} et~al.}{2016}]{mckinnon16}
{McKinnon} R.,  {Torrey} P.,   {Vogelsberger} M.,  2016, \mn@doi [\mnras]
  {10.1093/mnras/stw253}, \href
  {https://ui.adsabs.harvard.edu/abs/2016MNRAS.457.3775M} {457, 3775}

\bibitem[\protect\citeauthoryear{{McKinnon}, {Torrey}, {Vogelsberger},
  {Hayward}  \& {Marinacci}}{{McKinnon} et~al.}{2017}]{mckinnon17}
{McKinnon} R.,  {Torrey} P.,  {Vogelsberger} M.,  {Hayward} C.~C.,
  {Marinacci} F.,  2017, \mn@doi [\mnras] {10.1093/mnras/stx467}, \href
  {https://ui.adsabs.harvard.edu/abs/2017MNRAS.468.1505M} {468, 1505}

\bibitem[\protect\citeauthoryear{{McKinnon}, {Vogelsberger}, {Torrey},
  {Marinacci}  \& {Kannan}}{{McKinnon} et~al.}{2018}]{mckinnon18}
{McKinnon} R.,  {Vogelsberger} M.,  {Torrey} P.,  {Marinacci} F.,   {Kannan}
  R.,  2018, \mn@doi [\mnras] {10.1093/mnras/sty1248}, \href
  {https://ui.adsabs.harvard.edu/abs/2018MNRAS.478.2851M} {478, 2851}

\bibitem[\protect\citeauthoryear{{McKinnon}, {Kannan}, {Vogelsberger},
  {O'Neil}, {Torrey}  \& {Li}}{{McKinnon} et~al.}{2019}]{mckinnon19}
{McKinnon} R.,  {Kannan} R.,  {Vogelsberger} M.,  {O'Neil} S.,  {Torrey} P.,
  {Li} H.,  2019, arXiv e-prints, \href
  {https://ui.adsabs.harvard.edu/abs/2019arXiv191202825M} {p. arXiv:1912.02825}

\bibitem[\protect\citeauthoryear{{Narayanan}, {Conroy}, {Dav{\'e}}, {Johnson}
  \& {Popping}}{{Narayanan} et~al.}{2018}]{narayanan18}
{Narayanan} D.,  {Conroy} C.,  {Dav{\'e}} R.,  {Johnson} B.~D.,   {Popping} G.,
   2018, \mn@doi [\apj] {10.3847/1538-4357/aaed25}, \href
  {http://adsabs.harvard.edu/abs/2018ApJ...869...70N} {869, 70}

\bibitem[\protect\citeauthoryear{{Narayanan} et~al.,}{{Narayanan}
  et~al.}{2021}]{narayanan21}
{Narayanan} D.,  et~al., 2021, \mn@doi [\apjs] {10.3847/1538-4365/abc487},
  \href {https://ui.adsabs.harvard.edu/abs/2021ApJS..252...12N} {252, 12}

\bibitem[\protect\citeauthoryear{{Nelson} et~al.,}{{Nelson}
  et~al.}{2016a}]{nelson16_balmer}
{Nelson} E.~J.,  et~al., 2016a, \mn@doi [\apjl] {10.3847/2041-8205/817/1/L9},
  \href {http://adsabs.harvard.edu/abs/2016ApJ...817L...9N} {817, L9}

\bibitem[\protect\citeauthoryear{{Nelson} et~al.,}{{Nelson}
  et~al.}{2016b}]{nelson16_insideout}
{Nelson} E.~J.,  et~al., 2016b, \mn@doi [\apj] {10.3847/0004-637X/828/1/27},
  \href {http://adsabs.harvard.edu/abs/2016ApJ...828...27N} {828, 27}

\bibitem[\protect\citeauthoryear{{Nelson} et~al.,}{{Nelson}
  et~al.}{2021}]{nelson21}
{Nelson} E.~J.,  et~al., 2021, \mn@doi [\mnras] {10.1093/mnras/stab2131}, \href
  {https://ui.adsabs.harvard.edu/abs/2021MNRAS.508..219N} {508, 219}

\bibitem[\protect\citeauthoryear{{O'Donnell}}{{O'Donnell}}{1994}]{odonnell94}
{O'Donnell} J.~E.,  1994, \mn@doi [\apj] {10.1086/173713}, \href
  {http://adsabs.harvard.edu/abs/1994ApJ...422..158O} {422, 158}

\bibitem[\protect\citeauthoryear{{Oey} et~al.,}{{Oey} et~al.}{2007}]{oey07}
{Oey} M.~S.,  et~al., 2007, \mn@doi [\apj] {10.1086/517867}, \href
  {https://ui.adsabs.harvard.edu/abs/2007ApJ...661..801O} {661, 801}

\bibitem[\protect\citeauthoryear{{Osterbrock} \& {Ferland}}{{Osterbrock} \&
  {Ferland}}{2006}]{osterbrock06}
{Osterbrock} D.~E.,  {Ferland} G.~J.,  2006, {Astrophysics of gaseous nebulae
  and active galactic nuclei, 2nd ed.}.
University Science Books

\bibitem[\protect\citeauthoryear{{Pellegrini}, {Rahner}, {Reissl}, {Glover},
  {Klessen}, {Rousseau-Nepton}  \& {Herrera-Camus}}{{Pellegrini}
  et~al.}{2020a}]{pellegrini20emp}
{Pellegrini} E.~W.,  {Rahner} D.,  {Reissl} S.,  {Glover} S.~C.~O.,  {Klessen}
  R.~S.,  {Rousseau-Nepton} L.,   {Herrera-Camus} R.,  2020a, \mn@doi [\mnras]
  {10.1093/mnras/staa1473}, \href
  {https://ui.adsabs.harvard.edu/abs/2020MNRAS.496..339P} {496, 339}

\bibitem[\protect\citeauthoryear{{Pellegrini}, {Reissl}, {Rahner}, {Klessen},
  {Glover}, {Pakmor}, {Herrera-Camus}  \& {Grand}}{{Pellegrini}
  et~al.}{2020b}]{pellegrini20}
{Pellegrini} E.~W.,  {Reissl} S.,  {Rahner} D.,  {Klessen} R.~S.,  {Glover} S.
  C.~O.,  {Pakmor} R.,  {Herrera-Camus} R.,   {Grand} R. J.~J.,  2020b, \mn@doi
  [\mnras] {10.1093/mnras/staa2555}, \href
  {https://ui.adsabs.harvard.edu/abs/2020MNRAS.498.3193P} {498, 3193}

\bibitem[\protect\citeauthoryear{{Peters} et~al.,}{{Peters}
  et~al.}{2017}]{peters17}
{Peters} T.,  et~al., 2017, \mn@doi [\mnras] {10.1093/mnras/stw3216}, \href
  {https://ui.adsabs.harvard.edu/abs/2017MNRAS.466.3293P} {466, 3293}

\bibitem[\protect\citeauthoryear{{Planck Collaboration} et~al.,}{{Planck
  Collaboration} et~al.}{2016}]{planck-collaboration16}
{Planck Collaboration} et~al., 2016, \mn@doi [\aap]
  {10.1051/0004-6361/201525830}, \href
  {http://adsabs.harvard.edu/abs/2016A%26A...594A..13P} {594, A13}

\bibitem[\protect\citeauthoryear{{Poetrodjojo}, {D'Agostino}, {Groves},
  {Kewley}, {Ho}, {Rich}, {Madore}  \& {Seibert}}{{Poetrodjojo}
  et~al.}{2019}]{poetrodjojo19}
{Poetrodjojo} H.,  {D'Agostino} J.~J.,  {Groves} B.,  {Kewley} L.,  {Ho} I.~T.,
   {Rich} J.,  {Madore} B.~F.,   {Seibert} M.,  2019, \mn@doi [\mnras]
  {10.1093/mnras/stz1241}, \href
  {https://ui.adsabs.harvard.edu/abs/2019MNRAS.487...79P} {487, 79}

\bibitem[\protect\citeauthoryear{{Puglisi} et~al.,}{{Puglisi}
  et~al.}{2016}]{puglisi16}
{Puglisi} A.,  et~al., 2016, \mn@doi [\aap] {10.1051/0004-6361/201526782},
  \href {http://adsabs.harvard.edu/abs/2016A%26A...586A..83P} {586, A83}

\bibitem[\protect\citeauthoryear{{Rahmati}, {Pawlik}, {Rai{\v{c}}evi{\'c}}  \&
  {Schaye}}{{Rahmati} et~al.}{2013}]{rahmati13}
{Rahmati} A.,  {Pawlik} A.~H.,  {Rai{\v{c}}evi{\'c}} M.,   {Schaye} J.,  2013,
  \mn@doi [\mnras] {10.1093/mnras/stt066}, \href
  {https://ui.adsabs.harvard.edu/abs/2013MNRAS.430.2427R} {430, 2427}

\bibitem[\protect\citeauthoryear{{Rand}, {Kulkarni}  \& {Hester}}{{Rand}
  et~al.}{1990}]{rand90}
{Rand} R.~J.,  {Kulkarni} S.~R.,   {Hester} J.~J.,  1990, \mn@doi [\apjl]
  {10.1086/185679}, \href
  {https://ui.adsabs.harvard.edu/abs/1990ApJ...352L...1R} {352, L1}

\bibitem[\protect\citeauthoryear{{Reynolds}}{{Reynolds}}{1989}]{reynolds89}
{Reynolds} R.~J.,  1989, \mn@doi [\apjl] {10.1086/185412}, \href
  {https://ui.adsabs.harvard.edu/abs/1989ApJ...339L..29R} {339, L29}

\bibitem[\protect\citeauthoryear{{Reynolds}}{{Reynolds}}{1990}]{reynolds90}
{Reynolds} R.~J.,  1990, in {Bowyer} S.,  {Leinert} C.,  eds,  Proceedings of
  the 139th. Symposium of the International Astronomical Union Vol. 139, The
  Galactic and Extragalactic Background Radiation. p.~157

\bibitem[\protect\citeauthoryear{{Robitaille}, {Rice}, {Beaumont}, {Ginsburg},
  {MacDonald}  \& {Rosolowsky}}{{Robitaille} et~al.}{2019}]{robitaille19}
{Robitaille} T.,  {Rice} T.,  {Beaumont} C.,  {Ginsburg} A.,  {MacDonald} B.,
  {Rosolowsky} E.,  2019, {astrodendro: Astronomical data dendrogram creator}
  (\mn@eprint {ascl} {1907.016})

\bibitem[\protect\citeauthoryear{{Rosdahl} \& {Teyssier}}{{Rosdahl} \&
  {Teyssier}}{2015}]{rosdahl15}
{Rosdahl} J.,  {Teyssier} R.,  2015, \mn@doi [\mnras] {10.1093/mnras/stv567},
  \href {https://ui.adsabs.harvard.edu/abs/2015MNRAS.449.4380R} {449, 4380}

\bibitem[\protect\citeauthoryear{{Rosdahl}, {Blaizot}, {Aubert}, {Stranex}  \&
  {Teyssier}}{{Rosdahl} et~al.}{2013}]{rosdahl13_rt}
{Rosdahl} J.,  {Blaizot} J.,  {Aubert} D.,  {Stranex} T.,   {Teyssier} R.,
  2013, \mn@doi [\mnras] {10.1093/mnras/stt1722}, \href
  {https://ui.adsabs.harvard.edu/abs/2013MNRAS.436.2188R} {436, 2188}

\bibitem[\protect\citeauthoryear{{Rosdahl} et~al.,}{{Rosdahl}
  et~al.}{2018}]{rosdahl18}
{Rosdahl} J.,  et~al., 2018, \mn@doi [\mnras] {10.1093/mnras/sty1655}, \href
  {http://adsabs.harvard.edu/abs/2018MNRAS.479..994R} {479, 994}

\bibitem[\protect\citeauthoryear{{Salim} \& {Narayanan}}{{Salim} \&
  {Narayanan}}{2020}]{salim20}
{Salim} S.,  {Narayanan} D.,  2020, \mn@doi [\araa]
  {10.1146/annurev-astro-032620-021933}, \href
  {https://ui.adsabs.harvard.edu/abs/2020ARA&A..58..529S} {58, 529}

\bibitem[\protect\citeauthoryear{{Salim} et~al.,}{{Salim}
  et~al.}{2016}]{salim16}
{Salim} S.,  et~al., 2016, \mn@doi [\apjs] {10.3847/0067-0049/227/1/2}, \href
  {https://ui.adsabs.harvard.edu/abs/2016ApJS..227....2S} {227, 2}

\bibitem[\protect\citeauthoryear{{Salim}, {Boquien}  \& {Lee}}{{Salim}
  et~al.}{2018}]{salim18_curves}
{Salim} S.,  {Boquien} M.,   {Lee} J.~C.,  2018, \mn@doi [\apj]
  {10.3847/1538-4357/aabf3c}, \href
  {http://adsabs.harvard.edu/abs/2018ApJ...859...11S} {859, 11}

\bibitem[\protect\citeauthoryear{{Salpeter}}{{Salpeter}}{1955}]{salpeter55}
{Salpeter} E.~E.,  1955, \mn@doi [\apj] {10.1086/145971}, \href
  {http://adsabs.harvard.edu/abs/1955ApJ...121..161S} {121, 161}

\bibitem[\protect\citeauthoryear{{Sanders}, {Shapley}, {Zhang}  \&
  {Yan}}{{Sanders} et~al.}{2017}]{sanders17}
{Sanders} R.~L.,  {Shapley} A.~E.,  {Zhang} K.,   {Yan} R.,  2017, \mn@doi
  [\apj] {10.3847/1538-4357/aa93e4}, \href
  {http://adsabs.harvard.edu/abs/2017ApJ...850..136S} {850, 136}

\bibitem[\protect\citeauthoryear{{Schaerer}, {Fragos}  \& {Izotov}}{{Schaerer}
  et~al.}{2019}]{schaerer19}
{Schaerer} D.,  {Fragos} T.,   {Izotov} Y.~I.,  2019, \mn@doi [\aap]
  {10.1051/0004-6361/201935005}, \href
  {https://ui.adsabs.harvard.edu/abs/2019A&A...622L..10S} {622, L10}

\bibitem[\protect\citeauthoryear{{Senchyna}, {Stark}, {Mirocha}, {Reines},
  {Charlot}, {Jones}  \& {Mulchaey}}{{Senchyna} et~al.}{2020}]{senchyna20}
{Senchyna} P.,  {Stark} D.~P.,  {Mirocha} J.,  {Reines} A.~E.,  {Charlot} S.,
  {Jones} T.,   {Mulchaey} J.~S.,  2020, \mn@doi [\mnras]
  {10.1093/mnras/staa586}, \href
  {https://ui.adsabs.harvard.edu/abs/2020MNRAS.494..941S} {494, 941}

\bibitem[\protect\citeauthoryear{{Shivaei}, {Reddy}, {Steidel}  \&
  {Shapley}}{{Shivaei} et~al.}{2015}]{shivaei15_SFR}
{Shivaei} I.,  {Reddy} N.~A.,  {Steidel} C.~C.,   {Shapley} A.~E.,  2015,
  \mn@doi [\apj] {10.1088/0004-637X/804/2/149}, \href
  {http://adsabs.harvard.edu/abs/2015ApJ...804..149S} {804, 149}

\bibitem[\protect\citeauthoryear{{Smith}, {Safranek-Shrader}, {Bromm}  \&
  {Milosavljevi{\'c}}}{{Smith} et~al.}{2015}]{smith15}
{Smith} A.,  {Safranek-Shrader} C.,  {Bromm} V.,   {Milosavljevi{\'c}} M.,
  2015, \mn@doi [\mnras] {10.1093/mnras/stv565}, \href
  {https://ui.adsabs.harvard.edu/abs/2015MNRAS.449.4336S} {449, 4336}

\bibitem[\protect\citeauthoryear{{Smith}, {Ma}, {Bromm}, {Finkelstein},
  {Hopkins}, {Faucher-Gigu{\`e}re}  \& {Kere{\v{s}}}}{{Smith}
  et~al.}{2019}]{smith19}
{Smith} A.,  {Ma} X.,  {Bromm} V.,  {Finkelstein} S.~L.,  {Hopkins} P.~F.,
  {Faucher-Gigu{\`e}re} C.-A.,   {Kere{\v{s}}} D.,  2019, \mn@doi [\mnras]
  {10.1093/mnras/sty3483}, \href
  {https://ui.adsabs.harvard.edu/abs/2019MNRAS.484...39S} {484, 39}

\bibitem[\protect\citeauthoryear{{Smith}, {Kannan}, {Garaldi}, {Vogelsberger},
  {Pakmor}, {Springel}  \& {Hernquist}}{{Smith} et~al.}{2021a}]{smith21}
{Smith} A.,  {Kannan} R.,  {Garaldi} E.,  {Vogelsberger} M.,  {Pakmor} R.,
  {Springel} V.,   {Hernquist} L.,  2021a, arXiv e-prints, \href
  {https://ui.adsabs.harvard.edu/abs/2021arXiv211002966S} {p. arXiv:2110.02966}

\bibitem[\protect\citeauthoryear{{Smith} et~al.,}{{Smith}
  et~al.}{2021b}]{smith21_rt}
{Smith} A.,  et~al., 2021b, arXiv e-prints, \href
  {https://ui.adsabs.harvard.edu/abs/2021arXiv211113721S} {p. arXiv:2111.13721}

\bibitem[\protect\citeauthoryear{{Sparre}, {Hayward}, {Feldmann},
  {Faucher-Gigu{\`e}re}, {Muratov}, {Kere{\v s}}  \& {Hopkins}}{{Sparre}
  et~al.}{2017}]{sparre17_bursty}
{Sparre} M.,  {Hayward} C.~C.,  {Feldmann} R.,  {Faucher-Gigu{\`e}re} C.-A.,
  {Muratov} A.~L.,  {Kere{\v s}} D.,   {Hopkins} P.~F.,  2017, \mn@doi [\mnras]
  {10.1093/mnras/stw3011}, \href
  {http://adsabs.harvard.edu/abs/2017MNRAS.466...88S} {466, 88}

\bibitem[\protect\citeauthoryear{{Springel}}{{Springel}}{2010}]{springel10}
{Springel} V.,  2010, \mn@doi [\araa] {10.1146/annurev-astro-081309-130914},
  \href {http://adsabs.harvard.edu/abs/2010ARA%26A..48..391S} {48, 391}

\bibitem[\protect\citeauthoryear{{Springel} \& {Hernquist}}{{Springel} \&
  {Hernquist}}{2005}]{springel05}
{Springel} V.,  {Hernquist} L.,  2005, \mn@doi [\apjl] {10.1086/429486}, \href
  {http://adsabs.harvard.edu/abs/2005ApJ...622L...9S} {622, L9}

\bibitem[\protect\citeauthoryear{{Stanway}, {Eldridge}  \& {Becker}}{{Stanway}
  et~al.}{2016}]{stanway16}
{Stanway} E.~R.,  {Eldridge} J.~J.,   {Becker} G.~D.,  2016, \mn@doi [\mnras]
  {10.1093/mnras/stv2661}, \href
  {https://ui.adsabs.harvard.edu/abs/2016MNRAS.456..485S} {456, 485}

\bibitem[\protect\citeauthoryear{{Tacchella} et~al.,}{{Tacchella}
  et~al.}{2015}]{tacchella15}
{Tacchella} S.,  et~al., 2015, \mn@doi [\apj] {10.1088/0004-637X/802/2/101},
  \href {http://adsabs.harvard.edu/abs/2015ApJ...802..101T} {802, 101}

\bibitem[\protect\citeauthoryear{{Tacchella} et~al.,}{{Tacchella}
  et~al.}{2018}]{tacchella18_dust}
{Tacchella} S.,  et~al., 2018, \mn@doi [\apj] {10.3847/1538-4357/aabf8b}, \href
  {https://ui.adsabs.harvard.edu/abs/2018ApJ...859...56T} {859, 56}

\bibitem[\protect\citeauthoryear{{Tacchella}, {Forbes}  \&
  {Caplar}}{{Tacchella} et~al.}{2020}]{tacchella20}
{Tacchella} S.,  {Forbes} J.~C.,   {Caplar} N.,  2020, \mn@doi [\mnras]
  {10.1093/mnras/staa1838}, \href
  {https://ui.adsabs.harvard.edu/abs/2020MNRAS.497..698T} {497, 698}

\bibitem[\protect\citeauthoryear{{Thilker}, {Braun}  \& {Walterbos}}{{Thilker}
  et~al.}{2000}]{thilker00}
{Thilker} D.~A.,  {Braun} R.,   {Walterbos} R. A.~M.,  2000, \mn@doi [\aj]
  {10.1086/316852}, \href
  {https://ui.adsabs.harvard.edu/abs/2000AJ....120.3070T} {120, 3070}

\bibitem[\protect\citeauthoryear{{Trayford}, {Frenk}, {Theuns}, {Schaye}  \&
  {Correa}}{{Trayford} et~al.}{2019}]{trayford19}
{Trayford} J.~W.,  {Frenk} C.~S.,  {Theuns} T.,  {Schaye} J.,   {Correa} C.,
  2019, \mn@doi [\mnras] {10.1093/mnras/sty2860}, \href
  {http://adsabs.harvard.edu/abs/2019MNRAS.483..744T} {483, 744}

\bibitem[\protect\citeauthoryear{{Tsai} \& {Mathews}}{{Tsai} \&
  {Mathews}}{1995}]{tsai95}
{Tsai} J.~C.,  {Mathews} W.~G.,  1995, \mn@doi [\apj] {10.1086/175943}, \href
  {https://ui.adsabs.harvard.edu/abs/1995ApJ...448...84T} {448, 84}

\bibitem[\protect\citeauthoryear{{Vale Asari}, {Couto}, {Cid Fernandes},
  {Stasi{\'n}ska}, {de Amorim}, {Ruschel-Dutra}, {Werle}  \& {Florido}}{{Vale
  Asari} et~al.}{2019}]{vale-asari19}
{Vale Asari} N.,  {Couto} G.~S.,  {Cid Fernandes} R.,  {Stasi{\'n}ska} G.,  {de
  Amorim} A.~L.,  {Ruschel-Dutra} D.,  {Werle} A.,   {Florido} T.~Z.,  2019,
  \mn@doi [\mnras] {10.1093/mnras/stz2470}, \href
  {https://ui.adsabs.harvard.edu/abs/2019MNRAS.489.4721V} {489, 4721}

\bibitem[\protect\citeauthoryear{{Veilleux}, {Cecil}  \&
  {Bland-Hawthorn}}{{Veilleux} et~al.}{2005}]{veilleux05}
{Veilleux} S.,  {Cecil} G.,   {Bland-Hawthorn} J.,  2005, \mn@doi [\araa]
  {10.1146/annurev.astro.43.072103.150610}, \href
  {https://ui.adsabs.harvard.edu/abs/2005ARA&A..43..769V} {43, 769}

\bibitem[\protect\citeauthoryear{{Vogelsberger}, {Marinacci}, {Torrey}  \&
  {Puchwein}}{{Vogelsberger} et~al.}{2020}]{vogelsberger20_review}
{Vogelsberger} M.,  {Marinacci} F.,  {Torrey} P.,   {Puchwein} E.,  2020,
  \mn@doi [Nature Reviews Physics] {10.1038/s42254-019-0127-2}, \href
  {https://ui.adsabs.harvard.edu/abs/2020NatRP...2...42V} {2, 42}

\bibitem[\protect\citeauthoryear{{Walcher}, {Groves}, {Budav{\'a}ri}  \&
  {Dale}}{{Walcher} et~al.}{2011}]{walcher11}
{Walcher} J.,  {Groves} B.,  {Budav{\'a}ri} T.,   {Dale} D.,  2011, \mn@doi
  [\apss] {10.1007/s10509-010-0458-z}, \href
  {https://ui.adsabs.harvard.edu/abs/2011Ap%26SS.331....1W} {331, 1}

\bibitem[\protect\citeauthoryear{{Wang} \& {Lilly}}{{Wang} \&
  {Lilly}}{2020}]{wang20_p1}
{Wang} E.,  {Lilly} S.~J.,  2020, \mn@doi [\apj] {10.3847/1538-4357/ab7b7d},
  \href {https://ui.adsabs.harvard.edu/abs/2020ApJ...892...87W} {892, 87}

\bibitem[\protect\citeauthoryear{{Weinberger}, {Springel}  \&
  {Pakmor}}{{Weinberger} et~al.}{2020}]{weinberger20}
{Weinberger} R.,  {Springel} V.,   {Pakmor} R.,  2020, \mn@doi [\apjs]
  {10.3847/1538-4365/ab908c}, \href
  {https://ui.adsabs.harvard.edu/abs/2020ApJS..248...32W} {248, 32}

\bibitem[\protect\citeauthoryear{{Weingartner} \& {Draine}}{{Weingartner} \&
  {Draine}}{2001}]{weingartner01}
{Weingartner} J.~C.,  {Draine} B.~T.,  2001, \mn@doi [\apj] {10.1086/318651},
  \href {https://ui.adsabs.harvard.edu/abs/2001ApJ...548..296W} {548, 296}

\bibitem[\protect\citeauthoryear{{Weisz} et~al.,}{{Weisz}
  et~al.}{2012}]{weisz12}
{Weisz} D.~R.,  et~al., 2012, \mn@doi [\apj] {10.1088/0004-637X/744/1/44},
  \href {http://adsabs.harvard.edu/abs/2012ApJ...744...44W} {744, 44}

\bibitem[\protect\citeauthoryear{{Westfall} et~al.,}{{Westfall}
  et~al.}{2019}]{westfall19}
{Westfall} K.~B.,  et~al., 2019, \mn@doi [\aj] {10.3847/1538-3881/ab44a2},
  \href {https://ui.adsabs.harvard.edu/abs/2019AJ....158..231W} {158, 231}

\bibitem[\protect\citeauthoryear{{Wilkins} et~al.,}{{Wilkins}
  et~al.}{2020}]{wilkins20}
{Wilkins} S.~M.,  et~al., 2020, \mn@doi [\mnras] {10.1093/mnras/staa649}, \href
  {https://ui.adsabs.harvard.edu/abs/2020MNRAS.493.6079W} {493, 6079}

\bibitem[\protect\citeauthoryear{{Wood} \& {Reynolds}}{{Wood} \&
  {Reynolds}}{1999}]{wood99}
{Wood} K.,  {Reynolds} R.~J.,  1999, \mn@doi [\apj] {10.1086/307939}, \href
  {https://ui.adsabs.harvard.edu/abs/1999ApJ...525..799W} {525, 799}

\bibitem[\protect\citeauthoryear{{Yang} et~al.,}{{Yang} et~al.}{2020}]{yang20}
{Yang} Y.-L.,  et~al., 2020, \mn@doi [\apj] {10.3847/1538-4357/ab7201}, \href
  {https://ui.adsabs.harvard.edu/abs/2020ApJ...891...61Y} {891, 61}

\bibitem[\protect\citeauthoryear{{Zhang} et~al.,}{{Zhang}
  et~al.}{2017}]{zhang17_DIG}
{Zhang} K.,  et~al., 2017, \mn@doi [\mnras] {10.1093/mnras/stw3308}, \href
  {https://ui.adsabs.harvard.edu/abs/2017MNRAS.466.3217Z} {466, 3217}

\bibitem[\protect\citeauthoryear{{Zurita}, {Rozas}  \& {Beckman}}{{Zurita}
  et~al.}{2000}]{zurita00}
{Zurita} A.,  {Rozas} M.,   {Beckman} J.~E.,  2000, \aap, \href
  {https://ui.adsabs.harvard.edu/abs/2000A&A...363....9Z} {363, 9}

\bibitem[\protect\citeauthoryear{{da Silva}, {Fumagalli}  \& {Krumholz}}{{da
  Silva} et~al.}{2014}]{da-silva14}
{da Silva} R.~L.,  {Fumagalli} M.,   {Krumholz} M.~R.,  2014, \mn@doi [\mnras]
  {10.1093/mnras/stu1688}, \href
  {http://adsabs.harvard.edu/abs/2014MNRAS.444.3275D} {444, 3275}

\makeatother
\end{thebibliography}

\appendix

\section{Additional maps}
\label{appsec:maps}

\subsection{LMC maps}
\label{appsec:LMC_maps}

\begin{figure*}
\includegraphics[width=\linewidth]{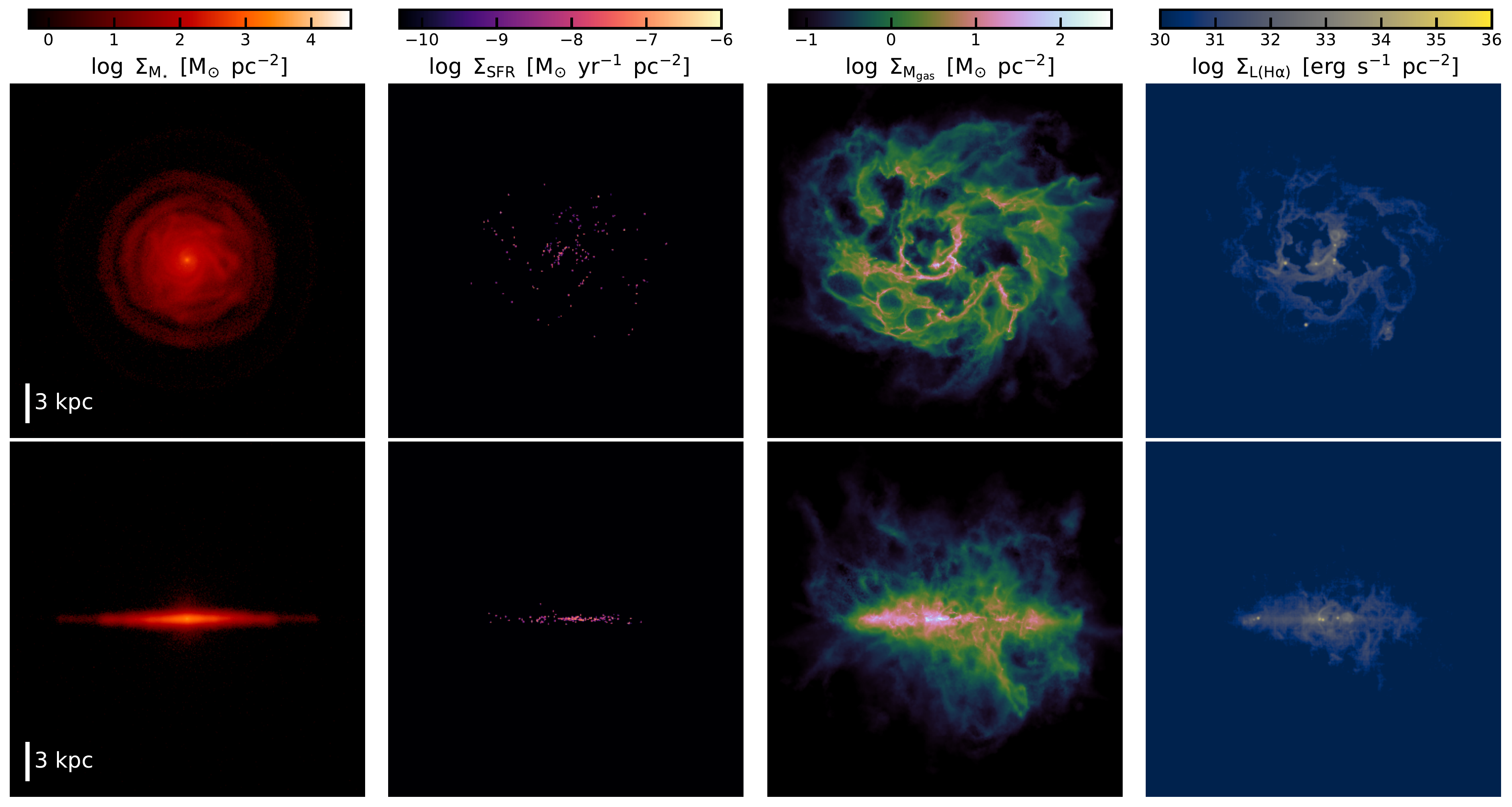}
\caption{LMC-BC03 simulation at 713 Myr. This figure follows the same layout as Fig.~\ref{fig:MW_maps}: maps of stellar mass, SFR (averaged over the past 100\,Myr), gas mass and H$\alpha$ luminosity are shown from left to right. The stellar mass, SFR, and H$\alpha$ are smoothed by 50 pc in order to increase visibility. The top and bottom panels show face-on and edge-on projections, respectively. The dimensions of each map are $30\times30$ kpc (the ruler in the bottom left indicates 3 kpc).}
\label{appfig:LMC_maps}
\end{figure*}

For completeness, we plot in Fig.~\ref{appfig:LMC_maps} to maps of stellar mass, SFR, gas mass and H$\alpha$ luminosity for the LMC-BC03 simulation.

\subsection{Effect of smoothing}
\label{appsec:smoothing}

\begin{figure*}
\includegraphics[width=0.47\linewidth]{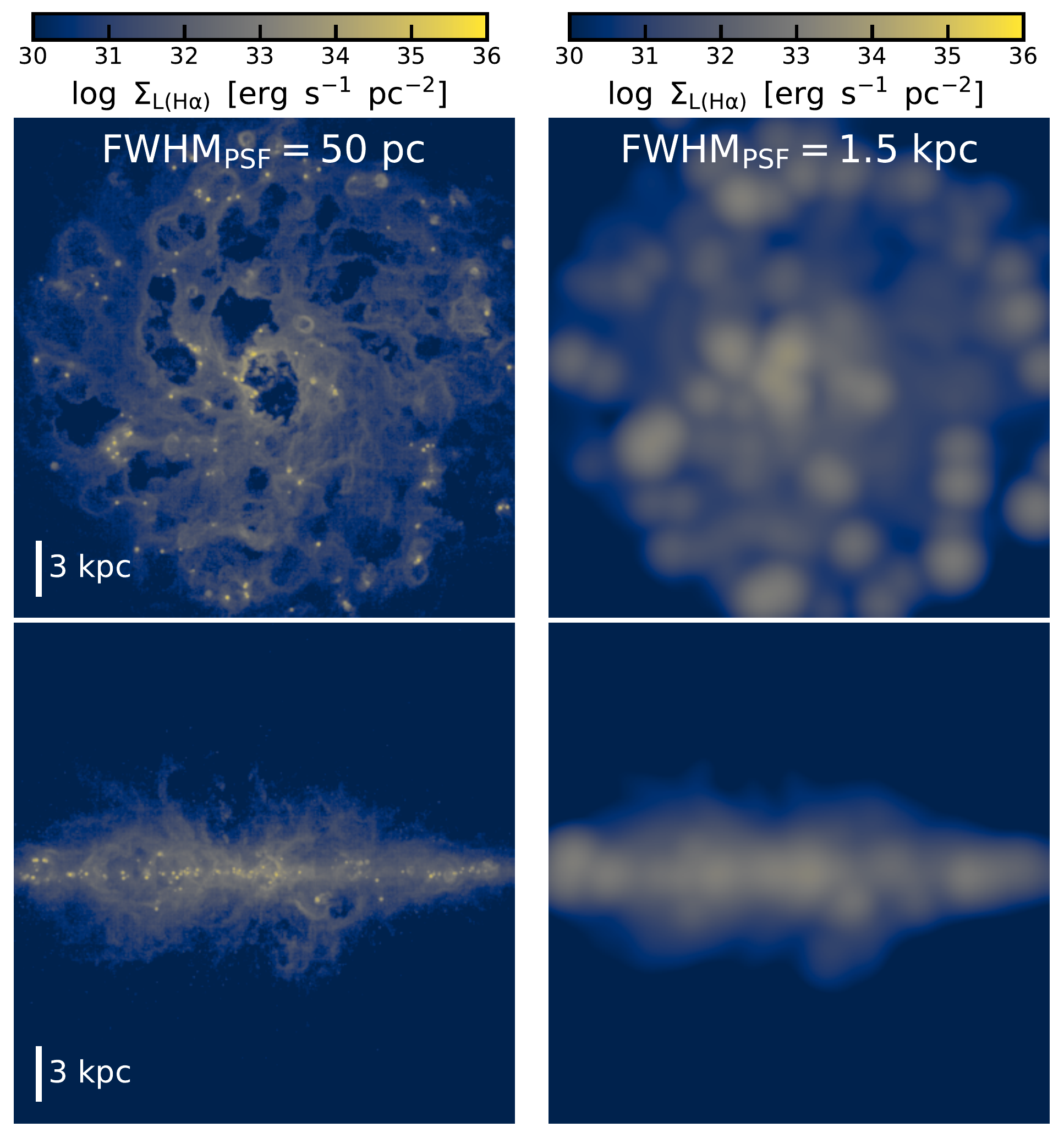}
\hspace{0.05\linewidth}
\includegraphics[width=0.47\linewidth]{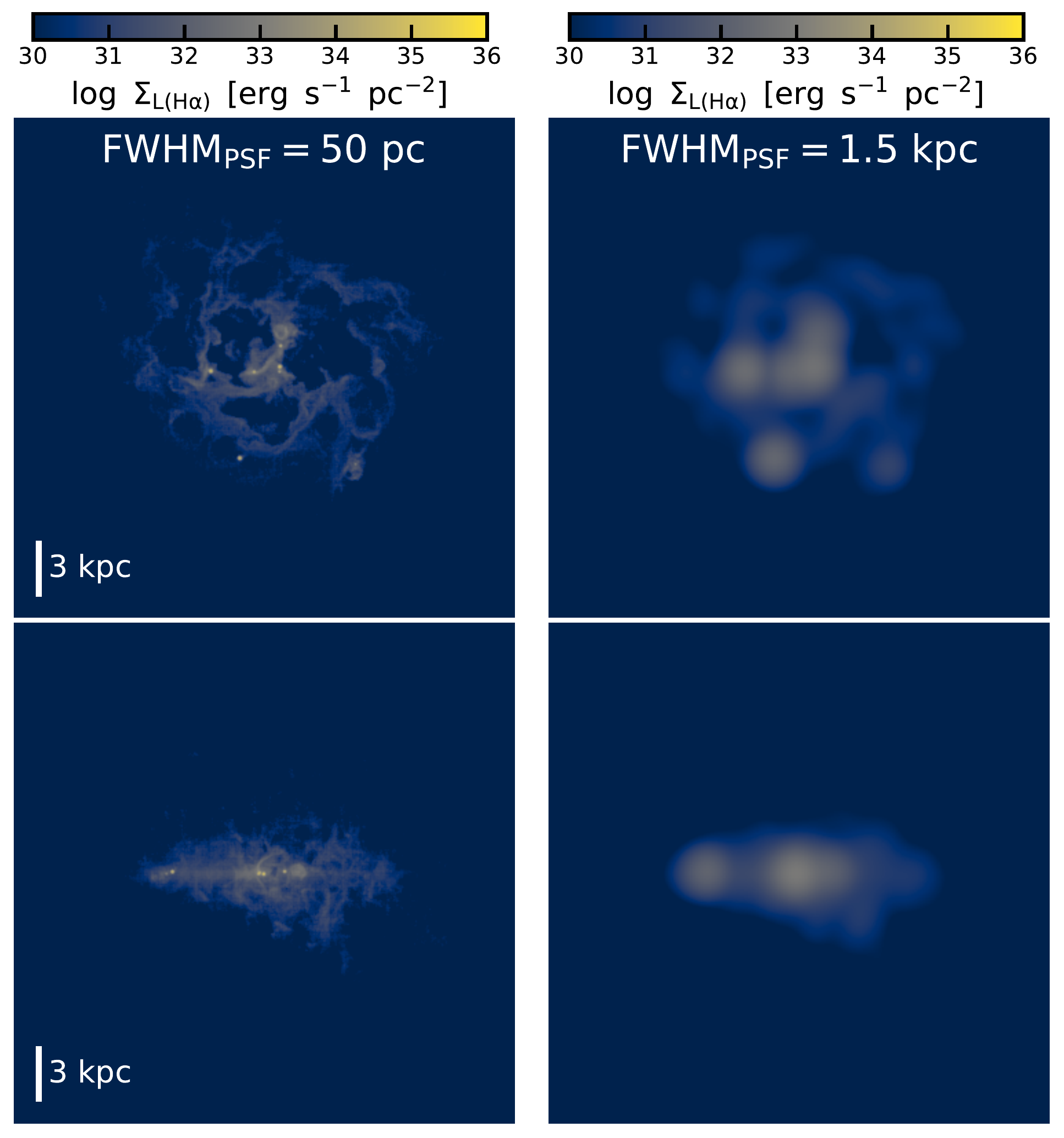}
\caption{Effect of smoothing. We plot from left to right the H$\alpha$ surface brightness maps of the MW with a PSF full width at half maximum (FWHM) of 50 pc, the MW with PSF FWHM of 1.5 kpc, the LMC-BC03 with PSF FWHM of 50 pc, and the LMC-BC03 with PSF FWHM of 1.5 kpc. The top panels show the face-on projections, while the bottom panels show the edge-on projections.}
\label{appfig:smoothing}
\end{figure*}

Fig.~\ref{appfig:smoothing} shows how different resolutions affect the visual appearance of the H$\alpha$ surface brightness maps. As discussed in Section~\ref{subsec:DIG_def}, we focus on a resolution of 50 pc and 1.5 kpc, meaning that the full width at half maximum (FWHM) of the PSF has a size of 50 pc and 1.5 kpc. This corresponds to roughly to resolution of MANGA and MUSE for nearby galaxies.

\section{Ionizing radiation from different stellar population models}
\label{appsec:stellar_pop}

\begin{figure*}
\includegraphics[width=\linewidth]{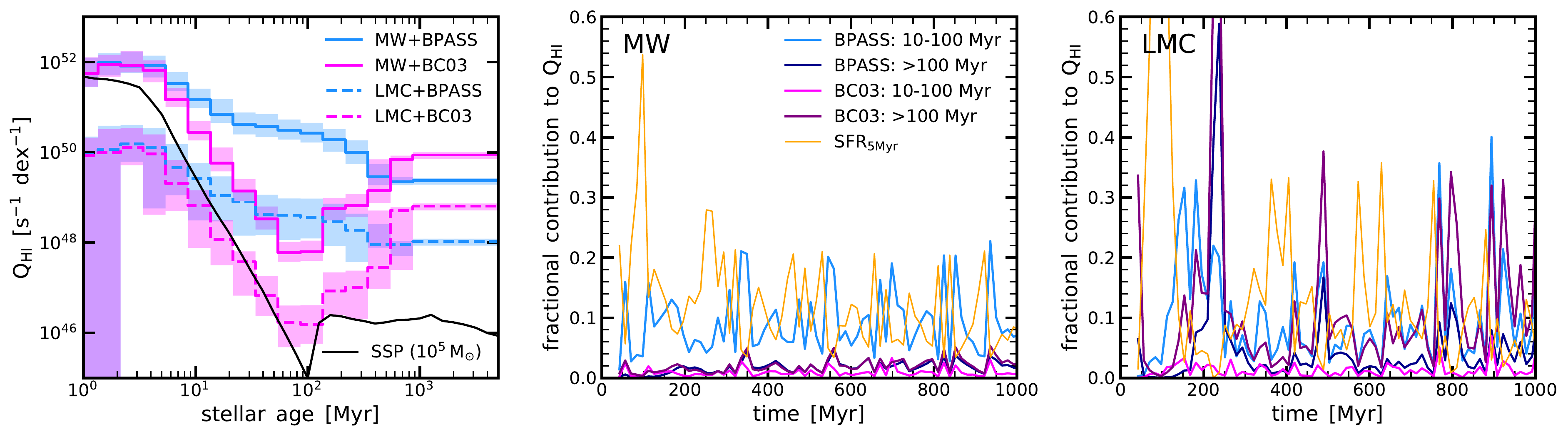}
\caption{Stellar population dependence of the H-ionizing Lyman continuum (LyC) photon production rate. \textit{Left}: The histogram shows the number of LyC photons as a function of stellar age, reflecting both the evolution of individual simple stellar populations (SSPs) as well as the star-formation history (SFH) of the MW simulation (solid lines) and LMC simulation (dashed lines). For reference, the black solid line marks the evolution of a BC03 SSP with an intial mass of $10^5~\text{M}_{\odot}$. The blue and pink lines show the BPASS \citep{eldridge09} and BC03 \citep{bruzual03} stellar synthesis models. Specifically, the solid line and shaded region mark the median and the $16$--$84^\text{th}$ percentile across all simulation snapshots. The constancy beyond $\sim500$\,Myr is caused by the initialisation of the simulation, where all old stars have a single age of 5\,Gyr. \textit{Middle} and \textit{right}: Fractional contribution of older stars to the H-ionizing LyC flux emission as a function of time for the MW SFH (middle panel) and LMC SFH (right panel). The pink and purple (light and dark blue) lines indicate the fraction of LyC photons that stem from $10$--$100$\,Myr and $>100$\,Myr old stars in the BC03 (BPASS) models. The orange curve shows the normalised SFR averaged over 5\,Myr. For the SFH of the MW simulation, we find that LyC photons are mainly produced by young stars ($<10$\,Myr old stars) both in BC03 and BPASS models, though in the BPASS model intermediate-age stars ($10$--$100\,\mathrm{Myr}$) contribute more significantly ($\sim10\%$) to the ionizing budget than in the BC03 model. The contribution fraction of intermediate-age stars peaks (up to $\sim20\%$) shortly after a burst of star formation. For the LMC SFH, the contribution of intermediate-age (and even older stars) stars varies more and peaks at larger values.}
\label{appfig:Nion}
\end{figure*}

The intrinsic production rate of H-ionizing Lyman continuum (LyC) photons of a galaxy depends on the IMF, the star-formation history (SFH), and the stellar population synthesis model. For the discussion here and throughout the paper, we fix the IMF to the \citet{chabrier03} IMF with an upper mass limit of $100~\text{M}_{\odot}$. We investigate the effect of the SFH and two stellar models: the Binary Population and Spectral Synthesis model \citep[BPASS;][]{eldridge09} and the \citet[][denoted as BC03]{bruzual03} model. Specifically, for each snapshot, we adopt the masses, ages, and metallicities of the stars of the MW and LMC simulations and convolve those with the BC03 and the BPASS models to compute the intrinsic production rate of LyC photons.

Fig.~\ref{appfig:Nion} investigates the stellar age dependence of the LyC photon production rate. Specifically, the left panel plots the histogram of the number of LyC photons ($\mathrm{Q}_{\HI}$) as a function of stellar age, reflecting both the evolution of individual simple stellar populations (SSPs) as well as the SFH of the MW simulation (solid lines) and LMC simulation (dashed lines). As a reference, we also indicate the evolution of a BC03 SSP with an initial mass of $10^5~\text{M}_{\odot}$ as a black line. The blue and pink lines show the BPASS and BC03 stellar synthesis models. Specifically, the solid line and shaded region mark the median and the $16$--$84^\text{th}$ percentile across all simulation snapshots. The constancy beyond $\sim500$\,Myr is caused by the initialisation of the simulation, where all old stars have a single age of 5\,Gyr.

The left panel of Fig.~\ref{appfig:Nion} clearly shows that young stars with ages of $<10$\,Myr dominate production of LyC photons for both BC03 and BPASS. However, $\mathrm{Q}_{\HI}$ declines faster with increasing stellar age for BC03 than for BPASS, implying that intermediate-age stars ($10$--$200\,\mathrm{Myr}$) contribute more significantly in BPASS to $\mathrm{Q}_{\HI}$ than in BC03. After the main sequence stars have died, post-AGB stars produce most of the ionizing flux in the older stellar populations ($>100\,\mathrm{Myr}$), which explains the upturn at older ages for the BC03 models and the constancy in the case of the BPASS model \citep[e.g.,][]{byler19}. 

Comparing the MW to the LMC simulation, we find a large ($1-2$ magnitudes) normalisation difference because of the higher stellar mass and SFR of the MW simulation  than the LMC simulation (Table~\ref{tab:simulation}). This normalisation difference is stellar age dependent mainly because of the different SFH. The ratio of younger versus older stars is larger for the MW than the LMC, which can be directly seen looking at the specific SFR ($4\times10^{-11}\,\mathrm{yr}^{-1}$ for the MW versus $1\times10^{-11}\,\mathrm{yr}^{-1}$ for the LMC). An additional effect is the metallicity difference between the MW and LMC simulation, though this effect plays only a minor role \citep[e.g.,][]{choi17, byler17}. 

The middle and right panels of Fig.~\ref{appfig:Nion} show the contribution of intermediate (10--100 Myr) and old ($>100$\,Myr) stars to $\mathrm{Q}_{\HI}$ for the BPASS and BC03 models. For reference, the SFHs (SFR$_5$) are shown as orange lines, linearly re-scaled by a factor of $0.1/\langle \mathrm{SFR}\rangle$ (see Table~\ref{tab:simulation} for the MW and LMC values of $\langle \mathrm{SFR}\rangle$). For the SFH of the MW (middle panel), we find that in the case of BC03 the contribution to $\mathrm{Q}_{\HI}$ of both intermediate and old stars is negligible with $0.6\%$ and $1.8\%$, respectively. In the case of BPASS, the fractional contribution of intermediate-age stars is significant with $7_{-3}^{+8}\%$, while the contribution of old stars remains low ($1.6\%$). The contribution fraction of intermediate-age stars peaks (up to $\sim20\%$) shortly after a burst of star formation.

The right panel of Fig.~\ref{appfig:Nion} shows the results for the SFH of the LMC. As discussed in Section~\ref{subsec:time_evolution}, the SFH is more variable for the LMC than the MW, i.e. there are periods in the SFH of the LMC where the SFR$_5$ is $<10^2~\mathrm{M}_{\odot}~\mathrm{yr}^{-1}$, typically proceeding a burst of star formation. In these phases of little-to-no star formation, the LyC production rate of young stars is very low (see also extended percentiles to low $\mathrm{Q}_{\HI}$ values in the left panel), leading to a boost in the relative contribution of intermediate and old stars. Specifically, in the BC03 case, we find a median contribution from old stars of $7_{-4}^{+12}\%$, but with short phases where the contributions amount to more than 30\%. intermediate-age stars still only contribution on the level of $\sim1\%$. For the BPASS model, the situation is flipped: intermediate-age stars contribute more prominently with $8_{-5}^{+10}\%$, while the contribution from old stars is smaller ($2_{-1}^{+6}\%$).

In summary, even when just studying the production rate of H-ionizing LyC photons without any radiative transfer effects, we see that changes in the stellar models (BC03 versus BPASS) and the SFH can lead to drastically different outcomes regarding the contribution of older stellar populations to the ionizing flux \citep[see also, e.g.,][]{choi17, rosdahl18, ma20}.


\bsp	
\label{lastpage}
\end{document}